\documentclass[aps,prd,showpacs,notitlepage,nofootinbib,preprintnumbers,amsmath,amssymb]{revtex4-1}

\usepackage{graphics,graphicx}
\usepackage{dcolumn}
\usepackage{bm}
\usepackage{epsfig}
\usepackage[usenames]{color}
\usepackage{hyperref} 
\usepackage{mathbbol}
\usepackage{mathtools}
\usepackage{epstopdf}
\usepackage{simplewick}
\usepackage[utf8]{inputenc} 	
\usepackage{slashed}
\usepackage{stackrel}
\usepackage[normalem]{ulem}

\def\eq#1{{Eq.~(\ref{#1})}}

\newcommand{\tr}{\mbox{tr}}

\newcommand{\bra}[1]{\left\langle #1 \right|}
\newcommand{\ket}[1]{\left| #1 \right\rangle}

\newcommand{\ul}[1]{\underline{#1}}

\newcommand{\ubar}[1]{\overline{U}_{#1}}

\newcommand{\ord}[1]{\mathcal{O}\left( #1 \right)}
\newcommand{\gPM}{g^{+-}}

\newcommand{\half}{\frac{1}{2}}

\begin{document}
\title{Quark branching in QCD matter to any order in opacity \\ beyond the soft gluon emission limit}
\author{Matthew D. Sievert}
  \email[Email: ]{sievertmd@lanl.gov}
  \affiliation{Theoretical Division, Los Alamos National Laboratory, Los Alamos, NM 87545, USA}
\author{Ivan Vitev}
  \email[Email: ]{ivitev@lanl.gov}
  \affiliation{Theoretical Division, Los Alamos National Laboratory, Los Alamos, NM 87545, USA}
\date{\today}
%
%
\begin{abstract}
Cold nuclear matter effects in reactions with nuclei at a future electron-ion collider (EIC) lead to a modification of semi-inclusive hadron production, jet cross sections, and  jet substructure  when compared to the vacuum.  At leading order in the strong coupling, a jet produced at an EIC is initiated as an energetic quark, and the process of this quark splitting into a quark-gluon system underlies experimental observables.  The spectrum of gluons associated with the branching of this quark jet is heavily modified by multiple scattering in a medium, allowing jet cross sections and jet substructure to be used as a probe of the medium's properties.  We present a formalism that allows us to compute the gluon spectrum of a quark jet  to an arbitrary order in opacity, the average number of scatterings in the medium.  This calculation goes beyond the simplifying limit in which the gluon radiation is soft and can be interpreted as energy loss of the quark, and it significantly extends previous work which computes the full gluon spectrum only to first order in opacity. The theoretical  framework demonstrated here applies equally well to light parton and heavy quark branching,  and is easily generalizable to all in-medium splitting processes. 
\end{abstract}
%
%
%
\maketitle
%

%
\section{Introduction}
%

The attenuation of the production cross section of energetic particles and jets in high energy reactions with nuclei is one of the primary signatures of inelastic parton scattering in dense nuclear matter~\cite{Gyulassy:1990ye,Gyulassy:2003mc}. The rapid development of heavy ion programs at fixed target and collider experiments fueled tremendous interest in medium-induced bremsstrahlung processes and radiative parton energy loss in Quantum Chromodynamics (QCD)~\cite{Gyulassy:1993hr}, often discussed in analogy with the Landau-Pomeranchuck-Migdal effect for photon emission in Quantum Electrodynamics (QED)~\cite{Landau:1953um,Migdal:1956tc}. To this effect, initial efforts have focused on the energy loss of energetic quarks and gluons as they propagate in the quark-gluon
plasma (QGP), a deconfined state of strongly-interacting matter that existed in the early universe and is recreated today in relativistic heavy ion collisions. Radiative energy loss in QCD is synonymous with soft gluon bremsstrahlung, a process in which hard quarks and gluons shed energy in small quanta during propagation through matter. As a result, the leading parton always remains the most energetic. This does not preclude the possibility  that it may  dissipate a sizable fraction of its  energy, but this is achieved through multiple gluon emission. All radiative parton energy loss approaches rely on perturbative techniques and treat the interactions of the jet with the quasi-particles of the medium primarily through $t$-channel  gluon exchanges~\cite{Gyulassy:1993hr}.

Important considerations, where various  approaches available in the   literature deviate,  are the kinematic regimes in which the parton system is produced and  the size of the nuclear medium.  To give an example, let us denote by $ l_f $ the quantum-mechanical time for the splitting process or gluon emission to occur, by $\lambda$ the scattering length of the parent parton, and by $L$ the size of the medium. In practice, it is well understood that none of those quantities are fixed. Scattering lengths depend on the density and dynamics of nuclear matter, and the size of the realistic medium cannot be defined in a model-independent fashion. The time for the splitting process to take place depends sensitively on the  kinematics and interactions in  the medium. If one considers the simultaneous limit   $l_f/\lambda \gg 1$ and $L / l_f  \gg 1 $, a continuous approximation to the process of parton system propagation can be made, where a Schroedinger-like equation can be solved based on a Gaussian approximation~\cite{Baier:1996sk,Baier:1998kq}. Similarly, this limit can be treated in a path integral approach~\cite{Zakharov:1996fv,Zakharov:1997uu} where a Gaussian approximation is again essential to obtain a tractable resummed result.  A hard thermal loop approach to the problem of gluon emission has also been developed~\cite{Arnold:2002ja}.

The situation where the number of scatterings  in an energetic parton system is very large is not often encountered in reactions with heavy nuclei~\cite{Gyulassy:1999zd}. Even more importantly, the splitting time  which is the inverse virtuality of the two parton system  
before and after the  branching  $ 0 <  l_f < \infty $ is different for different diagrams and does not obey any particular hierarchy with respect to fixed length scales. This motivates the development of formalisms where the exact kinematics of parton propagation in matter are kept~\cite{Gyulassy:2000fs,Gyulassy:2000er}. In this spirit, the result for gluon bremsstrahlung is written as a series in the correlations between multiple scattering 
centers, or the opacity.  In the calculation of more inclusive jet quenching observables, such as the suppression of the inclusive hadron spectrum, the contribution of higher orders in opacity is power suppressed. This is most easily seen
in the high twist approach~\cite{Wang:2001ifa,Guo:2006kz}. Differences can be expected, however, for observables that are more sensitive to the double differential splitting kernels,  such as the ones related to jet substructure~\cite{Chien:2015hda,Kang:2016ehg}. By this we mean more exclusive observables that depend on the longitudinal momentum sharing and transverse momentum or angular separation of the daughter partons in the splitting. The evaluation of higher orders in opacity can become numerically expensive, even in the simplifying limit of soft gluon emission. It was shown, however, that in certain  simplifying geometries corrections up 9$^{\rm th}$ order in opacity can be evaluated, albeit with decreasing numerical accuracy~\cite{Wicks:2008ta}.  It was pointed out that the path integral approach, without the approximations above, is formally equivalent to the opacity series and can be re-expanded into diagrams to recover the opacity  expansion~\cite{Wiedemann:2000za}.  Finite size effects have been introduced in the  hard thermal loop approach as  well~\cite{CaronHuot:2010bp}.

The popular energy loss approach has been extended to account for the mass of heavy partons~\cite{Dokshitzer:2001zm,Djordjevic:2003zk,Zhang:2003wk,Arleo:2012rs}.  Initial-state inelastic processes have also been discussed~\cite{Vitev:2007ve,Xing:2011fb}, first motivated by fixed target  experiments~\cite{Neufeld:2010dz}. Higher order corrections to the jet-medium  interactions~\cite{Ghiglieri:2015ala}  and  soft splitting interference have been studied~\cite{MehtarTani:2011tz,Casalderrey-Solana:2015bww}.   Last but not least, soft photon emission has been calculated using techniques similar  to the ones described  above~\cite{Arnold:2002ja,Zhang:2003bsa,Majumder:2007ne,Vitev:2008vk}.

Advances in the theoretical understanding and  experimental measurements of  reconstructed jets  necessitate  more precise control over in-medium branching processes. This requires the development of theories of hard-probe production in the presence of nuclear matter that transcend the thirty-year-old energy loss approach. An important step in this direction is to obtain in-medium splitting kernels beyond the soft gluon  approximation. The corrections to vacuum branching in the higher twist approach were discussed in Ref.~\cite{Wang:2009qb}.  A full set of in-medium splitting kernels,  $q\rightarrow qg$,  $g\rightarrow gg$, $q\rightarrow gq$ $g\rightarrow q \bar{q}$,  to first order in opacity was derived in Ref.~\cite{Ovanesyan:2011kn} in an effective theory of jet propagation using Glauber gluon interactions sourced by an external  potential~\cite{Idilbi:2008vm,Ovanesyan:2011xy}. The $\lambda \ll l_f \ll L$ limit has also been generalized to full longitudinal splitting kinematics for the  $g\rightarrow gg$  process~\cite{Blaizot:2012fh, Apolinario:2014csa}.  The energy loss limit has also been
relaxed in Ref.~\cite{Blagojevic:2018nve}. Multiple parton branching beyond the soft gluon approximation has been rigorously calculated in Ref.~\cite{Fickinger:2013xwa}.   To address heavy flavor observables in ultrarelativistic nuclear collisions, the heavy flavor splitting kernels  $Q\rightarrow Qg$, $Q\rightarrow gQ$ $g\rightarrow Q \bar{Q}$ were obtained to first order in opacity~\cite{Kang:2016ofv}.  It is important to realize that to bridge the gap between high energy particle physics and high energy nuclear physics and to reduce the systematic uncertainties inherent in phenomenological models of jet quenching,  resummation~\cite{Chang:2014fba,Kang:2014xsa,Chien:2015vja,Li:2017wwc} and  higher order calculations~\cite{Kang:2016ofv,Kang:2017frl}  are necessary.  These are not only facilitated by, but, in fact, require the implementation of the full in-medium splitting functions.

In this paper, we focus on the calculation of the gluon emission spectrum beyond the soft gluon approximation to an arbitrary order in opacity,  with semi-inclusive deep inelastic scattering  (SIDIS) processes in mind. We will illustrate the technique on the example of a 
quark jet. This is the dominant channel at leading order, and in the kinematic regions of interest the next-to-leading order corrections to the flavor composition of jets are small.  Our work sets the stage for the calculation of all in-medium splittings to high orders in opacity and for related phenomenology.  We write down explicitly the result for the quark to quark + gluon splitting channel, including light and heavy quarks,  to second order in opacity and expect that our results will find application in the description of 
semi-inclusive hadron  suppression~\cite{Chang:2014fba,Wang:2002ri,Accardi:2002tv,Arleo:2003jz,Kopeliovich:2003py} and further expand the program at a future EIC~\cite{Accardi:2012qut}.

Although we approach the problem in the standard framework of Feynman perturbation theory, the Fourier transformation to work with fixed longitudinal positions of the scattering centers naturally brings in many of the intuitive elements of Light-Front Perturbation Theory (LFPT) \cite{Brodsky:1997de, Lepage:1980fj}.  As such, the parton branching will be formulated in terms of light-front wave functions (LFWF) and the associated energy denominators which describe a fluctuation in the virtuality of an intermediate state; these quantities possess clear and intuitive kinematic dependences because of the Galilean invariance of the light-front Hamiltonian~\cite{Brodsky:1997de, Blaizot:2012fh}.  Light-front dynamics is a natural description for many high-energy processes, and in that context, the phases due to multiple scattering which stimulate radiative energy loss are also crucial for the construction of single spin asymmetries in QCD~\cite{Kovchegov:2012ga, Brodsky:2013oya, Collins:2002kn}.  Quark jet production at an EIC is also intimately related with the transverse-momentum-dependent parton distribution functions of a heavy nucleus, which have been previously explored~\cite{Kovchegov:2015zha, Kovchegov:2013cva}.

Our work is organized as follows:  in Sec.~\ref{sec:DIS} we set up the calculation of the gluon-in-quark-jet distribution in SIDIS, defining our frame and coordinates and laying out the origin of medium modifications to this distribution.  In Sec.~\ref{sec:Modification} we formulate the coupling of the jet to the medium at the Lagrangian level, using this to compute both the vacuum distribution and the modification from the medium at first order in opacity.  The results obtained here utilize the language of LFPT, reproducing and generalizing some previous results in the literature.  In Sec.~\ref{sec:React} we generalize the first order in opacity results to construct the kernel of a recursion relation, which is used to derive expressions for the gluon-in-quark-jet distribution at any arbitrary order in opacity.  To illustrate the use of this recursion relation, in Sec.~\ref{sec:Results} we explicitly compute for the first time the exact distribution to second order in opacity, verifying that it reproduces the known result in the literature in the soft-gluon approximation.  The primary new results of this work are the recursion relation \eq{e:reactmtx} with the accompanying kernel (the ``reaction operator'') \eq{e:reactker} and the explicit second order in opacity results Eqs.~\eqref{e:NLO1} and \eqref{e:NLOexact}.  Finally, in Sec.~\ref{sec:Concl} we conclude by summarizing our key results and outlining the applications which we intend to pursue in future work.

%
\section{Cold Nuclear Matter Effects in Deep Inelastic Scattering}
\label{sec:DIS}
%

Consider the process of  jet production in Semi-Inclusive Deep Inelastic Scattering (SIDIS) on a heavy nucleus, $\gamma^* + A \rightarrow (jet) + X$.  We will work in the Breit frame, in which the virtual photon travels along the $+z$ axis with a large momentum $p^+$ and the nucleus travels along the $-z$ axis with a large momentum $p_N^-$ per nucleon.  Here and throughout this paper, we employ light-front coordinates $v^{\pm} \equiv \sqrt{\frac{\gPM}{2}} ( v^0 \pm v^3)$ and denote transverse vectors as $\ul{v} \equiv (v_\bot^1 , v_\bot^2)$ with magnitudes $v_T \equiv | \ul{v} |$.  Different references adopt different conventions for the normalization of the light-front coordinates; here we will choose $\gPM = 1$, although it is not uncommon to see the convention $\gPM = 2$ in the literature.  In this frame, the momentum of the virtual photon and struck nucleon are
\begin{align}
q_\gamma^\mu &= \Big( p^+ \, , \, - x_B p_N^- \, , \, \ul{0} \Big) \, , \qquad 
p_N^\mu = \Big(0^+ \, , \, p_N^- \, , \, \ul{0} \Big) ,
\end{align}
with the Bjorken variable $x_B = \frac{Q^2}{2 p_N \cdot q_\gamma} = \frac{Q^2}{2 p^+ p_N^-}$.  The photon-nucleon center-of-mass energy $\sqrt{s_\gamma}$ is given by $s_\gamma = (p_N + q_\gamma)^2 = 2 (1-x_B) p^+ p_N^-$, and for power counting purposes we will consider $p^+ \sim p_N^- \sim \ord{Q}$; that is, a situation close to the center-of-mass frame.  
\footnote{We note that similar derivations in the rest frame of the medium can look quite different from the one we perform here.  In particular, the poles which appear and are enclosed during integration can differ significantly between the two frames.  However, the final results ultimately agree, and the fundamental ingredients, the LFWF and the phases $\Delta \phi \sim \Delta E^- \, z^+$ are manifestly boost invariant.}
The jet-nucleon center-of-mass energy $\sqrt{s}$ is related to the photon-nucleon center-of-mass energy by $s = 2 p^+ p_N^- = s_\gamma / (1-x_B)$.  We will also work in the light-cone gauge $A^+ = 0$ for which only physical gluon polarizations exist; this gauge is also convenient because it is equivalent to Feynman gauge with eikonal accuracy.

%
\begin{figure}[ht]
\begin{center}
\includegraphics[width= \textwidth]{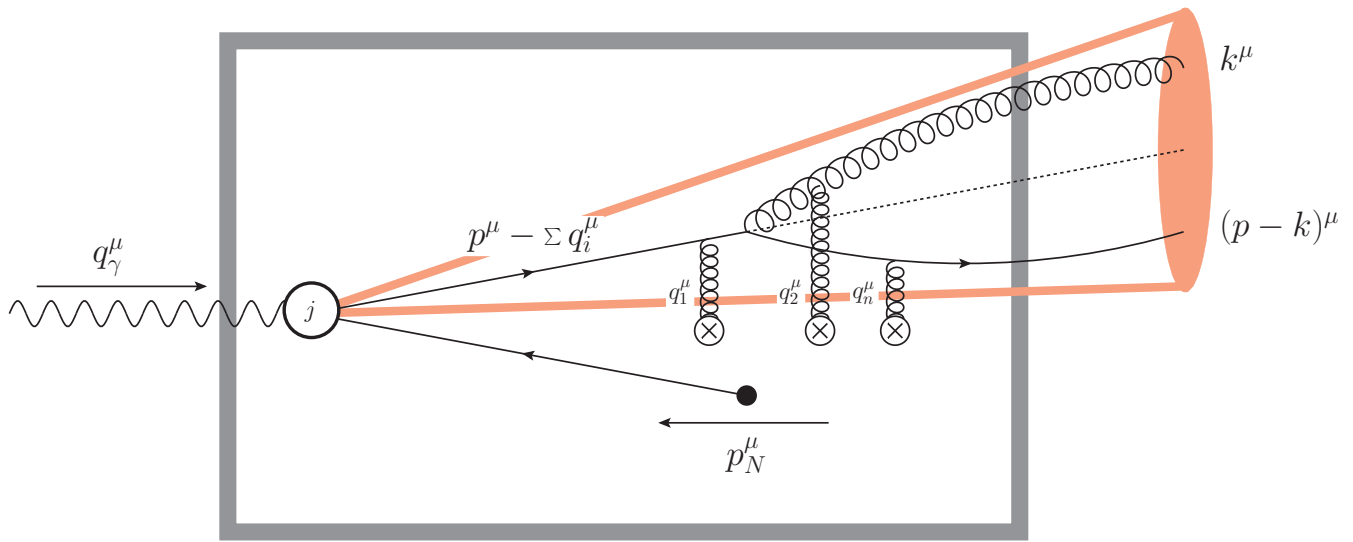} 
\caption{(Color online) Illustration of the indicated jet kinematics for SIDIS in the Breit frame.  The dark box represents the medium (nucleus) and the red cone represents the jet.}
\label{f:Jet_Kinematics}
\end{center}
\end{figure}
%

This setup is illustrated in Fig.~\ref{f:Jet_Kinematics}.  At leading order in the strong coupling, the virtual photon strikes a quark inside one of the nucleons, ejecting it with the large longitudinal momentum $p^+$ along the $+z$ axis.  This high-energy quark propagates through the rest of the medium, potentially undergoing multiple scattering in the process, before emerging in the final state as a quark jet with momentum $p$.  For a simple process like SIDIS, factorization theorems have been proven which express the inclusive jet production cross-section $p^+ \frac{d\sigma}{d^2 p \, dp^+}$ in terms of parton distribution functions (PDFs): TMD factorization \cite{Collins:1989gx, Collins:1981uk} at low $p_T$ which is sensitive to the nonperturbative transverse momentum distribution within the nucleus, and collinear factorization (see e.g. \cite{Collins:2011zzd}) at high $p_T$ which generates this momentum by a perturbative hard recoil.  Note that the jet transverse momentum $p_T$ has a significantly different interpretation for jets produced by SIDIS in the Breit frame than for heavy-ion collisions.  In the latter, the jet $p_T$ is a hard scale measuring the total energy of the jet, whereas in the former, the jet energy corresponds to $p^+$ and flows primarily in the longitudinal direction, with $p_T$ being a potentially soft scale.

In this work, we will focus on the effects of multiple scattering on the jet and on its substructure, so it is sufficient to regard the interaction of the virtual photon with the nucleus as a source current $j$ which creates a high-energy quark jet at some point in the medium.  For SIDIS, this current can be expressed directly in terms of the PDFs, but this abstraction also allows us to apply the general discussion of medium modification to other processes, such as heavy-ion collisions, in which the medium through which the jet propagates can be very different from the cold nuclear matter present in SIDIS.

Measurements of jet substructure  \cite{Chatrchyan:2013kwa,Sirunyan:2017bsd,Kauder:2017cvz,Sirunyan:2018jqr,Aaboud:2018hpb}  are more sensitive  to the  detailed QCD dynamics of  in-medium parton branching  than the overall inclusive or tagged  jet production cross-sections, though the latter are larger and more accurately measured at present~\cite{Aad:2012vca,Abelev:2013kqa,Aad:2014bxa,Khachatryan:2016jfl,Adamczyk:2017yhe,Sirunyan:2017jic,Aaboud:2018twu}.  At leading order in the strong coupling, the first contribution to the substructure of the quark jet comes from its splitting into a quark-gluon system.  Thus, we will consider in particular the distribution of gluons within the quark jet 
\begin{align}
x p^+ \, \frac{dN}{d^2 k \, dx \, d^2 p \, dp^+} 
\hspace{1cm} \mathrm{for} \hspace{1cm}
\gamma^* + A \rightarrow \Big( g (k) \in (q-jet)(p) \Big) + X \, , 
\end{align}
with $k^\mu$ the momentum of the gluon, $p^\mu$ the total momentum of the quark jet, and $x \equiv k^+ / p^+$ the longitudinal momentum fraction of the gluon within the jet.  At leading twist in the $1/Q$ expansion, the gluon will be emitted as in the vacuum, without modification from the medium.  Such splittings, if integrated out, contribute to the leading-logarithmic Dokshitzer-Gribov-Lipatov-Altarelli-Parisi~\cite{Altarelli:1977zs}  or Collins-Soper-Sterman evolution  of the underlying quark PDF~\cite{Collins:1989gx, Kovchegov:2015zha}.  For this vacuum substructure of the quark jet to be modified by the medium, the gluon must be emitted from an interaction {\it{inside}} the medium itself.  Because the medium is highly Lorentz-contracted in Bjorken kinematics, these modifications are necessarily suppressed by the Lorentz gamma factor and are intrinsically higher-twist effects. 

Let us, for the sake of discussion, take nuclear matter of fixed size and transport properties. If the length a parton propagates in the medium before escaping is $L$ and its mean free path is $\lambda \equiv \frac{1}{\rho \sigma_{el}}$, then the ratio of these quantities $\chi \equiv L / \lambda$ is referred to as the opacity: the average number of  scatterings which that parton will undergo.  (Equivalently, one can boost all these scales from the rest frame of the medium to the Breit frame, writing $\chi = L^+ / \lambda^+$.)  The mechanisms by which scattering in the medium will modify the jet substructure are generally classified as collisional versus radiative energy loss.  Collisional energy loss refers to the higher-twist corrections which generate a direct transfer of energy between the jet and the medium; these effects, and other higher-twist corrections, generically enter at $\ord{\frac{\bot^2}{Q^2}} \sim \ord{\frac{\bot^2}{s}}  $ with $\bot^2$ some associated transverse momentum scale and $Q^2=x_B s$ in DIS.  We note that there are other sub-eikonal  or higher twist effects such as the u-channel  knockout of a parton from the medium (jet conversion) or s-channel jet annihilation into  a pair of partons. In the high energy kinematics that we consider here,  these are also suppressed by powers of $s$. 

Radiative energy loss refers to parton branching and  the stimulation of gluon emission. In the limit where the gluon becomes very soft ($x \rightarrow 0$) or very hard ($x \rightarrow 1$) there exist additional constraints from the eikonal limit discussed above. For soft quarks and gluons the u-channel and s-channel interactions with the medium can become comparable to the t-channel and lead to   $\ord{1}$ corrections. Thus, the additional condition to ensure that the in-medium splitting kernels are accurate is $\min( x , 1-x ) \gg \frac{ \perp^2 }{ s }$, which cuts out the endpoints of phase space where $x$ goes to $0$ or $1$.  Gluon emission, and parton splitting in general, is associated with the accumulation of phases as the jet propagates through the medium.  As we will see, a typical phase $\Delta \phi$ is accumulated as the product of a wave number $\frac{1}{l_f^+} = \frac{\bot^2}{x(1-x) p^+}$ and the mean free path $\lambda^+$ between scatterings:
\begin{align}
\Delta \phi \sim \frac{\lambda^+}{l_f^+} = \frac{L^+}{\chi \: l_f^+} 
 = \left( \frac{1}{p_N^- l_f^+} \right) \frac{p_N^- L^+}{\chi}  =    \frac{x_B}{x(1-x)} \frac{\bot^2}{Q^2}  \frac{m_N L}{\chi}
\sim \left(\frac{\bot^2}{Q^2}\right) \left( \frac{A^{1/3}}{\chi} \right) .
\end{align}
In the equation above we have used $p_N^- L^+ = m_N L$ in the rest frame of the medium, the relations between DIS variables, and the fact that the length of the medium is proportional to the number of nucleons (or scattering centers, more generally) along the jet path that goes as  $A^{1/3}$.  If the opacity $\chi$ is not extremely large ($\chi \sim A^{1/3}$ such that every available nucleon / scattering center is struck) then the phases associated with radiative energy loss are enhanced by the length of the medium and dominate over collisional energy loss and other higher-twist effects.  

Finally, the average number of scatterings which takes place in the medium is controlled by the opacity $\chi$, with the inclusion of correlations between multiple scatterings in the medium contributing to higher powers of the opacity.  A perturbative approach to the computation of medium modification thus generically translates the perturbation series in the strong coupling $\alpha_s$, here denoting the interaction between the jet and the medium,  into a series in the opacity $\chi$ for each correlated rescattering, referred to as the opacity expansion.  At relatively small values of $\chi$, the opacity expansion can be truncated at finite order, while for very large values it must be resummed.  The precise value of the opacity at which resummation is mandatory depends on the observable, but in general it need not be restricted to $\chi \ll 1$.  Indeed, for radiative energy loss in the soft-gluon approximation, rapid convergence is seen up to $\ord{\chi^3}$ even for opacities as large as $\chi \sim 5$ \cite{Gyulassy:2000er}.  

In this work, we will extend the opacity expansion approach to relax the soft-gluon approximation up to any finite order in opacity; this will yield the full solution for gluon radiation from both light and heavy quarks to any desired accuracy.  For practical purposes,  we implicitly assume that we have the ability to truncate the opacity expansion, such that $\chi = \frac{L^+}{\lambda^+} \sim \ord{\mathrm{few}}$.  Together with the assumption that the accumulated phases associated with gluon radiation may be large $\Delta \phi \sim \frac{\lambda^+}{l_f^+} \sim \ord{1}$ but that collisional losses and other higher-twist effects are small $\frac{\bot^2}{Q^2} \sim \frac{1}{l_f^+ p_N^-} \ll 1$, this defines the parametric regime that motivates  our calculation:
\begin{align}
\frac{1}{p_N^-} \ll l_f^+ \sim \lambda^+ \sim L^+ .
\end{align}
The leftmost inequality enforces the eikonal approximation (twist expansion) $\frac{\bot^2}{Q^2} \ll 1$; the middle comparison permits the accumulated phases to be large $\Delta \phi \sim \ord{1}$; and the rightmost comparison reflects the ability to terminate the opacity expansion $1 < \chi < \mathrm{few}$.  For comparison, the continuous Schroedinger-like description is valid in the limit of small phases and high opacities, $\frac{1}{p_N^-} \ll \lambda^+ \ll l_f^+ \ll L^+$ \cite{Baier:1996sk,Baier:1998kq}.  Because we assume no particular strong hierarchy among the scales $l_f^+ , \lambda^+ , L^+$, our results should be general enough to reproduce the other known limits under the appropriate assumptions.
%

%
\section{Medium Modification in the Opacity Expansion Approach}
\label{sec:Modification}
%

%
\subsection{Scattering in an External Potential}
%

The coupling of a jet to the external vector potential $A_{ext}$ sourced by the constituents of a nuclear medium can be introduced directly at the Lagrangian level \cite{Ovanesyan:2011kn}. To linear order in $A_{ext}$ and the coupling of the jet to the medium $g_{eff}$ we write
\begin{subequations} \label{e:Lagr1}
\begin{align}
\mathcal{L}_{opac.} &= \mathcal{L}_{QCD} + \mathcal{L}_{ext}^{qG} + \mathcal{L}_{ext}^{gG} + \mathcal{L}_{G.F.} + \cdots   \, , 
\\
\mathcal{L}_{ext}^{q G} &= + g_{eff} \: \bar{\psi} \, \slashed{A}_{ext}^a \, t^a \, \psi \, , 
\\
\mathcal{L}_{ext}^{gG} &= - g_{eff} f^{a b c} 
\Big[ 
(A_{ext})_\mu^b \, A_\nu^c \, (\partial^\mu A^{\nu \, a}) +
A_\mu^b \, (A_{ext})_\nu^c \, (\partial^\mu A^{\nu \, a}) + 
A_\mu^b \, A_\nu^c \, (\partial^\mu (A^{\nu \, a}_{ext}))
\Big] \,  .
\end{align}
\end{subequations}
The added terms introduce ordinary quark/gluon Feynman rules for single scatterings in an external field, and we keep the effective coupling $g_{eff}$ a free parameter (not necessarily fixed to be the same QCD coupling $g$ in $\mathcal{L}_{QCD}$).  Jets arise from hard scattering processes, the Feynman rules for which are contained in $\mathcal{L}_{QCD}$; still it is useful to consider for our case a quark source term $j$ with dimensions of $m^{5/2}$ and 
\begin{equation}
\mathcal{L}_{source} =  \bar{\psi} j + \bar{j} \psi .
\end{equation}

The external field $A_{ext}^{\mu \, a} (x)$ is a superposition of the color fields $a_i^{\mu \, a} (x - x_i)$  of a large number $N$ of scattering centers distributed throughout the medium with spacetime positions $\{ x_i \}$:
\begin{subequations} \label{e:Aext1}
\begin{align}
A_{ext}^{\mu \, a} (x) &= \sum_i a_i^{\mu \, a} (x - x_i) \; ,
\\
A_{ext}^{\mu \, a} (q) &\equiv \int d^4 x \, e^{i q \cdot x} A_{ext}^{\mu \, a} (x) 
= \sum_i e^{i q \cdot x_i} \, a_i^{\mu a} (q) \; .
\end{align}
\end{subequations}
The external potential in the $A^+ = 0$ gauge from a single scattering center is given in the eikonal approximation by
\begin{align} \label{e:Aext2}
g_{eff} a_{i}^{\mu \, a} (q) = g^{\mu +} \: (t^a)_i \left[ 2\pi \delta(q^+) \right] \: v(q_T^2) \; , 
\end{align}
with $(t^a)_i$ a matrix in the color space of the $i^{th}$ scattering center.  The scattering potential $v(q_T^2)$ is proportional to the propagator of the exchanged gluon, which may acquire a thermal mass $\mu$ in a hot medium:
\begin{align} \label{e:vext1}
v(q_T^2) &\rightarrow \frac{- g_{eff}^2}{q_T^2 + \mu^2} \; ,
\end{align}
and its square is proportional to the elastic scattering cross-section of a quark on the scattering center:
\begin{align} \label{e:elast1}
 \frac{d\sigma^{el}}{d^2 q} = \frac{1}{(2\pi)^2} \frac{C_F}{2 N_c} [ v(q_T^2) ]^2 .
\end{align}
Note that the color factors given here treat the scattering centers in the medium as being the fundamental representation. If the scattering centers are
in the adjoint representation $ {C_F}/{2 N_c}  \rightarrow 1/2 $.

These external potentials will need to be averaged in the medium; for this, we will use Gaussian averaging as described below~\footnote{The Gaussian averaging of fields discussed here should not be confused, for example,  with the  Gaussian approximation employed to make a Schroedinger equation in the vary large number if scattering centers limit solvable.}.  This assumption corresponds to limiting the interaction with a given scattering center to two gluons each, and as such all target fields are averaged pairwise.  For scattering centers which are locally color neutral, such as nucleons in cold nuclear matter, two-gluon exchange is the leading interaction, with higher-order field correlations suppressed by at least a factor of $\alpha_s$.  Using Eqs.~\eqref{e:Aext1} and \eqref{e:Aext2}, the average of two color fields is
\begin{align} \label{e:Gauss1}
\left\langle g_{eff} A_{ext}^{\mu a} (x) \: \Big( g_{eff} A_{ext}^{\nu b} (y) \Big)^*  \right\rangle_{med} &=
g^{\mu +} g^{\nu +}
\sum_{i j} \int \frac{d^2 q}{(2\pi)^2} \frac{d^2 q'}{(2\pi)^2} \frac{dq^-}{2\pi} 
\frac{dq^{\prime \, -}}{2\pi} \: 
e^{-i q^- (x^+ - x_i^+)} \, e^{i q^{\prime \, -} (y^+ - x_j^+)} \, 
\notag \\ &\times
 e^{i \ul{q} \cdot (\ul{x} - \ul{x}_i)} \, e^{-i \ul{q}' \cdot (\ul{y} - \ul{x}_j)} \, v(q_T^2) \, v(q_T^{\prime \, 2}) \:
\Big\langle (t^a)_i \: (t^b)_j \Big\rangle_{med} \, .
\end{align}
The color averaging of the scattering centers yields zero unless $i = j$:
\begin{align} \label{e:coloravg}
\Big\langle (t^a)_i \: (t^b)_j \Big\rangle_{med} \equiv 
\begin{cases}
\frac{1}{N_c} \tr[ (t^a) ] \: \frac{1}{N_c} \tr[ (t^b) ] & \mathrm{for} \: i \neq j \\
\frac{1}{N_c} \tr[ (t^a) \: (t^b) ] & \mathrm{for} \: i = j
\end{cases}
\:\: = \:\: \frac{1}{2 N_c} \delta^{a b} \delta_{i j} \; ,
\end{align}
giving
\begin{align} \label{e:Gauss2}
\left\langle g_{eff} A_{ext}^{\mu a} (x) \: \Big( g_{eff} A_{ext}^{\nu b} (y) \Big)^*  \right\rangle_{med} &=
g^{\mu +} g^{\nu +} \, \delta^{a b} \: \sum_{i = 1}^N 
\int \frac{d^2 q}{(2\pi)^2} \frac{d^2 q'}{(2\pi)^2} \frac{dq^-}{2\pi} \frac{dq^{\prime \, -}}{2\pi} \: 
e^{-i q^- x^+} \, e^{i q^{\prime \, -} y^+} \, e^{i \ul{q} \cdot \ul{x}} \, e^{-i \ul{q}' \cdot \ul{y}}
\notag \\ &\times
e^{i (q^- - q^{\prime \, -}) x_i^+} \, e^{- i (\ul{q} - \ul{q}') \cdot \ul{x}_i} \: 
\frac{1}{2 N_c} v(q_T^2) \, v(q_T^{\prime \, 2}) \, .
\end{align}
Following Ref.~\cite{Gyulassy:2000er}, we convert the remaining sum over sites into an average, $\sum_i (\cdots) = N \langle \cdots \rangle$, which we evaluate as an ensemble average over the parameters of the medium.  In this case, that corresponds to averaging over the transverse position $\ul{x_i}$ and the longitudinal position $x_i^+$ of the scattering center:
\begin{align} \label{e:ensavg}
\sum_i f(x_i^+ , \ul{x_i}) &= N \times \int \frac{d^2 x_i}{A_\bot} 
\int\limits_{- R^+}^{R^+} \frac{d x_i^+}{2 R^+} \: f(x_i^+ , \ul{x_i})
=
\int \frac{d^2 x_i}{\sigma_{el}} 
\int\limits_{- R^+}^{R^+} \frac{d x_i^+}{\lambda^+} \: f(x_i^+ , \ul{x_i})  \; ,
\end{align}
where we take the medium to have transverse area $A_\bot$ and longitudinal extent $2 R^+$ centered around zero, and in the last step we have introduced the mean free path $\lambda = \frac{1}{\rho \sigma_{el}}$ with $\rho$ the number density of scattering centers and $\sigma_{el}$ given by the integral of Eq.~\eqref{e:elast1}.  Using \eq{e:ensavg} one readily obtains (c.f. Eq. (2.2) of Ref.~\cite{Blaizot:2012fh})
\begin{align} \label{e:Gauss3}
\left\langle g_{eff} A_{ext}^{\mu a} (x) \: \Big( g_{eff} A_{ext}^{\nu b} (y) \Big)^*  \right\rangle_{med} &= g^{\mu +} g^{\nu +} \, \delta^{a b} \, \delta(x^+ - y^+) \, \gamma(\ul{x} - \ul{y}) \, ,
\end{align}
with
\begin{align} \label{e:Gauss4}
\gamma(\ul{x} - \ul{y}) = \frac{1}{\lambda^+ \, C_F} 
\int\frac{d^2 q}{(2\pi)^2} \, e^{i \ul{q} \cdot (\ul{x} - \ul{y})} \, 
\left[ \frac{(2\pi)^2}{\sigma_{el}} \, \frac{d\sigma^{el}}{d^2 q} \right] \, .
\end{align}
We have written  Eq.~\eqref{e:Gauss4} as if the medium had a uniform density, such that $\rho$ and hence $\lambda^+$ were constants, resulting in a correlation $\gamma$ which depends only on $\ul{x} - \ul{y}$.  For more realistic phenomenology, one can generalize this straightforwardly to make these quantities dependent upon the longitudinal position $x^+$ or the impact parameter $\tfrac{\ul{x} + \ul{y}}{2}$; this latter generalization is more difficult with path-integral techniques than with the opacity formalism presented here \cite{Blaizot:2012fh}.

%
\begin{figure}[ht]
\begin{center}
\includegraphics[width= 0.5 \textwidth]{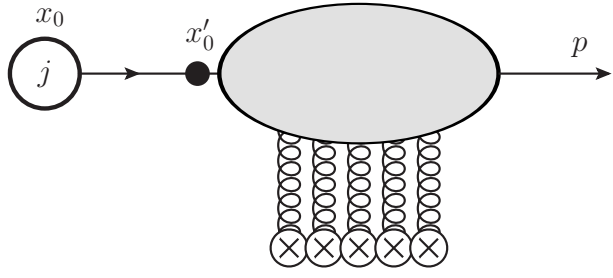} 
\caption{
Diagrammatic representation of the matrix element \eq{e:0broad0}.  A source at point $x_0$ creates a quark, which propagates in the external field \eq{e:Aext2} before emerging in the final state with momentum $p$.  
}
\label{f:Source_Ampl}
\end{center}
\end{figure}
%

With the Lagrangian Eq.~\eqref{e:Lagr1} and the Gaussian averaging \eq{e:Gauss3} it is straightforward to calculate scattering amplitudes in perturbation theory.  For inclusive quark jet production from a source $j$ the amplitude is
\begin{align} \label{e:0broad0}
M &\equiv \int d^4 x_0^\prime \:\: {}_{out} \! \bra{q(p)} \bar{\psi}(x_0^\prime) \ket{\Omega} \: j(x_0^\prime - x_0) \, .
\end{align}
As illustrated in Fig.~\ref{f:Source_Ampl}, this denotes a source at point $x_0$ which is used to create a quark at point $x_0^\prime$, which then evolves until a quark $q(p)$ is detected in the final state.  The state $\ket{\Omega}$ denotes the interacting vacuum of the Lagrangian Eq.~\eqref{e:Lagr1}.  The amplitude \eq{e:0broad0} is then straightforward to calculate using the standard techniques of Lehmann-Symanzik-Zimmermann (LSZ)  reduction \cite{Lehmann:1954rq} and the Gell-Mann-Low theorem \cite{GellMann:1951rw}.  At $0^{th}$ order in $g_{eff}$ we have the vacuum amplitude, unmodified by rescattering in the medium,
\begin{align} \label{e:0broad1}
M_0 &= i \int d^4 x_0^\prime \ d^4 x_f \: e^{i p \cdot x_f} \: \ubar{}(p) (i \slashed{\partial}_{x_f} + m)
\: \left[ i S_F (x_f - x_0^\prime) \right] \: j(x_0^\prime - x_0)
\notag \\ &=
e^{i p \cdot x_0} \: \left[ \ubar{}(p) \: j(p) \right] ,
\end{align}
with $i S_F$ the free fermion propagator and we note that, since $j(\Delta x)$ has dimensions of $m^{5/2}$, its Fourier transform $j(p)$ has dimensions of $m^{-3/2}$.  Squaring the amplitude Eq.~\eqref{e:0broad1}, averaging over the quantum numbers, and integrating over the phase space of the final state gives
\begin{align}
N_0 &\equiv \int \frac{d^2 p \, dp^+}{2 (2\pi)^3 p^+} \: \langle |M_0^2| \rangle
=
\int \frac{d^2 p \, dp^+}{2 (2\pi)^3 p^+} \: \left[ \bar{j}(p) \, \slashed{p} \, j(p) \right].
\end{align}
Given that $j(p)$ has dimensions of $m^{-3/2}$, we see that $N_0$ has dimensions of $m^0$ and can be interpreted as the number of produced particles (quark jets) in the final state.  Equivalently, this gives the phase space distribution of quarks generated by the source current as
\begin{align} \label{e:0broad2}
p^+ \frac{d N_0}{d^2 p \, dp^+} = \frac{1}{2 (2\pi)^3} \: \left[ \bar{j}(p) \, \slashed{p} \, j(p) \right].
\end{align}
%

%
\begin{figure}[ht]
\begin{center}
\includegraphics[width= 0.5 \textwidth]{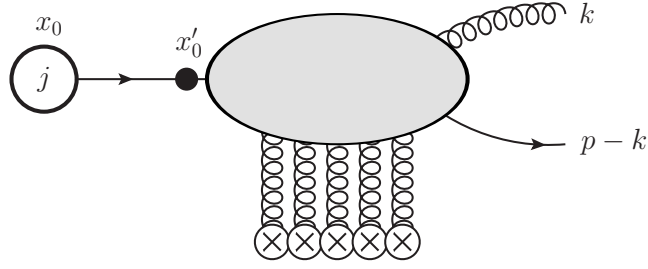} 
\caption{
Diagrammatic representation of the radiation amplitude Eq.~\eqref{e:0rad0}.  A source at point $x_0$ creates an energetic quark, which propagates through the external field Eq.~\eqref{e:Aext2}, radiating a gluon in the process.  
}
\label{f:Rad_Ampl}
\end{center}
\end{figure}
%

\begin{figure}[ht]
\begin{center}
\includegraphics[width= 0.5 \textwidth]{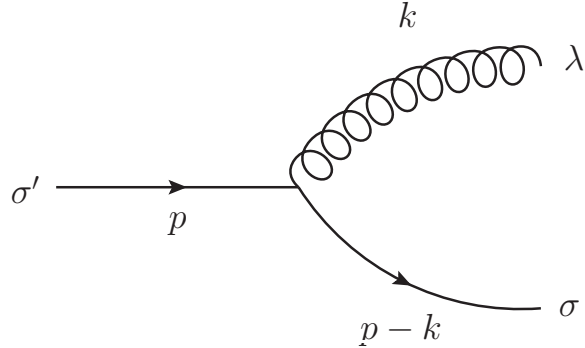} 
\caption{
Quark-gluon splitting wave function of \eq{e:LFWF}, calculated in the conventions of Ref.~\cite{Kovchegov:2012mbw}.
}
\label{f:LFWF}
\end{center}
\end{figure}

Similarly, we can use the Lagrangian Eq.~\eqref{e:Lagr1} to compute the first contribution to jet substructure: the distribution of gluons within the quark jet.  For this, we want a similar scattering amplitude to \eq{e:0broad0}, but with a quark and a gluon in the final state (see Fig.~\ref{f:Rad_Ampl}):
\begin{align} \label{e:0rad0}
R &\equiv \int d^4 x_0^\prime \:\: {}_{out} \! \bra{ q(p-k) \: G(k) } \bar\psi(x_0^\prime) \ket{\Omega} \: j(x_0^\prime - x_0) .
\end{align}
Again, at $0^{th}$ order in $g_{eff}$ we have the unmodified vacuum spectrum of gluon radiation:
\begin{align}
R_0 &= \int d^4 x_0^\prime \, d^4 z \, e^{i k \cdot z} e^{i (p-k) \cdot z} \, \ubar{}(p-k) 
\Big[ i g t^a \slashed{\epsilon}^*(k) \Big] \Big[ i S_F (z - x_0^\prime) \Big] j(x_0^\prime - x_0)
\notag \\ &=
\int dz^+ \frac{d \ell^-}{2\pi} \, e^{-i [\ell^- - (p-k)^- - k^-] z^+} \, 
\left[ e^{+ i \ell^- x_0^+} \, e^{+ i p^+ x_0^-} \, e^{- i \ul{p} \cdot \ul{x_0}} \right] \: 
\ubar{}(p-k) \Big[ i g t^a \slashed{\epsilon}^*(k) \Big] 
\left[ \frac{i}{2 p^+} \frac{\slashed{\ell} }{\ell^- - p^- + i \epsilon} \right] \: j(\ell).
\end{align}
We can perform the $d\ell^-$ integral by residues, closing the contour below for $z^+ - x_0^+ > 0$ and setting $\ell^- = p^- - i\epsilon$:
\begin{align}
R_0 &= \int dz^+ \, e^{-i [p^- - (p-k)^- - k^- - i \epsilon] z^+} \, e^{i p \cdot x_0} \: 
\left[ \ubar{}(p-k) \: i g t^a \slashed{\epsilon}^*(k) \: U(p) \right]
\: \bigg[ \ubar{}(p) j(p) \bigg] \:\left[ \frac{1}{2 p^+} \theta(z^+ - x_0^+)\right] ,
\end{align}
where we have replaced the numerator of the quark propagator by a sum (implied) over spinors $\slashed{\ell} \rightarrow \slashed{p} = U(p) \: \ubar{}(p)$ after putting the quark momentum fully on shell.  At this stage, all propagators have been put on mass shell, and we have an explicit dependence on the ``time'' coordinate $z^+$; this situation is equivalent to formulating the scattering amplitude directly in LFPT.  As is natural in any time-ordered perturbation theory, the emission time $z^+$ is bounded causally from below by the source time $x_0^+$, but it is unbounded above.  Note that the $i\epsilon$ regulator from the denominator of the quark propagator automatically regulates the convergence of the integral at the upper limit $z^+ \rightarrow +\infty$.  Integrating over the emission time $z^+$ yields
\begin{align} \label{e:0rad1}
R_0 &= e^{i p \cdot x_0} \, \left[0 - e^{-i [p^- - (p-k)^- - k^-] x_0^+} \right] \: 
\Big( \frac{1}{2 p^+} \frac{1}{p^- - (p-k)^- - k^-} \, 
\ubar{}(p-k) \left[ - g t^a \slashed{\epsilon}^*(k) \right] U(p) \Big)
\, \left[ \ubar{}(p) j(p) \right] ,
\end{align}
and we recognize the combined factor in parentheses as the light-front wave function (LFWF) for a quark to split into a quark + gluon system, normalized in the conventions of Ref.~\cite{Kovchegov:2012mbw}  (although note that there, they use the metric $\gPM = 2$ rather than $\gPM = 1$ as used here) and depicted in Fig.~\ref{f:LFWF},
%
%
%
%
%
%
\begin{align} \label{e:LFWF}
\psi( x , \ul{k - x p} ) &\equiv \frac{1}{2 p^+} \frac{1}{p^- - (p-k)^- - k^-} \, 
\ubar{\sigma}(p-k) \left[ - g \slashed{\epsilon}_\lambda^*(k) \right] U_{\sigma'} (p)
\notag \\ &=
\frac{g x (1-x)}{(k - x p)_T^2 + x^2 m^2} \: 
\bigg\{ \frac{2-x}{x\sqrt{1-x}} \left( \ul{\epsilon}^*_\lambda \cdot (\ul{k} - x \ul{p}) \right)
\Big[ \, \mathbb{1} \, \Big]_{\sigma \sigma'}
+ 
\frac{\lambda}{\sqrt{1-x}} \left( \ul{\epsilon}^*_\lambda \cdot (\ul{k} - x \ul{p}) \right)
\Big[ \, \tau_3 \, \Big]_{\sigma \sigma'}
\notag \\ & \hspace{4cm} +
\frac{i m \, x}{\sqrt{1-x}} \ul{\epsilon}^*_\lambda \times 
\Big[ \, \ul{\tau_\bot} \, \Big]_{\sigma \sigma'}
\bigg\} .
\end{align}
Here, $x = \frac{k^+}{p^+}$ is the longitudinal momentum fraction of the radiated gluon, $\lambda$ is its spin, $m$ is the mass of the quark, and it is convenient to express the dependence on the spins $\sigma , \sigma'$ of the outgoing and incoming quarks, respectively, in terms of the unit and Pauli matrices $[ \mathbb{1} ] , [ \vec{\tau} ]$ \cite{Kovchegov:2015zha}.  Note that the LFWF depends on transverse momentum only through the intrinsic transverse momentum of the splitting $\ul{\kappa} = \ul{k - x p}$.  In a frame in which $\ul{p} = 0$, the transverse momentum of the splitting is $\ul{\kappa} = \ul{k}$, but after a transverse boost to a frame in which $\ul{p} \neq 0$, this intrinsic momentum becomes $\ul{\kappa} = \ul{k - xp}$.  This characteristic form of the transverse momentum dependence in a LFWF is the origin of the ``mysterious'' structure in Ref.~\cite{Baier:1998kq} and reflects the Galilean symmetry discussed in Ref.~\cite{Blaizot:2012fh}.  We note also that the LFWF's are boost-invariant (since they depend only on $x$ and not on $p^+$) and explicitly conserve angular momentum Fock state by Fock state (since the orbital factor $\ul{\epsilon_\lambda^*} \cdot \ul{\kappa}$ is an eigenfunction of longitudinal angular momentum $\hat{L}_z$ with eigenvalue $(-\lambda)$).  Thus our final expressions are manifestly boost-invariant, even though we started this calculation with the center-of-mass frame in mind.

In terms of the light-front wave function Eq.~\eqref{e:LFWF}, the vacuum radiation amplitude Eq.~\eqref{e:0rad1} takes the simple form
\begin{align} \label{e:0rad2}
R_0 &= e^{i p \cdot x_0} \, \left[0 - e^{-i [p^- - (p-k)^- - k^-] x_0^+} \right] \: 
(t^a) \, \psi (x, \ul{k - xp}) \, \left[ \ubar{}(p) j(p) \right].
\end{align}
Before continuing to compute the gluon phase space distribution, we note the origin of the two phases in \eq{e:0rad2}.  There is a trivial ``production phase'' $e^{i p \cdot x_0}$ associated with the source at point $x_0$ emitting a quark with momentum $p$, and there is an ``emission phase'' $\left[0 - e^{-i [p^- - (p-k)^- - k^-] x_0^+} \right]$ associated with the interval $z^+ \in (x_0^+, \infty)$ in which the gluon radiation can occur.  Note that the energy scale entering the emission phase is exactly the same as the energy denominator entering the LFWF Eq.~\eqref{e:LFWF}.
\begin{align} \label{e:Edenom1}
\Delta E^- (\ul{k} - x \ul{p}) \equiv p^- - (p-k)^- - k^- = - \frac{ (\ul{k} - x \ul{p})_T^2 + x^2 m^2}{2 x (1-x) p^+}\, .
\end{align}
Because the upper limit of the emission time is unbounded, the first term in the emission factor is zero; when an in-medium scattering occurs after the splitting, this will introduce an upper bound on the $z^+$ interval, and the first term will be nonzero.

Squaring \eq{e:0rad2} and averaging over the quantum numbers, we obtain
\begin{align}
\langle | R_0 |^2 \rangle = C_F  \Big| \psi(x, \ul{k - xp}) \Big|^2 \,
\Big[ \bar{j} (p) \, \slashed{p} \, j(p) \Big] \, .
\end{align}
Noting that the source current $j(p)$ has dimensions of $m^{-3/2}$, the spinors $U(p)$ has dimensions of $m^{+1/2}$, and the LFWF in Eq.~\eqref{e:LFWF} has dimensions of $m^{-1}$, we see that $\langle |R_0|^2 \rangle$ has net dimensions of $m^{-4}$.  This implies that, after integrating over the on-shell phase space for both the final-state quark and gluon, the result is dimensionless and can be interpreted as a number distribution:
\begin{align} \label{e:PS}
\left. N \right|_{\ord{\chi^0}} = \int \frac{d^2 k \, dk^+}{2 (2\pi)^3 k^+} \, \frac{d^2 (p-k) \, d(p-k)^+}{2 (2\pi)^3 (p-k)^+} \, 
\langle | R_0 |^2 \rangle \, ,
\end{align}
such that the phase space distribution of gluons within the quark jet is
\begin{align} \label{e:0rad3}
\left. x p^+ \, \frac{d N}{d^2 k \, dx \, d^2 p \, dp^+} \right|_{\ord{\chi^0}} = \frac{1}{2(2\pi)^3} \frac{C_F}{1-x} \, \Big| \psi(x, \ul{k - xp}) \Big|^2 \: \times \left( p^+ \frac{dN_0}{d^2 p \, dp^+} \right) \, .
\end{align}
After summing over the quantum numbers of the outgoing particles and averaging over the spins of the initiating quark, the LFWF's are given by
\begin{align} \label{e:LFWFsq}
\left\langle \psi(x , \ul{\kappa}) \, \psi^* (x , \ul{\kappa'}) \right\rangle &\equiv
\sum_{\lambda = \pm 1} \, \half \tr\bigg[ \psi(x , \ul{\kappa}) \, \psi^* (x , \ul{\kappa'}) \bigg]
\notag \\ &=
\frac{8\pi\alpha_s \, (1-x)}{ [\kappa_T^2 + x^2 m^2] \, [\kappa_T^{\prime \, 2} + x^2 m^2]}
\bigg[ (\ul{\kappa} \cdot \ul{\kappa}') \, \left[ 1 + (1-x)^2 \right] + x^4 m^2 \bigg]\, ,
\end{align}
which are just the massive generalizations of the Altarelli-Parisi splitting functions.  This gives the vacuum distribution as 
\begin{align}
\left. x p^+ \, \frac{d N}{d^2 k \, dx \, d^2 p \, dp^+} \right|_{\ord{\chi^0}} = 
\frac{\alpha_s \, C_F}{2 \pi^2} \, 
\frac{(k - x p)_T^2\, \left[ 1 + (1-x)^2 \right] + x^4 m^2}{ [(k - x p)_T^2 + x^2 m^2]^2}
\: \times \left( p^+ \frac{dN_0}{d^2 p \, dp^+} \right) \, ,
\end{align}
which, after some algebra, is seen to be equivalent to Eq.~(2.29) of Ref.~\cite{Kang:2016ofv}.
\footnote{In this reference, the authors set $p_T = 0$ and implicitly divide the left-hand side by 
$p^+ \frac{dN_0}{d^2 p \, dp^+}$, writing the left-hand side as $x \frac{dN}{d^2 k \, dx}$.}
In the massless limit the connection to the Altarelli-Parisi splitting functions becomes explicit:
\begin{align}
P_{G/q} (x) &= C_F \frac{1 + (1-x)^2}{x} = C_F \frac{1}{2 g^2} \frac{\kappa_T^2}{x (1-x)} |\psi(x, \ul{\kappa})|^2,
\end{align}
such that \eq{e:0rad3} can be rewritten as
\begin{align} \label{e:0rad4}
\left. x p^+ \, \frac{d N}{d^2 k \, dx \, d^2 p \, dp^+} \right|_{\ord{\chi^0}} &= 
\frac{\alpha_s}{2 \pi^2} \,
\frac{x \, P_{G/q} (x)}{(k-xp)_T^2} \: \times \left( p^+ \frac{dN_0}{d^2 p \, dp^+} \right)
\notag \\ &=
\frac{\alpha_s C_F}{2 \pi^2} \,
\frac{1 + (1-x)^2}{(k-xp)_T^2} \: \times \left( p^+ \frac{dN_0}{d^2 p \, dp^+} \right) .
\end{align}

%
\subsection{Leading Order Jet Substructure in the Opacity Expansion}
\label{sec:LO}
%

\begin{figure}[ht]
\begin{center}
\includegraphics[width=\textwidth]{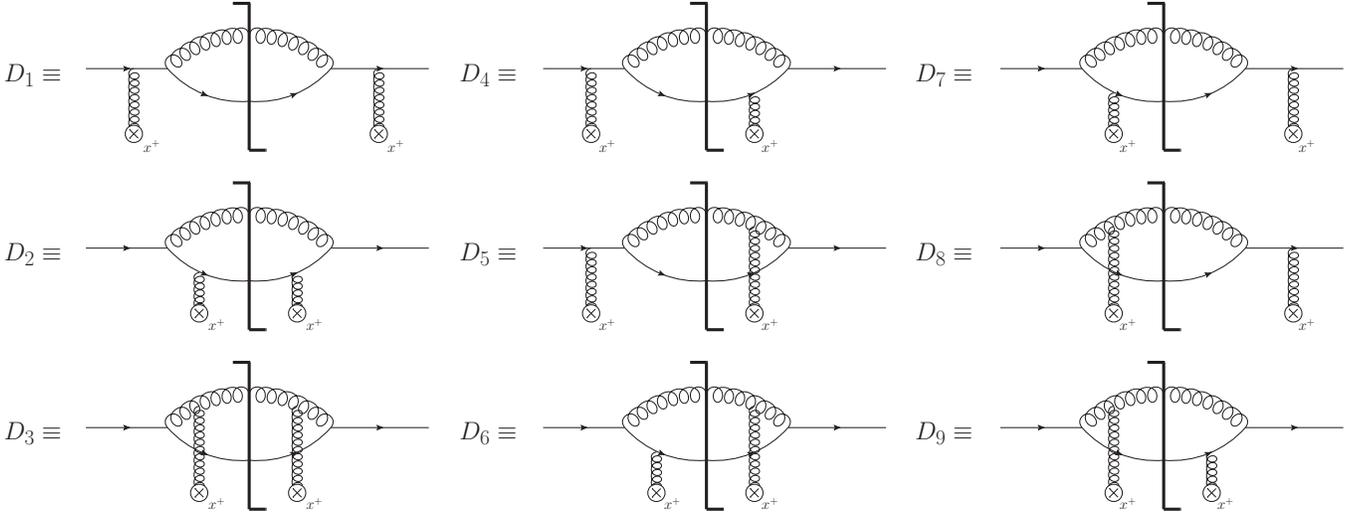} 
\caption{Single-Born ``direct'' scattering diagrams on a scattering center at position $x^+$.}
\label{f:Born}
\end{center}
\end{figure}

\begin{figure}[ht]
\begin{center}
\includegraphics[width=0.6\textwidth]{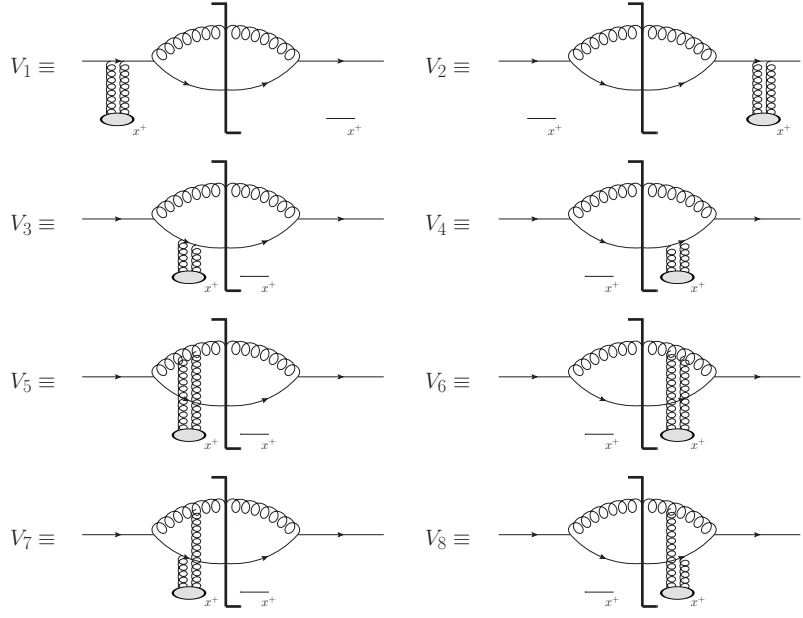} 
\caption{Double-Born ``virtual'' scattering diagrams on a scattering center at position $x^+$.}
\label{f:dblBorn}
\end{center}
\end{figure}

Although the distribution Eq.~\eqref{e:0rad4} of gluons within a quark jet represents the leading contribution to jet substructure, this contribution consists of gluon emission in the vacuum.  The mechanism by which the modification of this substructure occurs is through scattering in the medium. In this process, phases associated with the scattering in the medium accumulate, and the endpoints of the emission phases in Eq.~\eqref{e:0rad2}  are fixed to be located in the medium.  
Using the Gaussian averaging Eq.~\eqref{e:Gauss3}, the first contribution from scattering in the medium comes from the 17 diagrams shown in Figs.~\ref{f:Born} and \ref{f:dblBorn}.  If one gluon is exchanged with the jet in the amplitude (and another in the complex-conjugate amplitude), there are $3^2 = 9$ ``single-Born'' or ``direct'' diagrams $D_{1 - 9}$ as illustrated in Fig.~\ref{f:Born}.  At the same order, there are contributions in which the scattering center exchanges two gluons with the jet on the same side of the cut, generating the $2 \times 4 = 8$ ``double-Born'' or ``virtual'' diagrams $V_{1-8}$ shown in Fig.~\ref{f:dblBorn}.

%
%

%
%

%
\begin{figure}[ht]
\begin{center}
\includegraphics[width=\textwidth]{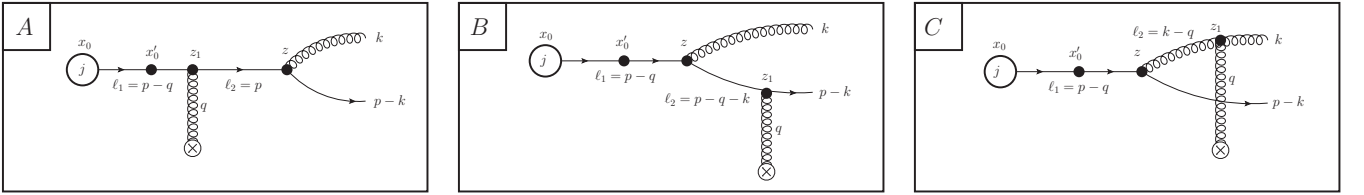}
\caption{
First-order rescattering corrections in the medium.  There are three distinct diagrams, in which the rescattering occurs before the splitting ($R_1^A$ , left panel), after the splitting on the quark ($R_1^B$ , center panel), and after the splitting on the gluon ($R_1^C$ , right panel)
}
\label{f:Rad_Ampl_1}
\end{center}
\end{figure}
%

To illustrate the calculation, consider the three building blocks of the 9 ``direct'' diagrams at the amplitude level shown in Fig.~\ref{f:Rad_Ampl_1}, starting with the case in which the rescattering happens before the splitting, as shown in the left panel of Fig.~\ref{f:Rad_Ampl_1}:
\begin{align}
R_1^A &= \int d^4 x_0^\prime \, d^4 z_1 \, d^4 z \: e^{i k \cdot z} \, e^{i (p-k) \cdot z} \:
\ubar{}(p-k) \Big[ i g t^a \slashed{\epsilon}^* (k) \Big] \Big[ i S_F (z - z_1) \Big] 
\Big[ i g_{eff} \hat{\slashed{A}}_{ext} (z_1) \Big] 
\notag \\ &\hspace{1cm} \times
\Big[ i S_F (z_1 - x_0^\prime) \Big] j(x_0^\prime - x_0)
\notag \\ \notag \\ 
&= \sum_i \int \frac{d^4 q}{(2\pi)^4} \, \frac{d \ell_2^-}{2\pi} \, dz^+ \: 
e^{i q \cdot (x_i - x_0)} \, e^{i p \cdot x_0} e^{-i \ell_2^- (z^+ - x_0^+)} \, e^{- i p^- x_0^+} \, 
e^{+i [ (p-k)^- + k^-] z^+}
 \:
\notag \\ &\hspace{1cm} \times
\ubar{}(p-k) \Big[ g t^a \slashed{\epsilon}^* (k) \Big] 
\Big[ \frac{1}{2 p^+} \frac{\slashed{\ell_2}}{\ell_2^- - p^- + i \epsilon} \Big] 
\Big[ g_{eff} \hat{\slashed{a}}_{i} (q) \Big] 
\Big[ \frac{1}{2 (p^+ - q^+)} 
\frac{(\slashed{\ell_2} -  \slashed{q})}{\ell_2^- - q^- - (p-q)^- + i \epsilon} \Big] 
j(\ell_2 - q) ,
\end{align}
where $\hat{A} \equiv A^a t^a$.  The manipulations in going from coordinate to momentum space are exact and involve only conservation of momentum.  The rest of the calculation proceeds along the lines of the amplitude $R_0$ for radiation in vacuum: we first perform the $dq^-$ integral by residues, enclosing the pole in the upper half-plane only if $x_i^+ - x_0^+ > 0$ which sets $q^- = \ell_2^- - (p-q)^- + i \epsilon$ and gives
\begin{align}
R_1^A &= -i \sum_i \int \frac{d^2q \, dq^+}{(2\pi)^3} \, \frac{d \ell_2^-}{2\pi} \, dz^+ \: 
\left[ e^{i [\ell_2^- - (p-q)^- + i \epsilon] (x_i^+ - x_0^+)} \, e^{i q^+ \cdot (x_i^- - x_0^-)} \, 
e^{-i \ul{q} \cdot (\ul{x_i} - \ul{x_0})} \right] 
\, e^{i p \cdot x_0} \, e^{-i \ell_2^- (z^+ - x_0^+)} \, e^{- i p^- x_0^+} \, 
\notag \\ &\hspace{1cm} \times
e^{+i [ (p-k)^- + k^-] z^+} \:
\ubar{}(p-k) \Big[ g t^a \slashed{\epsilon}^* (k) \Big] 
\Big[ \frac{1}{2 p^+} \frac{\slashed{\ell_2}}{\ell_2^- - p^- + i \epsilon} \Big] 
\Big[ g_{eff} \hat{\slashed{a}}_{i} (q) \Big] U(\ell_2 - q)
\notag \\ &\hspace{1cm} \times
\Big[ \ubar{}(\ell_2 - q) j(\ell_2 - q) \Big] \: 
\Big[ \frac{1}{2 (p^+ - q^+)} \theta(x_i^+ - x_0^+) \Big] \, .
\end{align}
We again replaced the numerator $\slashed{\ell_2} - \slashed{q}$ by a spinor sum after putting the propagator on shell.  Next we perform the $d\ell_2^-$ integral by residues, picking up the pole at $\ell_2^- = p^- - i\epsilon$ only if $z^+ - x_i^+ > 0$ to obtain
\begin{align}
R_1^A &= - \sum_i \theta(x_i^+ - x_0^+) \int \frac{d^2q \, dq^+}{(2\pi)^3} \, dz^+ \: 
\left[ e^{i [p^- - (p-q)^-] (x_i^+ - x_0^+)} \, e^{i q^+ \cdot (x_i^- - x_0^-)} \, 
e^{-i \ul{q} \cdot (\ul{x_i} - \ul{x_0})} \right] 
\, e^{i p \cdot x_0} \, 
\notag \\ &\hspace{1cm} \times
e^{-i [p^- - (p-k)^- - k^- - i \epsilon] z^+} \:
\Big[ \ubar{}(p-k) \left[ g t^a \slashed{\epsilon}^* (k) \right] U(p) \Big] \:
\left[ \ubar{}(p) \left[ g_{eff} \hat{\slashed{a}}_{i} (q) \right] U(p - q) \right]
\notag \\ &\hspace{1cm} \times
\Big[ \ubar{}(p - q) j(p - q) \Big] \: 
 \frac{1}{2 (p^+ - q^+)} 
\Big[ \frac{1}{2 p^+} \theta(z^+ - x_i^+) \Big].
\end{align}
The $d z^+$ integral provides the energy denominator to complete the LFWF Eq.~\eqref{e:LFWF} and leaves the boundary phases from
 $z \in ( x_i^+,+\infty)$, giving
\begin{align} \label{e:1Arad1}
R_1^A &= +i \, (t^a t^b) \sum_i (t^b)_i \, \theta(x_i^+ - x_0^+) \int \frac{d^2q}{(2\pi)^2} \: 
e^{-i [ (p-q)^- - p^- ] x_i^+} \, \left[ 0 - e^{-i \Delta E^- (\ul{k} - x \ul{p}) x_i^+} \right] \, e^{i (p-q) \cdot x_0}
\notag \\ &\hspace{1cm} \times
e^{-i \ul{q} \cdot \ul{x_i}} \, v(q_T^2) \: \psi(x, \ul{k - x p}) \: 
\Big[ \ubar{}(p - q) j(p - q) \Big] ,
\end{align}
where we used the eikonal form Eq.~\eqref{e:Aext2} of the external potential.  As before, we note that the emission phase can be written in terms of the energy denominator Eq.~\eqref{e:Edenom1}: $p^- - k^- - (p-k)^- = \Delta E^- (\ul{k} - x \ul{p})$.  We also note the appearance of a new type of phase arising from scattering in the medium: an ``impulse phase'' $e^{-i [(p-q)^- - p^-] x_i^+}$ which reflects the instantaneous change in the wavelength (inverse formation time) of the quark due to a change in its transverse momentum.  After multiplying by one of the complex-conjugate amplitudes, these impulse phases will also cancel or assemble into energy denominators, as we will show.  

In the same way, we compute the amplitudes $B$ and $C$ from Fig.~\ref{f:Rad_Ampl_1} to be:
\begin{align} \label{e:1Brad1}
R_1^B &= + i (t^b t^a) \sum_i (t^b)_i \, \theta(x_i^+ - x_0^+)  \int \frac{d^2q}{(2\pi)^2} \: 
\left[ e^{- i \Delta E^- (\ul{k} - x \ul{p} + x \ul{q}) x_i^+} - e^{- i \Delta E^- (\ul{k} - x \ul{p} + x \ul{q}) x_0^+} \right] 
\notag \\ & \hspace{1cm} \times
e^{i (p - q) \cdot x_0} \, e^{-i [ (p-k-q)^- - (p-k)^- ] x_i^+} \: e^{-i \ul{q} \cdot \ul{x_i}} \: v(q_T^2) \:
\psi(x,\ul{k - x p + x q}) \: \Big[ \ubar{}(p-q) j(p-q) \Big]  \, , 
\\ \notag \\ \label{e:1Crad1}
R_1^C &= - f^{a b c} (t^c) \sum_i (t^b)_i \, \theta(x_i^+ - x_0^+) \int \frac{d^2q}{(2\pi)^2} \: 
\left[ e^{-i \Delta E^- (\ul{k} - x \ul{p} - (1-x) \ul{q}) x_i^+} - e^{-i (\ul{k} - x \ul{p} - (1-x) \ul{q}) x_0^+} \right]
\notag \\ &\hspace{1cm}\times
\, e^{i (p-q) \cdot x_0} \, e^{-i [(k-q)^- - k^-] x_i^+} \, e^{- i \ul{q} \cdot \ul{x_i}} \: v(q_T^2) \:\:
\psi(x, \ul{ k - x p - (1-x) q }) \: \left[ \ubar{}(p-q) \, j(p - q) \right] .
\end{align}
Now consider the combination of impulse phases that arises from the interference of any two such amplitudes.  Noting that the averaging of the target fields as in Eq.~\eqref{e:ensavg} and Eq.~\eqref{e:Gauss3} yields a delta function setting the momentum transfer $q$ equal in the amplitude and complex conjugate amplitude, we readily see that the impulse phases cancel exactly for the square of any amplitude.  For the nontrivial interferences $R_1^A (R_1^B)^*$ , $R_1^A (R_1^C)^*$ , $R_1^B (R_1^C)^*$, the net effect of the impulse phases can always be expressed in terms of energy denominators in LFPT:
\begin{align}
R_1^A (R_1^B)^*:&  [(p-q)^- - p^-] - [(p-k-q)^- - (p-k)^-] 
\notag \\ & \hspace{1cm} =
[(p-q)^- - k^- - (p-k-q)^-] - [p^- - k^- - (p-k)^- ] 
\notag \\ & \hspace{1cm} =
\Delta E^- (\ul{k} - x \ul{p} + x \ul{q}) - \Delta E^- (\ul{k} - x \ul{p}) \, ,
\\ \notag \\
R_1^A (R_1^C)^*:&  [(p-q)^- - p^-] - [(k-q)^- - k^-] 
\notag \\ & \hspace{1cm} =
[(p-q)^- - (k-q)^- - (p-k)^-] - [p^- - k^- - (p-k)^-]
\notag \\ & \hspace{1cm} =
\Delta E^- (\ul{k} - x \ul{p} - (1-x) \ul{q}) - \Delta E^- (\ul{k} - x \ul{p}) \, , 
\\ \notag \\
R_1^B (R_1^C)^*:&  [(p-k-q)^- - (p-k)^-] - [(k-q)^- - k^-] 
\notag \\ & \hspace{1cm} =
[(p-q)^- - (k-q)^- - (p-k)^-] - [(p-q)^- - k^- - (p-k-q)^- ] 
\notag \\ & \hspace{1cm} =
\Delta E^- (\ul{k} - x \ul{p} - (1-x) \ul{q}) - \Delta E^- (\ul{k} - x \ul{p} + x \ul{q})  \, .
\end{align}
This general feature can be understood as follows.  The amplitude squared is related by the optical theorem to the imaginary part of a forward scattering amplitude, in which zero net momentum is transferred between the initial state in the amplitude and the ``final state'' - the initial state of the complex-conjugate amplitude.  At this level, the momentum transfer from the medium flows into the jet, through various partons, and back out.  Depending on that momentum routing in a given diagram, the momentum flow may pass through the parton branching, modifying the intrinsic transverse momentum $\ul{\kappa} = \ul{k} - x \ul{p}$ which enters the arguments of the LFWF and energy denominators.  Depending on the diagram, the result may be a shift to the jet center of mass $\ul{p} \rightarrow \ul{p} - \ul{q}$ leading to $\ul{\kappa} \rightarrow \ul{\kappa} + x \ul{q}$; the relative gluon momentum $\ul{k} \rightarrow \ul{k} - \ul{q}$ leading to $\ul{\kappa} \rightarrow \ul{\kappa} - \ul{q}$; or to both, leading to $\ul{\kappa} \rightarrow \ul{\kappa} - (1-x) \ul{q}$.  This general feature indicates that the kinematic effect of a scattering in the medium can always be incorporated in a shift by one of $-\ul{q}$, $+ x \ul{q}$, or $- (1-x) \ul{q}$ in the ingredients of LFPT.

Combining these amplitudes, we form the 9 ``direct'' diagrams $D_{1-9}$; together with the two-particle phase space \eqref{e:PS}, they are:
\begin{align}
\left. x p^+ \frac{dN}{d^2 k \, dx \, d^2 p \, dp^+} \right|_{\ord{\chi^1}}^{\mathrm{direct}} &= 
\frac{C_F}{2(2\pi)^3 (1-x)} \int\limits_{x_0^+}^{R^+} \frac{dz^+}{\lambda^+} \int\frac{d^2 q}{\sigma_{el}} \frac{d\sigma^{el}}{d^2 q} 
\: \left[ \:\sum_{i=1}^9 D_i \: \right]
\times \left( p^+ \frac{dN_0}{d^2 (p-q) \, dp^+} \right)
\end{align}
with
\begin{subequations} \label{e:direct1}
\begin{align}
D_1 &= \psi(x , \ul{k} - x \ul{p}) 
\left[ 0 - e^{-i \Delta E^- (\ul{k} - x \ul{p}) z^+} \right]
\left[ 0 - e^{+i \Delta E^- (\ul{k} - x \ul{p}) z^+} \right] 
\psi^* (x , \ul{k} - x \ul{p}) \, , 
\\
D_2 &= \psi(x , \ul{k} - x \ul{p} + x \ul{q}) 
\left[ e^{-i \Delta E^- (\ul{k} - x \ul{p} + x \ul{q}) z^+} - e^{-i \Delta E^- (\ul{k} - x \ul{p} + x \ul{q}) x_0^+} \right]
\notag \\ &\hspace{1cm} \times
\left[ e^{+i \Delta E^- (\ul{k} - x \ul{p} + x \ul{q}) z^+} - e^{+i \Delta E^- (\ul{k} - x \ul{p} + x \ul{q}) x_0^+} \right] 
\psi^* (x , \ul{k} - x \ul{p} + x \ul{q})\, , 
\\
D_3 &= \frac{N_c}{C_F} \, \psi(x , \ul{k} - x \ul{p} -(1-x) \ul{q}) 
\left[ e^{-i \Delta E^- (\ul{k} - x \ul{p} - (1-x) \ul{q}) z^+} - e^{-i \Delta E^- (\ul{k} - x \ul{p} - (1-x) \ul{q}) x_0^+} \right]
\notag \\ &\hspace{1cm} \times
\left[ e^{+i \Delta E^- (\ul{k} - x \ul{p} - (1-x) \ul{q}) z^+} - e^{+i \Delta E^- (\ul{k} - x \ul{p} - (1-x) \ul{q}) x_0^+} \right] 
\psi^* (x , \ul{k} - x \ul{p} -(1-x) \ul{q})\, , 
\\
D_4 &= \left[\frac{-1}{2 N_c C_F} \, e^{+i [ \Delta E^- (\ul{k} - x \ul{p}) - \Delta E^- (\ul{k} - x \ul{p} + x \ul{q}) ] z^+} \right] 
\, \psi(x , \ul{k} - x \ul{p}) 
\left[ 0 - e^{-i \Delta E^- (\ul{k} - x \ul{p}) z^+} \right]
\notag \\ &\hspace{1cm} \times
\left[ e^{+i \Delta E^- (\ul{k} - x \ul{p} + x \ul{q}) z^+} - e^{+i \Delta E^- (\ul{k} - x \ul{p} +x \ul{q}) x_0^+} \right] 
\psi^* (x , \ul{k} - x \ul{p} + x \ul{q})\, , 
\\
D_5 &= \left[\frac{N_C}{2 C_F} \, e^{+i [ \Delta E^- (\ul{k} - x \ul{p}) - \Delta E^- (\ul{k} - x \ul{p} - (1-x) \ul{q}) ] z^+} \right] 
\, \psi(x , \ul{k} - x \ul{p}) 
\left[ 0 - e^{-i \Delta E^- (\ul{k} - x \ul{p}) z^+} \right]
\notag \\ &\hspace{1cm} \times
\left[ e^{+i \Delta E^- (\ul{k} - x \ul{p} - (1-x) \ul{q}) z^+} - e^{+i \Delta E^- (\ul{k} - x \ul{p} - (1-x) \ul{q}) x_0^+} \right] 
\psi^* (x , \ul{k} - x \ul{p} - (1-x) \ul{q})\, , 
\\
D_6 &= \left[\frac{-N_C}{2 C_F} \, e^{+i [ \Delta E^- (\ul{k} - x \ul{p} + x \ul{q}) - \Delta E^- (\ul{k} - x \ul{p} - (1-x) \ul{q}) ] z^+} \right] 
\, \psi(x , \ul{k} - x \ul{p} + x \ul{q}) 
\left[ e^{-i \Delta E^- (\ul{k} - x \ul{p} + x \ul{q}) z^+} - e^{-i \Delta E^- (\ul{k} - x \ul{p} + x \ul{q}) x_0^+} \right]
\notag \\ &\hspace{1cm} \times
\left[ e^{+i \Delta E^- (\ul{k} - x \ul{p} - (1-x) \ul{q}) z^+} - e^{+i \Delta E^- (\ul{k} - x \ul{p} - (1-x) \ul{q}) x_0^+} \right] 
\psi^* (x , \ul{k} - x \ul{p} - (1-x) \ul{q})\, , 
\\
D_7 &= \left[\frac{-1}{2 N_c C_F} \, e^{+i [ \Delta E^- (\ul{k} - x \ul{p} + x \ul{q}) - \Delta E^- (\ul{k} - x \ul{p}) ] z^+} \right] 
\, \psi (x , \ul{k} - x \ul{p} + x \ul{q}) 
\left[ e^{-i \Delta E^- (\ul{k} - x \ul{p} + x \ul{q}) z^+} - e^{-i \Delta E^- (\ul{k} - x \ul{p} +x \ul{q}) x_0^+} \right] 
\notag \\ &\hspace{1cm} \times
\left[ 0 - e^{+i \Delta E^- (\ul{k} - x \ul{p}) z^+} \right]
\psi^* (x , \ul{k} - x \ul{p}) \, , 
\\
D_8 &= \left[\frac{N_C}{2 C_F} \, e^{+i [ \Delta E^- (\ul{k} - x \ul{p} - (1-x) \ul{q}) - \Delta E^- (\ul{k} - x \ul{p}) ] z^+} \right] 
\, \psi (x , \ul{k} - x \ul{p} - (1-x) \ul{q}) 
\notag \\ &\hspace{1cm} \times
\left[ e^{-i \Delta E^- (\ul{k} - x \ul{p} - (1-x) \ul{q}) z^+} - e^{-i \Delta E^- (\ul{k} - x \ul{p} - (1-x) \ul{q}) x_0^+} \right] 
\left[ 0 - e^{+i \Delta E^- (\ul{k} - x \ul{p}) z^+} \right]
\psi^* (x , \ul{k} - x \ul{p}) \, , 
\\
D_9 &= \left[\frac{-N_C}{2 C_F} \, e^{+i [ \Delta E^- (\ul{k} - x \ul{p} - (1-x) \ul{q}) - \Delta E^- (\ul{k} - x \ul{p} + x \ul{q}) ] z^+} \right] 
\, \psi (x , \ul{k} - x \ul{p} - (1-x) \ul{q})
\notag \\ &\hspace{1cm} \times
\left[ e^{-i \Delta E^- (\ul{k} - x \ul{p} - (1-x) \ul{q}) z^+} - e^{-i \Delta E^- (\ul{k} - x \ul{p} - (1-x) \ul{q}) x_0^+} \right] 
\left[ e^{+i \Delta E^- (\ul{k} - x \ul{p} + x \ul{q}) z^+} - e^{+i \Delta E^- (\ul{k} - x \ul{p} + x \ul{q}) x_0^+} \right]
\notag \\ &\hspace{1cm} \times
\psi^* (x , \ul{k} - x \ul{p} + x \ul{q}) \, .
\end{align}
\end{subequations}
Note that the terms $D_{4 - 9}$ are interferences of the various single-Born diagrams; as such, they contain nontrivial impulse phases in the bracketed prefactors.  Note also that all of the phases here $\Delta E^- \, z^+$ are formally higher twist as seen in \eq{e:Edenom1}, but that, for relatively small opacities, the differences in positions can be parametrically large: $z^+ - x_0^+ \sim L^+$, such that these higher-twist effects are length-enhanced as discussed in Sec.~\ref{sec:DIS}.

\begin{figure}[ht]
\begin{center}
\includegraphics[width=\textwidth]{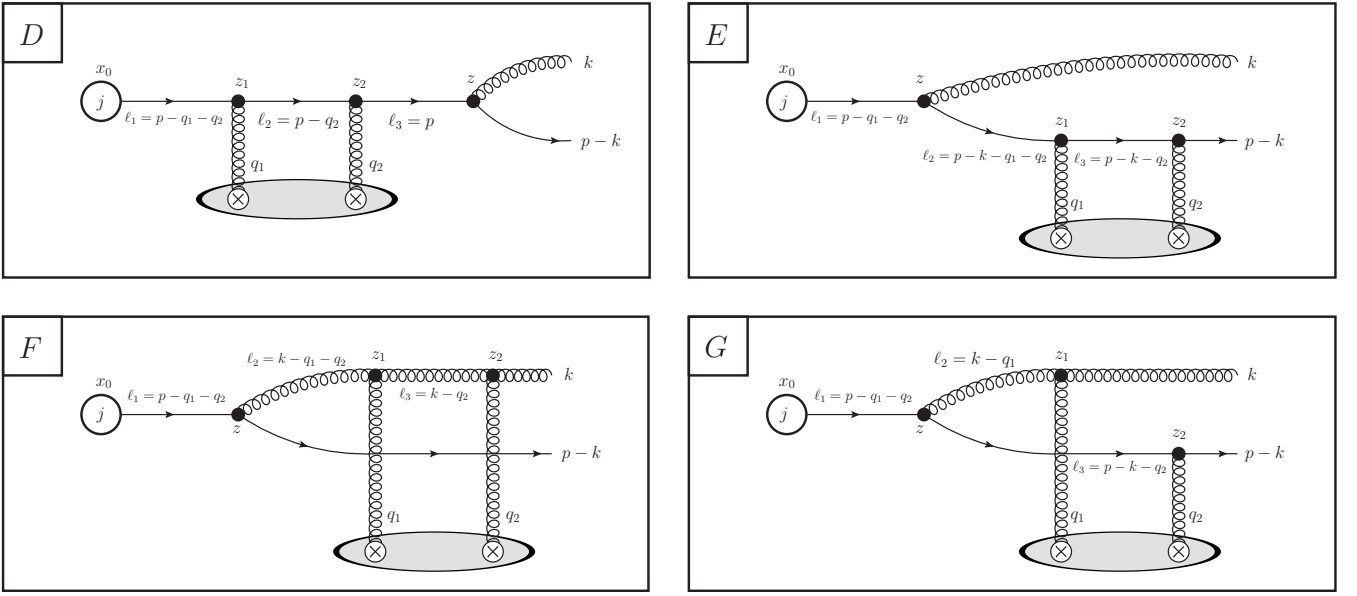}
\caption{
Double-Born rescattering corrections in the medium.  There are four non-zero diagrams, in which the double scattering occurs before the splitting ($R_2^D$ , top left panel), after the splitting on the quark ($R_2^E$ , top right panel), after the splitting on the gluon ($R_2^F$ , bottom left panel), and after the splitting on both quark and gluon ($R_2^G$).
}
\label{f:Rad_Ampl_2DEFG}
\end{center}
\end{figure}

Similarly, we can calculate the ``double-Born'' or ``virtual'' amplitudes shown in Fig.~\ref{f:Rad_Ampl_2DEFG}; the interference of these amplitudes with the vacuum amplitude Eq.~\eqref{e:0rad2} (along with the overall complex conjugates) generate the diagrams shown in Fig.~\ref{f:dblBorn}.  Note that, in addition to the amplitudes shown in Fig.~\ref{f:Rad_Ampl_2DEFG}, there are two others which could be drawn: amplitudes in which the scattering center interacts with the jet by exchanging one gluon before the splitting and another gluon after.  These diagrams vanish in the eikonal approximation, because it is impossible to simultaneously put the intermediate propagators on shell, leading to poles which cannot be enclosed and hence to zero.  Physically, the scattering is instantaneous (within the eikonal approximation) so that there is no time to radiate the gluon during that scattering.

%
%

The double-Born contributions are calculated in the same way as the single-Born ``direct'' contributions, with the only subtlety arising from the treatment of the pole of the intermediate propagator.  For concreteness, consider the amplitude $R_2^F$ in which the rescattering takes place on the radiated gluon:
\begin{align}
R_2^F &= \sum_{i j} \int d^4 z_1 \, d^4 z_2 \, d^4 z \, \int \frac{d^4 \ell_1}{(2\pi)^4}
\frac{d^4 \ell_2}{(2\pi)^4} \frac{d^4 \ell_3}{(2\pi)^4} \frac{d^4 q_1}{(2\pi)^4} 
\frac{d^4 q_2}{(2\pi)^4}
\notag \\ & \hspace{1cm}
\times
e^{i k \cdot z_2} \, e^{i (p-k) \cdot z} \, e^{-i \ell_1 \cdot (z-x_0)} \, e^{-i \ell_2 \cdot (z_1 - z)} \, 
e^{- i \ell_3 \cdot (z_2 - z_1)} \, e^{-i q_1 \cdot (z_1 - x_i)} \, e^{-i q_2 \cdot (z_2 - x_j)}
\notag \\ & \hspace{1cm}
\times
\ubar{}(p-k) \Big[ i g t^c \gamma_\mu \Big] 
\Big[ \frac{i \slashed{\ell_1}}{\ell_1^2 + i\epsilon} \Big] j(\ell_1) \:
\left( \frac{- i N^{\mu \mu'} (\ell_2) }{\ell_2^2 + i\epsilon} \right) \: 
\left( \frac{- i N^{\nu \nu'} (\ell_3) }{\ell_3^2 + i\epsilon} \right) \: 
\Big( \epsilon^* (k) \Big)^\beta
\notag \\ & \hspace{1cm}
\times
g_{eff} f^{c d b_1} \, a_i^{b_1 \, \alpha_1} (q_1) \Big[ (- \ell_3 - q_1)_{\mu'} \: g_{\nu \alpha_1} +
(q_1 - \ell_2)_\nu \: g_{\alpha_1 \mu'} + (\ell_2 + \ell_3)_{\alpha_1} \: g_{\mu' \nu} \Big]
\notag \\ & \hspace{1cm}
\times
g_{eff} f^{d a b_2} \, a_j^{b_2 \, \alpha_2} (q_2) \Big[ (-k - q_2)_{\nu'} \: g_{\beta \alpha_2} +
(q_2 - \ell_3)_\beta \: g_{\alpha_2 \nu'} + (\ell_3 + k)_{\alpha_2} \: g_{\nu' \beta} \Big] \, .
\end{align}
In the $A^+ = 0$ light-cone gauge, the numerator of the gluon propagator is given by
\begin{align} \label{e:LCprop}
N^{\mu \nu} (\ell) = g^{\mu \nu} - \frac{1}{\ell^+} g^{+ \mu} \ell^\nu - \frac{1}{\ell^+} \ell^{\mu} g^{+ \nu}
\end{align}
with $N^{\mu +} = N^{+ \nu} = 0$ exactly, along with the corresponding component of the polarization vector $\epsilon^{+ \, *} = 0$.  Because the eikonal external potential only has nonzero component $a_i^-$, this eliminates $2/3$ of the terms from the triple-gluon vertices, with only the terms contracting the polarization vector with the gluon numerators remaining:
\begin{align}
R_2^F &= (t^c) f^{d b_1 c} f^{d a b_2}  \sum_{i j} \int d^4 z \, \frac{d^4 \ell_1}{(2\pi)^4} 
\frac{d^4 q_1}{(2\pi)^4} \frac{d^4 q_2}{(2\pi)^4} e^{i (p-k) \cdot z} \, e^{-i \ell_1 \cdot (z-x_0)} \, 
e^{i [k - q_1 - q_2] \cdot z} \, e^{i q_1 \cdot x_i} \, e^{i q_2 \cdot x_j}
\notag \\ & \hspace{1cm}
\times
\ubar{}(p-k) \Big[ g \gamma_\mu \Big] 
\Big[ \frac{\slashed{\ell_1}}{\ell_1^2 + i\epsilon} \Big] j(\ell_1) \:
\left( \frac{ {N^\mu}_\nu (k - q_1 - q_2) \: {N^\nu}_\beta (k - q_2) \: ( \epsilon^* (k) )^\beta }
{[(k - q_1 - q_2)^2 + i\epsilon] \: [(k - q_2)^2 + i\epsilon]} \right) \: 
\notag \\ & \hspace{1cm}
\times
\Big[ g_{eff}  (2k - q_1 - 2 q_2) \cdot a_i^{b_1} (q_1) \Big] 
\Big[ g_{eff} (2k - q_2) \cdot a_j^{b_2} (q_2) \Big] \,  ,
\end{align}
where we have performed the $z_1 , z_2 , \ell_2 , \ell_3$ integrals to enforce momentum conservation.  The remaining effect of the numerators $N^{\mu \nu}$ is to shift the argument of the polarization vector; using the fact that $q_1^+ = q_2^+ = 0$ in the external potential and $\epsilon^{+ \, *} = 0$ in this gauge, we find
\begin{align}
{N^\mu}_\nu (k - q_1 - q_2) \: {N^\nu}_\beta (k - q_2) \: ( \epsilon^* (k) )^\beta =
( \epsilon^* (k - q_1 - q_2) )^\mu \, .
\end{align}
This evaluation of the numerator algebra also applies to the single-Born amplitude $R_1^C$ Eq.~\eqref{e:1Crad1} and in general to the rescattering of the radiated gluon.  Using the averaging Eq.~\eqref{e:Gauss3} of the target fields (note here that the second field is not complex-conjugated, leading to a relative minus sign in momentum space) and collecting the poles of $\ell_1^-$ and $q_1^-$ as usual gives
\begin{align}
R_2^F &=  -i \frac{N_c}{C_F} (t^a) \int\limits_{x_0^+}^{R^+} \frac{dx_i^+}{\lambda^+}
\int \frac{d^2 q_1}{\sigma_{el}} \, \frac{d\sigma^{el}}{d^2 q_1} \: 
\left[ e^{-i \Delta E^- (\ul{k} - x \ul{p}) x_i^+} - e^{-i \Delta E^- (\ul{k} - x \ul{p}) x_0^+} \right]
\, e^{i p \cdot x_0}
\notag \\ & \hspace{1cm}
\times
\psi(x , \ul{k} - x \ul{p}) \: \Big[ \ubar{} (p) j(p) \Big] \: 
\int \frac{d q_2^-}{2\pi} \, \frac{1}{k^- - q_2^- - (k - q_2)^- + i\epsilon} \, .
\end{align}
The subtlety associated with the double-Born amplitudes $R_2^D , R_2^E , R_2^F$ resides in the remaining integral over $q_2^-$: while there is still the pole of the $(k - q_2)$ propagator between the two exchanged gluons, there is no longer a Fourier factor regulating the convergence of the integral at infinity.  As such, the integral is logarithmic at large $| q_2^-|$ and must be carefully regulated.  There are a variety of ways to do this, such as regulating the magnitude of the minus momentum to enforce the eikonal approximation $|q_2^-| < p_N^-$ or by symmetrizing the integrand under $q_2^- \rightarrow - q_2^-$; either way leads to the result $- i /2$ for this ``contact'' limit of the integral:
\begin{align}
R_2^F &=  - \frac{N_c}{2 C_F} (t^a) \, e^{i p \cdot x_0} 
\int\limits_{x_0^+}^{R^+} \frac{dx_i^+}{\lambda^+}
\int \frac{d^2 q}{\sigma_{el}} \, \frac{d\sigma^{el}}{d^2 q} \: 
\left[ e^{-i \Delta E^- (\ul{k} - x \ul{p}) x_i^+} - e^{-i \Delta E^- (\ul{k} - x \ul{p}) x_0^+} \right]
\, \psi(x , \ul{k} - x \ul{p}) \: \Big[ \ubar{} (p) j(p) \Big] \, .
\end{align}
The calculation of the other double-Born amplitudes proceeds along the same lines with amplitudes $R_2^D , R_2^E , R_2^F$ all possessing a ``contact'' integral, while the amplitude $R_2^G$ has poles that can be collected normally:
\begin{subequations}
\begin{align}
R_2^D &= -\half \, (t^a) \, e^{i p \cdot x_0}
\int\limits_{x_0^+}^{R^+} \frac{dx_i^+}{\lambda^+} 
\int \frac{d^2 q}{\sigma_{el}} \, \frac{d\sigma^{el}}{d^2 q} \: 
\left[ 0 -  e^{- i \Delta E^- (\ul{k} - x \ul{p}) x_i^+} \right]
\psi(x, \ul{k-xp}) \: \Big[ \ubar{}(p) j(p) \Big] \, , 
\\ \notag \\
R_2^E &= -\half (t^a) \, e^{i p \cdot x_0}
\int\limits_{x_0^+}^{R^+} \frac{dx_i^+}{\lambda^+}
\int \frac{d^2 q}{\sigma_{el}} \, \frac{d\sigma^{el}}{d^2 q} \: 
\left[ e^{-i \Delta E^- (\ul{k} - x \ul{p}) x_i^+} - e^{-i \Delta E^- (\ul{k} - x \ul{p}) x_0^+} \right]
\psi(x, \ul{k-xp}) \: \Big[ \ubar{}(p) j(p) \Big] \, , 
\\ \notag \\
R_2^G &= \frac{N_C}{2 C_F} (t^a) \, e^{i p \cdot x_0}
\int\limits_{x_0^+}^{R^+} \frac{dx_i^+}{\lambda^+} \:
\int \frac{d^2 q}{\sigma_{el}} \, \frac{d\sigma^{el}}{d^2 q} \: 
\left[ e^{-i \Delta E^-(\ul{k} - x \ul{p} - \ul{q}) x_i^+}  - e^{-i \Delta E^-(\ul{k} - x \ul{p} - \ul{q}) x_0^+}\right]
\, e^{ i [\Delta E^-(\ul{k} - x \ul{p} - \ul{q}) - \Delta E^-(\ul{k} - x \ul{p})] x_i^+} \, 
\notag \\ & \hspace{1cm}
\times
\psi(x, \ul{k} - x \ul{p} - \ul{q}) \:\Big[ \ubar{}(p) j(p) \Big] \, .
\end{align}
\end{subequations}
Again, these amplitudes can be written entirely in terms of the elements of LFPT: the wave functions and the energy denominators.  Note the presence of a nontrivial impulse phase for amplitude $R_2^G$ due to a rescattering on the quark-gluon system which redistributes the transverse momentum between the partons, thus altering the virtuality (formation time) of the state.

Combining these amplitudes with the vacuum amplitude Eq.~\eqref{e:0rad2} and including the complex conjugates, we form the 8 ``virtual'' diagrams; together with the two-particle phase space Eq.~\eqref{e:PS}, they are:
\begin{align}
\left. x p^+ \frac{dN}{d^2 k \, dx \, d^2 p \, dp^+} \right|_{\ord{\chi^1}}^{\mathrm{virtual}} &= 
\frac{C_F}{2(2\pi)^3 (1-x)} \int\limits_{x_0^+}^{R^+} \frac{dz^+}{\lambda^+} \int\frac{d^2 q}{\sigma_{el}} \frac{d\sigma^{el}}{d^2 q} 
\: \left[ \:\sum_{i=1}^8 V_i \: \right]
\times \left( p^+ \frac{dN_0}{d^2 p \, dp^+} \right) \, ,
\end{align}
with
\begin{subequations} \label{e:virt1}
\begin{align}
V_1 &= -\half \: \psi(x, \ul{k} - x \ul{p}) 
\left[ 0 - e^{-i \Delta E^- (\ul{k} - x \ul{p}) z^+} \right]
\left[ 0 - e^{+i \Delta E^- (\ul{k} - x \ul{p}) x_0^+} \right] \psi^*(x, \ul{k} - x \ul{p})  \, , 
\\ \notag \\
V_2 &= -\half \: \psi(x, \ul{k} - x \ul{p}) 
\left[ 0 - e^{-i \Delta E^- (\ul{k} - x \ul{p}) x_0^+} \right]
\left[ 0 - e^{+i \Delta E^- (\ul{k} - x \ul{p}) z^+} \right] \psi^*(x, \ul{k} - x \ul{p})  \, , 
\\ \notag \\
V_3 &= -\half \: \psi(x, \ul{k} - x \ul{p}) 
\left[ e^{-i \Delta E^- (\ul{k} - x \ul{p}) z^+} - e^{-i \Delta E^- (\ul{k} - x \ul{p}) x_0^+} \right]
\left[ 0 - e^{+i \Delta E^- (\ul{k} - x \ul{p}) x_0^+} \right] \psi^*(x, \ul{k} - x \ul{p})  \, , 
\\ \notag \\
V_4 &= -\half \: \psi(x, \ul{k} - x \ul{p}) 
\left[ 0 - e^{-i \Delta E^- (\ul{k} - x \ul{p}) x_0^+} \right]
\left[ e^{+i \Delta E^- (\ul{k} - x \ul{p}) z^+} - e^{+i \Delta E^- (\ul{k} - x \ul{p}) x_0^+} \right]
\psi^*(x, \ul{k} - x \ul{p})  \, , 
\\ \notag \\
V_5 &= \frac{- N_c}{2 C_F} \: \psi(x, \ul{k} - x \ul{p}) 
\left[ e^{-i \Delta E^- (\ul{k} - x \ul{p}) z^+} - e^{-i \Delta E^- (\ul{k} - x \ul{p}) x_0^+} \right]
\left[ 0 - e^{+i \Delta E^- (\ul{k} - x \ul{p}) x_0^+} \right] \psi^*(x, \ul{k} - x \ul{p})  \, , 
\\ \notag \\
V_6 &= \frac{- N_c}{2 C_F} \: \psi(x, \ul{k} - x \ul{p}) 
\left[ 0 - e^{-i \Delta E^- (\ul{k} - x \ul{p}) x_0^+} \right]
\left[ e^{+i \Delta E^- (\ul{k} - x \ul{p}) z^+} - e^{+i \Delta E^- (\ul{k} - x \ul{p}) x_0^+} \right]
\psi^*(x, \ul{k} - x \ul{p})  \, , 
\\ \notag \\
V_7 &= \left[ \frac{N_c}{2 C_F} \, e^{i [ \Delta E^- (\ul{k} - x \ul{p} - \ul{q}) - \Delta E^- (\ul{k} - x \ul{p}) ] z^+} \right]
\psi(x, \ul{k} - x \ul{p} - \ul{q}) 
\left[ e^{-i \Delta E^- (\ul{k} - x \ul{p} - \ul{q}) z^+} - e^{-i \Delta E^- (\ul{k} - x \ul{p} - \ul{q}) x_0^+} \right]
\notag \\ & \hspace{1cm} \times
\left[ 0 - e^{+i \Delta E^- (\ul{k} - x \ul{p}) x_0^+} \right] \psi^*(x, \ul{k} - x \ul{p})  \, , 
\\ \notag \\
V_8 &= \left[ \frac{N_c}{2 C_F} \, e^{i [ \Delta E^- (\ul{k} - x \ul{p}) - \Delta E^- (\ul{k} - x \ul{p} - \ul{q}) ] z^+} \right]
\psi (x, \ul{k} - x \ul{p})  
\left[ 0 - e^{-i \Delta E^- (\ul{k} - x \ul{p}) x_0^+} \right]
\notag \\ & \hspace{1cm} \times
\left[ e^{+i \Delta E^- (\ul{k} - x \ul{p} - \ul{q}) z^+} - e^{+i \Delta E^- (\ul{k} - x \ul{p} - \ul{q}) x_0^+} \right]
\psi^*(x, \ul{k} - x \ul{p} - \ul{q})  \, .
\end{align}
\end{subequations}
As before, we note the presence of nontrivial impulse phases in the bracketed prefactors of $V_{7 - 8}$, arising this time from double-Born scattering on the quark-gluon system which redistributes transverse momentum between them. 

Finally, we combine the direct and virtual contributions, with the various phases adding to their complex conjugates to generate cosines.  This gives the first order result in the opacity expansion with exact kinematics:
\begin{align} \label{e:LOexact}
\left. x p^+ \frac{dN}{d^2 k \, dx \, d^2 p \, dp^+} \right|_{\ord{\chi^1}} &= \frac{C_F}{(2\pi)^3 (1-x)} \int\limits_0^{L^+} \frac{d (\delta z^+)}{\lambda^+} \int\frac{d^2 q}{\sigma_{el}}\frac{d\sigma^{el}}{d^2 q} 
\notag \\ \times \Bigg\{ 
\left( p^+ \frac{dN_0}{d^2 (p-q) \, dp^+} \right) &\Bigg[
\half \left| \psi (\ul{k} - x \ul{p}) \right|^2 +
\left| \psi (\ul{k} - x \ul{p} + x \ul{q}) \right|^2
\Bigg( 1 - \cos\Big( \Delta E^- (\ul{k} - x \ul{p} + x \ul{q}) \delta z^+ \Big) \Bigg)
\notag \\ &+
\frac{N_c}{C_F} \left| \psi (\ul{k} - x \ul{p} - (1-x) \ul{q}) \right|^2
\Bigg( 1 - \cos\Big( \Delta E^- (\ul{k} - x \ul{p} - (1-x) \ul{q}) \delta z^+ \Big) \Bigg)
\notag \\ &-
\frac{N_c}{2 C_F} \psi(\ul{k} - x \ul{p} + x \ul{q}) \, \psi^* (\ul{k} - x \ul{p} - (1-x) \ul{q}) 
\Bigg( 1 - \cos\Big( \Delta E^- (\ul{k} - x \ul{p} +x \ul{q}) \delta z^+ \Big) 
\notag \\ & \hspace{1cm}
- \cos\Big( \Delta E^- (\ul{k} - x \ul{p} - (1-x) \ul{q}) \delta z^+ \Big) 
\notag \\ & \hspace{1cm}
+ \cos\Big(  \left[ \Delta E^- (\ul{k} - x \ul{p} +x \ul{q}) - \Delta E^- (\ul{k} - x \ul{p} - (1-x) \ul{q}) \right] \delta z^+ \Big) \Bigg)
\notag \\ &+
\frac{1}{2 N_c C_F} \psi (\ul{k} - x \ul{p}) \, \psi^* (\ul{k} - x \ul{p} + x \ul{q}) 
\Bigg( 1 - \cos\Big( \Delta E^- (\ul{k} - x \ul{p} + x \ul{q}) \delta z^+ \Big) \Bigg)
\notag \\ &-
\frac{N_c}{2 C_F} \psi (\ul{k} - x \ul{p}) \, \psi^* (\ul{k} - x \ul{p} - (1-x) \ul{q}) 
\Bigg( 1 - \cos\Big( \Delta E^- (\ul{k} - x \ul{p} - (1-x) \ul{q}) \delta z^+ \Big) \Bigg) \Bigg]
\notag \\ +
\left( p^+ \frac{dN_0}{d^2 p \, dp^+} \right) &\Bigg[
\left[ - \half - \frac{N_c}{2 C_F} \right] \left| \psi(\ul{k} -x  \ul{p}) \right|^2
\Bigg( 1 - \cos\Big( \Delta E^- (\ul{k} - x \ul{p}) \delta z^+ \Big) \Bigg)
\notag \\ &-
\frac{N_c}{2 C_F} \psi (\ul{k} - x \ul{p} - \ul{q}) \, \psi^* (\ul{k} - x \ul{p}) 
\Bigg( \cos\Big( \Delta E^- (\ul{k} - x \ul{p}) \delta z^+ \Big) 
\notag \\ & \hspace{1cm}
- \cos\Big( \left[ \Delta E^- (\ul{k} - x \ul{p} - \ul{q}) - \Delta E^- (\ul{k} - x \ul{p}) \right] \delta z^+ \Big) \Bigg)
\notag \\ &-
\half \left| \psi(\ul{k} -x  \ul{p}) \right|^2
\cos\Big( \Delta E^- (\ul{k} - x \ul{p}) \delta z^+ \Big) \Bigg] \Bigg\} \, ,
\end{align} 
where $\delta z^+ \equiv z^+ - x_0^+$ and $L^+ \equiv R^+ - x_0^+$ and we will suppress the explicit $x$ dependence of the wave functions for brevity going forward.  Products of wave functions are also implied to be summed / averaged over quantum numbers as in Eq.~\eqref{e:LFWFsq}.

This general result at first order in opacity can be compared with a number of others in the literature.  Many of these use the ``broad source approximation,'' which assumes that the initial distribution $p^+ \frac{dN_0}{d^2 p \, dp^+}$ is insensitive to shifts $\ul{p} \rightarrow \ul{p} - \ul{q}$ in the momentum of the initial jet of order $q_T \sim \ord{\mu}$.  This assumption corresponds to 
\begin{align} \label{e:brdsrc}
p^+ \frac{dN_0}{d^2 p \, dp^+} = \frac{1}{\mu^2} \, \left( p^+ \frac{dN_0}{dp^+} \right) \, ,
\end{align}
where we have introduced the constant scale $\frac{1}{\mu^2}$ for dimensional consistency.
\footnote{In various references, this distribution is set entirely to $1$ or implicitly divided out from the left-hand side.}
Additionally, we introduce the short-hand notation
\begin{subequations}
\begin{align}
\ul{A} &\equiv \ul{k} - x \ul{p} \, , \\
\ul{B} &\equiv \ul{k} - x \ul{p} + x \ul{q}  \, ,\\
\ul{C} &\equiv \ul{k} - x \ul{p} - (1-x) \ul{q}  \, ,\\
\ul{D} &\equiv \ul{k} - x \ul{p} - \ul{q}  \, ,\\
\Omega_1 - \Omega_2 &\equiv \frac{B_T^2 + x^2 m^2}{2 x (1-x) p^+} 
= - \Delta E^- (\ul{B})  \, ,\\
\Omega_1 - \Omega_3 &\equiv \frac{C_T^2 + x^2 m^2}{2 x (1-x) p^+} 
= - \Delta E^- (\ul{C})  \, ,\\
\Omega_2 - \Omega_3 &= \frac{C_T^2 - B_T^2}{2 x (1-x) p^+} =
- \Delta E^- (\ul{C}) + \Delta E^- (\ul{B})  \, ,\\
\Omega_4 &\equiv \frac{A_T^2 + x^2 m^2}{2 x (1-x) p^+} 
= - \Delta E^- (\ul{A})  \, ,\\
\Omega_5 &\equiv \frac{A_T^2 - D_T^2}{2 x (1-x) p^+} =
- \Delta E^- (\ul{A}) + \Delta E^- (\ul{D})  \, ,
\end{align}
\end{subequations}
and the gluon mean free path $\lambda_g^+ \equiv \frac{C_F}{N_c} \lambda^+$.  (Note also that various references specify $p_T = 0$, which we have not assumed here.)  In terms of these quantities, the exact first order in opacity  LO result Eq.~\eqref{e:LOexact} becomes in the broad source approximation
\begin{align}
\left. x p^+ \frac{dN}{d^2 k \, dx \, d^2 p \, dp^+} \right|_{\ord{\chi^1}} &= \frac{C_F}{2(2\pi)^3 (1-x)} \int\limits_0^{L^+} \frac{d (\delta z^+)}{\lambda_g^+} \int\frac{d^2 q}{\sigma_{el}}\frac{d\sigma^{el}}{d^2 q} \times
\frac{1}{\mu^2} \, \left( p^+ \frac{dN_0}{dp^+} \right)
\notag \\ \times & \Bigg\{ 
\Bigg( 
\left[ \left| \psi (B) \right|^2 - \psi(B) \, \psi^* (C) \right]
+ \frac{1}{N_c^2} \left[ \psi (A) \, \psi^* (B) - \left| \psi (B) \right|^2 \right]
\Bigg)
\Bigg( 1 - \cos\Big( (\Omega_1 - \Omega_2) \delta z^+ \Big) \Bigg)
\notag \\ &+
\Bigg( 2 \left| \psi (C) \right|^2 - \psi (A) \, \psi^* (C) - \psi(B) \, \psi^* (C) \Bigg)
\Bigg( 1 - \cos\Big( (\Omega_1 - \Omega_3) \delta z^+ \Big) \Bigg) 
\notag \\ &+
\psi(B) \, \psi^* (C) 
\Bigg(1 - \cos\Big(  (\Omega_2 - \Omega_3) \delta z^+ \Big) \Bigg)
\notag \\ &+
\Bigg( \psi (D) \, \psi^* (A) - \left| \psi(A) \right|^2 \Bigg)
\Bigg(1 - \cos\Big( \Omega_4 \delta z^+ \Big) \Bigg)
\notag \\ &-
\psi (D) \, \psi^* (A) 
\Bigg(1 - \cos\Big( \Omega_5 \delta z^+ \Big) \Bigg)
\Bigg\} \, ,
\end{align}
with the explicit form
\begin{align}
x p^+ & \left. \frac{dN}{d^2 k \, dx \, d^2 p \, dp^+} \right|_{\ord{\chi^1}} = 
\frac{\alpha_s \, C_F}{2 \pi^2} \int\limits_0^{L^+} \frac{d (\delta z^+)}{\lambda_g^+} \int\frac{d^2 q}{\sigma_{el}}\frac{d\sigma^{el}}{d^2 q} \times
\frac{1}{\mu^2} \, \left( p^+ \frac{dN_0}{dp^+} \right)
\notag \\ \times & \Bigg\{ 
\Bigg( 
\left[ \frac{B_T^2 \, \left[ 1 + (1-x)^2 \right] + x^4 m^2}{ [B_T^2 + x^2 m^2]^2}
- \frac{(\ul{B} \cdot \ul{C}) \, \left[ 1 + (1-x)^2 \right] + x^4 m^2}
{ [B_T^2 + x^2 m^2] \, [C_T^2 + x^2 m^2]} \right]
\notag \\ & \hspace{1cm} +
\frac{1}{N_c^2} \left[ 
\frac{(\ul{B} \cdot \ul{A}) \, \left[ 1 + (1-x)^2 \right] + x^4 m^2}
{ [B_T^2 + x^2 m^2] \, [A_T^2 + x^2 m^2]}
- \frac{B_T^2 \, \left[ 1 + (1-x)^2 \right] + x^4 m^2} { [B_T^2 + x^2 m^2]^2} \right]
\Bigg) 
\Bigg( 1 - \cos\Big( (\Omega_1 - \Omega_2) \delta z^+ \Big) \Bigg)
\notag \\ &+
\Bigg( 2 \frac{C_T^2 \, \left[ 1 + (1-x)^2 \right] + x^4 m^2}{ [C_T^2 + x^2 m^2]^2}
- \frac{(\ul{C} \cdot \ul{A}) \, \left[ 1 + (1-x)^2 \right] + x^4 m^2}
{ [C_T^2 + x^2 m^2] \, [A_T^2 + x^2 m^2]}
- \frac{(\ul{C} \cdot \ul{B}) \, \left[ 1 + (1-x)^2 \right] + x^4 m^2}
{ [C_T^2 + x^2 m^2] \, [B_T^2 + x^2 m^2]}
\Bigg)
\notag \\ & \hspace{1cm} \times
\Bigg( 1 - \cos\Big( (\Omega_1 - \Omega_3) \delta z^+ \Big) \Bigg) 
\notag \\ &+
\frac{(\ul{B} \cdot \ul{C}) \, \left[ 1 + (1-x)^2 \right] + x^4 m^2}
{ [B_T^2 + x^2 m^2] \, [C_T^2 + x^2 m^2]}
\Bigg(1 - \cos\Big(  (\Omega_2 - \Omega_3) \delta z^+ \Big) \Bigg)
\notag \\ &+
\Bigg( 
\frac{(\ul{A} \cdot \ul{D}) \, \left[ 1 + (1-x)^2 \right] + x^4 m^2}
{ [A_T^2 + x^2 m^2] \, [D_T^2 + x^2 m^2]}
- 
\frac{A_T^2 \, \left[ 1 + (1-x)^2 \right] + x^4 m^2}
{ [A_T^2 + x^2 m^2]^2}
\Bigg)
\Bigg(1 - \cos\Big( \Omega_4 \delta z^+ \Big) \Bigg)
\notag \\ &-
\frac{(\ul{A} \cdot \ul{D}) \, \left[ 1 + (1-x)^2 \right] + x^4 m^2}
{ [A_T^2 + x^2 m^2] \, [D_T^2 + x^2 m^2]}
\Bigg(1 - \cos\Big( \Omega_5 \delta z^+ \Big) \Bigg)
\Bigg\} ,
\end{align}
in exact agreement with Eq.~(2.51) of Ref.~\cite{Kang:2016ofv}.  Having obtained agreement in the massive case, we also subsequently agree with the massless limit of Eq. (9.43) of Ref.~\cite{Ovanesyan:2011kn} and even with the expressions for the single- and double-Born contributions separately in Eqs. (9.15) - (9.18), before the broad source approximation is employed.  Finally, in the soft-gluon limit $x \ll 1$ and also dropping the leading $x^2m^2$ mass dependence  we have $\ul{A} = \ul{B} = \ul{k}$ and $\ul{C} = \ul{D} = (\ul{k} - \ul{q})$ such that
\begin{align} \label{e:LOsmallx}
x p^+  \left. \frac{dN}{d^2 k \, dx \, d^2 p \, dp^+} \right|_{\ord{\chi^1}} &
\overset{(x \ll 1)}{\approx} 
\frac{\alpha_s \, C_F}{\pi^2} \int\limits_0^{L^+} \frac{d (\delta z^+)}{\lambda_g^+} \int\frac{d^2 q}{\sigma_{el}}\frac{d\sigma^{el}}{d^2 q} \times
\frac{1}{\mu^2} \, \left( p^+ \frac{dN_0}{dp^+} \right)
\notag \\ & \hspace{1cm} \times
\frac{2 \, \ul{k} \cdot \ul{q}}{k_T^2 \, (k-q)_T^2} 
\Bigg[ 1 - \cos\left( \frac{(k-q)_T^2}{2 x p^+} \, \delta z^+ \right) \Bigg] ,
\end{align}
which can be compared with, e.g., Eq.~(113) of Ref.~\cite{Gyulassy:2000er}, again in the broad source approximation.  Thus, we see that the general form Eq.~\eqref{e:LOexact} contains all of these previous results, encoded into the ingredients of light-front perturbation theory: the light-front wave functions Eq.~\eqref{e:LFWF} and Eq.~\eqref{e:LFWFsq} and the energy denominators Eq.~\eqref{e:Edenom1}.  The next step will be to construct a general recursion relation (the ``reaction operator'' \cite{Gyulassy:2000er}) to construct higher orders in opacity in terms of these ingredients.

%
\section{The Recursion Relation at Finite $x$}
\label{sec:React}
%

Having constructed the first order in opacity contribution to the gluonic substructure of quark jets, we will now proceed to generalize this form to construct higher orders in opacity.  The strategy will be to use the 17 diagrams shown in Figs.~\ref{f:Born} and \ref{f:dblBorn} as the kernel of a diagrammatic recursion relation (the ``reaction operator''), generalizing the expressions in Eqs.~\eqref{e:direct1} and \eqref{e:virt1}.  In principle, one may worry that, in going to higher orders in opacity, the color structures associated with multiple scattering may become increasingly complex and not reduce down to a form of the building blocks Eqs.~\eqref{e:direct1} and \eqref{e:virt1}.  As a preliminary exercise, let us show that this does not occur, at least at the level of the Gaussian averaging Eq.~\eqref{e:Gauss3}: the color factors remain multiplicative at all orders in opacity.

%
\begin{figure}[ht]
\begin{center}
\includegraphics[width= \textwidth]{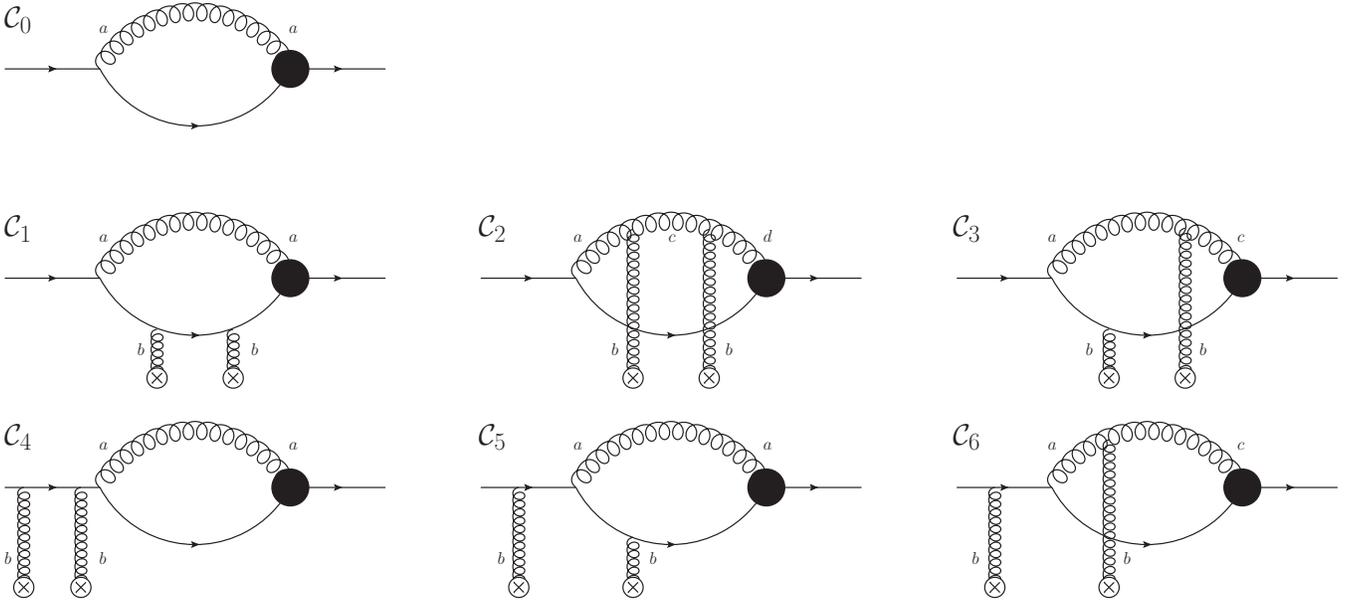} 
\caption{
Color structure of the 17 diagrams shown in Figs.~\ref{f:Born} and \ref{f:dblBorn}.  The placement of the final-state cut is not shown, as it is irrelevant for the determination of the color structure.  For a given order in opacity with color structure $\mathcal{C}_0$, the insertion of one extra two-gluon scattering in the external field can have one of the 6 topologies $\mathcal{C}_{1 - 6}$, which are calculated in Eqs.~\eqref{e:Color_Mult}.
}
\label{f:Color_Mult}
\end{center}
\end{figure}
%

The color structure for a given order in opacity can always be written in the form
\begin{align}
\mathcal{C}_0 \equiv \frac{1}{N_c} \tr[ t^a M^a] \, ,
\end{align}
for some color matrix $M^a$, as illustrated in Fig.~\ref{f:Color_Mult}.  At the next order in opacity under the approximation of Gaussian averaging Eq.~\eqref{e:Gauss3}, two additional gluons in a color-singlet configuration are inserted into the color trace in one of the 6 topologies shown in Fig.~\ref{f:Color_Mult}.  For clarity, here we do not concern ourselves with the placement of the final-state cut or the kinematic factors associated with the scattering amplitude; our goal is to show that the color structure remains multiplicative at any order in opacity, without generating fundamentally new structures.  In addition to the $1/N_c$ present in $\mathcal{C}_0$, the insertion of an extra two-gluon rescattering generates a factor of $1 / 2 N_c$ (from the color averaging Eq.~\eqref{e:coloravg} in the medium) and a factor of $2 N_c / C_F$ (from the conversion to the elastic scattering cross-section Eq.~\eqref{e:elast1}), leading to an overall prefactor of $1 / N_c C_F$ for the color structures $\mathcal{C}_{1-6}$.  The evaluation of these color structures is straightforward using the Fierz identity, yielding
\begin{subequations} \label{e:Color_Mult}
\begin{align}
\mathcal{C}_1 &= \frac{1}{N_c C_F} \tr[ t^b t^b \, t^a M^a ] = \mathcal{C}_0 \, ,  \\
\mathcal{C}_2 &= \frac{1}{N_c C_F} f^{a c b} f^{c d b} \tr[ t^a M^d ] = - \frac{N_c}{C_F} \: \mathcal{C}_0  \, , \\
\mathcal{C}_3 &= \frac{1}{N_c C_F} f^{a c b} \tr[ t^b t^a M^c ] = \frac{i N_c}{2 C_F} \: \mathcal{C}_0  \, , \\
\mathcal{C}_4 &= \frac{1}{N_c C_F} \tr[ t^a \, t^b t^b M^a ] = \mathcal{C}_0  \, , \\
\mathcal{C}_5 &= \frac{1}{N_c C_F} \tr[ t^b t^a t^b M^a ] = \frac{-1}{2 N_c C_F} \: \mathcal{C}_0  \, , \\
\mathcal{C}_6 &= \frac{1}{N_c C_F} f^{a c b} \tr[ t^a t^b M^c ] = \frac{-i N_c}{2 C_F} \: \mathcal{C}_0 \, .
\end{align}
\end{subequations}
The particular color factors obtained in \eq{e:Color_Mult} will be combined with various factors from the scattering amplitude, such as compensating factors of $\pm i$ and factors of $\half$ from the contact limit of double-Born interactions.  But the important feature is that the color structure $\mathcal{C}_0$ is only corrected multiplicatively by additional rescatterings.  Since for the vacuum structure $M^a = t^a$ and $\mathcal{C}_0 = C_F$, we can simply read off the color factors for each type of rescattering from the LO contributions in the opacity expansion, which were already computed in Eqs.~\eqref{e:direct1} and \eqref{e:virt1}.  These color factors are therefore preserved as we iterate the scatterings to higher orders in opacity.

Given the above inductive proof regarding the color structure, we will now proceed to promote the 17 LO diagrams Eqs.~\eqref{e:direct1} and \eqref{e:virt1} to the kernel of a recursion relation which expresses distribution at $N^{th}$ order in opacity in terms of the distribution at $(N-1)^{st}$ order.  We will do this by starting with the last scattering to take place on the jet and evolving the jet history backwards in (light-front) time in both the amplitude and complex-conjugate amplitude, which is possible because the procedure laid out in Sec.~\ref{sec:LO} is manifestly ordered in the positions of the rescatterings.  Because the rescatterings can take place differently in the amplitude and complex-conjugate amplitude, the extension of Eqs.~\eqref{e:direct1} and \eqref{e:virt1} will in general be off-forward, possessing partons of different momenta, or even different partons entirely at a given time in the jet history.  Keeping track of the upper limits of the gluon emission times separately in the amplitude and complex-conjugate amplitude is therefore very important in constructing the phases accumulated by the jet.

Let us define a set of four off-forward amplitudes at $N^{th}$ order in opacity, designating the partonic content of the amplitude and complex conjugate by $F = $ ``Final'' for the quark-gluon system after splitting and $I = $ ``Initial'' for the quark jet before splitting.  The Final/Final , Initial/Final , Final/Initial , and Initial/Initial amplitudes are then written
\begin{align}
&f^{(N)}_{F / F} (\ul{k} , \ul{k}' , \ul{p} \, ; \, x^+ , y^+) \, ,  \\
&f^{(N)}_{I / F} (\ul{k}' , \ul{p} \, ; \, x^+ , y^+)  \, , \\
&f^{(N)}_{F / I} (\ul{k} , \ul{p} \, ; \, x^+ , y^+)  \, , \\
&f^{(N)}_{I / I} (\ul{p} \, ; \, x^+ , y^+) \, , 
\end{align}
where $\ul{k}$ ($\ul{k}'$) is the relative transverse momentum of the gluon with respect to the quark, $x^+$ ($y^+$) is the upper limit on the gluon emission time in the amplitude (complex-conjugate amplitude), and $\ul{p}$ is the transverse center-of-mass momentum of the jet (quark before splitting, or quark + gluon after splitting).  The final measurement of the gluon distribution within the quark jet is given by the Final/Final amplitude with forward kinematics, along with the two-particle phase space:
\begin{align} \label{e:reactobs}
\left. x p^+ \frac{dN}{d^2 k \, dx \, d^2 p \, dp^+} \right|_{\ord{\chi^N}} = 
\frac{C_F}{2 (2\pi)^3 (1-x)} \, f_{F / F}^{(N)} (\ul{k} , \ul{k} , \ul{p} \, ; \, \infty^+ , \infty^+) \, ,
\end{align}
with the understanding that $e^{\pm i \Delta E^- \, \infty^+} \rightarrow 0$ due to the $i\epsilon$ regulator from the Feynman propagator, as in Eq.~\eqref{e:0rad1}.  We have also factored out the vacuum color factor $C_F$, so that any additional color factors are the standard ones written in Eqs.~\eqref{e:direct1} and \eqref{e:virt1}.

With this, we can write down the off-forward generalizations of Eqs.~\eqref{e:direct1} and \eqref{e:virt1} as the kernel of a recursion relation.  We emphasize that different diagrams can mix the different Initial/Final sectors with each other.  For instance, the direct diagram $D_2$ takes place entirely in the Final/Final sector, such that preceding scatterings can also take place in the Final/Final sector.  On the other hand, the direct diagram $D_1$ resolves the gluon emission in both the amplitude and complex-conjugate amplitude, such that preceding scatterings take place in the Initial/Initial sector.  

%
\begin{figure}[ht]
\begin{center}
\includegraphics[width= \textwidth]{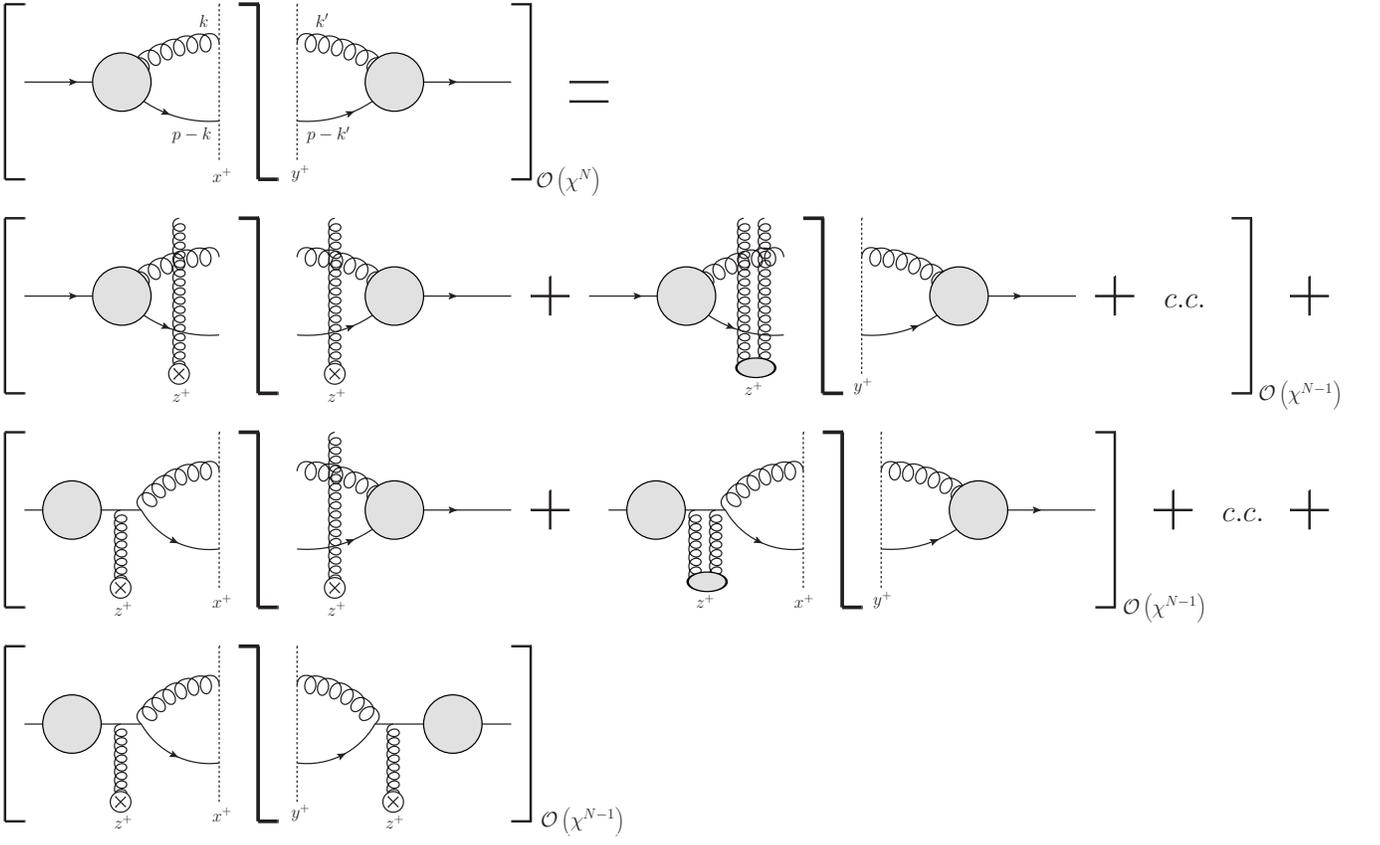} 
\caption{Diagrammatic representation of the recursion relation Eq.~\eqref{e:react1} for the Final/Final sector.  Dashed vertical lines with positions noted beneath indicate the evolution of the upper bound on the gluon emission time, and dangling gluons indicate that they can be attached in all possible ways to the available partons.  The first line of diagrams represents the diagonal Final/Final $\rightarrow$ Final/Final evolution, given by the first 8 lines of Eq.~\eqref{e:react1}.  The second line of diagrams represents the Final/Final $\rightarrow$ Initial/Final transition, given by the next 3 lines of Eq.~\eqref{e:react1}, with the similar transition to the Final/Initial sector denoted by $+ c.c.$.  The last diagram represents the transition to the Initial/Initial sector, given by the last line of Eq.~\eqref{e:react1}.  The recursion relations for the other sectors are similar, coupling only to subsets of these diagrams.
}
\label{f:ReactFF}
\end{center}
\end{figure}
%

The recursion relation for the Final/Final sector, shown in Fig.~\ref{f:ReactFF} is given by
\begin{align} \label{e:react1}
f_{F / F}^{(N)} & (\ul{k} , \ul{k}' , \ul{p} \, ; \, x^+ , y^+) = 
\int\limits_{x_0^+}^{\min[ x^+ , y^+ , R^+]} \hspace{-0.6cm} \frac{dz^+}{\lambda^+} \:
\int \frac{d^2q}{\sigma_{el}} \frac{d\sigma^{el}}{d^2 q}
\: \times \Bigg\{
\notag \\ & 
e^{i [ \Delta E^- (\ul{k} - x \ul{p} + x \ul{q}) - \Delta E^- (\ul{k} - x \ul{p}) ] z^+} \:
e^{i [ \Delta E^- (\ul{k}' - x \ul{p}) - \Delta E^- (\ul{k}' - x \ul{p} + x \ul{q}) ] z^+}
f_{F / F}^{(N-1)} (\ul{k} , \ul{k}' , \ul{p} - \ul{q} \, ; \, z^+ , z^+)
\notag \\ +\frac{N_c}{C_F} &
e^{i [ \Delta E^- (\ul{k} - x \ul{p} - (1-x) \ul{q}) - \Delta E^- (\ul{k} - x \ul{p}) ] z^+} \:
e^{i [ \Delta E^- (\ul{k}' - x \ul{p}) - \Delta E^- (\ul{k}' - x \ul{p} - (1-x) \ul{q}) ] z^+}
f_{F / F}^{(N-1)} (\ul{k} - \ul{q} , \ul{k}' - \ul{q} , \ul{p} - \ul{q} \, ; \, z^+ , z^+)
\notag \\ -\frac{N_c}{2 C_F} &
e^{i [ \Delta E^- (\ul{k} - x \ul{p} + x \ul{q}) - \Delta E^- (\ul{k} - x \ul{p}) ] z^+} \:
e^{i [ \Delta E^- (\ul{k}' - x \ul{p}) - \Delta E^- (\ul{k}' - x \ul{p} - (1-x) \ul{q}) ] z^+}
f_{F / F}^{(N-1)} (\ul{k} , \ul{k}' - \ul{q} , \ul{p} - \ul{q} \, ; \, z^+ , z^+)
\notag \\ -\frac{N_c}{2 C_F} &
e^{i [ \Delta E^- (\ul{k} - x \ul{p} - (1-x) \ul{q}) - \Delta E^- (\ul{k} - x \ul{p}) ] z^+} \:
e^{i [ \Delta E^- (\ul{k}' - x \ul{p}) - \Delta E^- (\ul{k}' - x \ul{p} + x \ul{q}) ] z^+}
f_{F / F}^{(N-1)} (\ul{k} - \ul{q} , \ul{k}' , \ul{p} - \ul{q} \, ; \, z^+ , z^+)
\notag \\ & \hspace{-1cm} + \left[ - \half - \frac{N_c}{2 C_F} \right] 
f_{F / F}^{(N-1)} (\ul{k} , \ul{k}' , \ul{p} \, ; \, z^+ , y^+)
\notag \\ & \hspace{-1cm} + \left[ - \half - \frac{N_c}{2 C_F} \right] 
f_{F / F}^{(N-1)} (\ul{k} , \ul{k}' , \ul{p} \, ; \, x^+ , z^+)
\notag \\ +\frac{N_c}{2 C_F} &
e^{i [ \Delta E^- (\ul{k} - x \ul{p} - \ul{q}) - \Delta E^- (\ul{k} - x \ul{p}) ] z^+} 
f_{F / F}^{(N-1)} (\ul{k} - \ul{q} , \ul{k}' , \ul{p} \, ; \, z^+ , y^+)
\notag \\ +\frac{N_c}{2 C_F} &
e^{i [ \Delta E^- (\ul{k}' - x \ul{p}) - \Delta E^- (\ul{k}' - x \ul{p} - \ul{q}) ] z^+} 
f_{F / F}^{(N-1)} (\ul{k} , \ul{k}' - \ul{q} , \ul{p} \, ; \, x^+ , z^+)
\notag \\ \notag \\ +&
\psi(\ul{k} - x \ul{p}) 
\left[ e^{-i \Delta E^- (\ul{k} - x \ul{p}) x^+} - e^{-i \Delta E^- (\ul{k} - x \ul{p}) z^+} \right]
\left[ \frac{-1}{2 N_c C_F} e^{i [ \Delta E^- (\ul{k}' - x \ul{p}) - \Delta E^- (\ul{k}' - x \ul{p} + x \ul{q}) ] z^+} \right]
f_{I / F}^{(N-1)} (\ul{k}'  , \ul{p} - \ul{q} \, ; \, z^+ , z^+)
\notag \\ +&
\psi(\ul{k} - x \ul{p}) 
\left[ e^{-i \Delta E^- (\ul{k} - x \ul{p}) x^+} - e^{-i \Delta E^- (\ul{k} - x \ul{p}) z^+} \right]
\left[ \frac{N_c}{2 C_F} e^{i [ \Delta E^- (\ul{k}' - x \ul{p}) - \Delta E^- (\ul{k}' - x \ul{p} - (1-x) \ul{q}) ] z^+} \right]
f_{I / F}^{(N-1)} (\ul{k}' - \ul{q} , \ul{p} - \ul{q} \, ; \, z^+ , z^+)
\notag \\ +&
\psi(\ul{k} - x \ul{p}) 
\left[ e^{-i \Delta E^- (\ul{k} - x \ul{p}) x^+} - e^{-i \Delta E^- (\ul{k} - x \ul{p}) z^+} \right]
\left[ - \half \right]
f_{I / F}^{(N-1)} (\ul{k}'  , \ul{p} \, ; \, z^+ , y^+)
\notag \\ \notag \\ +&
\psi^*(\ul{k}' - x \ul{p}) 
\left[ e^{i \Delta E^- (\ul{k}' - x \ul{p}) y^+} - e^{i \Delta E^- (\ul{k}' - x \ul{p}) z^+} \right]
\left[ \frac{-1}{2 N_c C_F} e^{i [ \Delta E^- (\ul{k} - x \ul{p} + x \ul{q}) - \Delta E^- (\ul{k} - x \ul{p}) ] z^+} \right]
f_{F / I}^{(N-1)} (\ul{k}  , \ul{p} - \ul{q} \, ; \, z^+ , z^+)
\notag \\ +&
\psi^*(\ul{k}' - x \ul{p}) 
\left[ e^{i \Delta E^- (\ul{k}' - x \ul{p}) y^+} - e^{i \Delta E^- (\ul{k}' - x \ul{p}) z^+} \right]
\left[ \frac{N_c}{2 C_F} e^{i [ \Delta E^- (\ul{k} - x \ul{p} - (1-x) \ul{q}) - \Delta E^- (\ul{k} - x \ul{p}) ] z^+} \right]
f_{F / I}^{(N-1)} (\ul{k} - \ul{q} , \ul{p} - \ul{q} \, ; \, z^+ , z^+)
\notag \\ +&
\psi^*(\ul{k}' - x \ul{p})
\left[ e^{i \Delta E^- (\ul{k}' - x \ul{p}) y^+} - e^{i \Delta E^- (\ul{k}' - x \ul{p}) z^+} \right]
\left[ - \half \right]
f_{F / I}^{(N-1)} (\ul{k} , \ul{p} \, ; \, x^+ , z^+)
\notag \\ \notag \\ +&
\psi(\ul{k} - x \ul{p})
\left[ e^{-i \Delta E^- (\ul{k} - x \ul{p}) x^+} - e^{-i \Delta E^- (\ul{k} - x \ul{p}) z^+} \right]
\left[ e^{i \Delta E^- (\ul{k}' - x \ul{p}) y^+} - e^{i \Delta E^- (\ul{k}' - x \ul{p}) z^+} \right]
\psi^*(\ul{k}' - x \ul{p})
f_{I / I}^{(N-1)} (\ul{p} - \ul{q} \, ; \, z^+ , z^+) \Bigg\}\, ,
\end{align}
and is the most general expression, receiving contributions from all 17 diagrams.  The terms which resolve a splitting in the amplitude (complex-conjugate amplitude) include an explicit factor of the wave function $\psi$ ($\psi^*$) during the transition from a Final-state to an Initial-state sector.  If all of the medium scatterings in either amplitude occur in the final state, without resolving the splitting, then the corresponding wave function is contained in the initial conditions Eqs.~\eqref{e:initconds} instead of the evolution kernel.  We also note that the upper limit of the $z^+$ integration depends on the minimum of the times $x^+$ and $y^+$, and since $z^+ \leq x^+ , y^+$, it is $z^+$ which will set the upper limit on the next iteration of this recursion relation.

The recursion relation for the Initial/Final sector is given by a subset of these contributions:
\begin{align} \label{e:react2}
f_{I / F}^{(N)} (\ul{k}' , \ul{p} \, ; \, x^+ , y^+) &= 
\int\limits_{x_0^+}^{\min[ x^+ , y^+ , R^+]} \hspace{-0.6cm} \frac{dz^+}{\lambda^+} \:
\int \frac{d^2q}{\sigma_{el}} \frac{d\sigma^{el}}{d^2 q}
\: \times \Bigg\{
\notag \\ +&
\left[ \frac{-1}{2 N_c C_F} e^{i [ \Delta E^- (\ul{k}' - x \ul{p}) - \Delta E^- (\ul{k}' - x \ul{p} + x \ul{q}) ] z^+} \right]
f_{I / F}^{(N-1)} (\ul{k}'  , \ul{p} - \ul{q} \, ; \, z^+ , z^+)
\notag \\ +&
\left[ \frac{N_c}{2 C_F} e^{i [ \Delta E^- (\ul{k}' - x \ul{p}) - \Delta E^- (\ul{k}' - x \ul{p} - (1-x) \ul{q}) ] z^+} \right]
f_{I / F}^{(N-1)} (\ul{k}' - \ul{q} , \ul{p} - \ul{q} \, ; \, z^+ , z^+)
\notag \\ +&
\left[ - \half \right]
f_{I / F}^{(N-1)} (\ul{k}'  , \ul{p} \, ; \, z^+ , y^+)
\notag \\ +& \left[ - \half - \frac{N_c}{2 C_F} \right] 
f_{I / F}^{(N-1)} (\ul{k}' , \ul{p} \, ; \, x^+ , z^+)
\notag \\ +&
\left[ \frac{N_c}{2 C_F}
e^{i [ \Delta E^- (\ul{k}' - x \ul{p}) - \Delta E^- (\ul{k}' - x \ul{p} - \ul{q}) ] z^+} \right]
f_{I / F}^{(N-1)} (\ul{k}' - \ul{q} , \ul{p} \, ; \, x^+ , z^+)
\notag \\ \notag \\ +&
\left[ e^{i \Delta E^- (\ul{k}' - x \ul{p}) y^+} - e^{i \Delta E^- (\ul{k}' - x \ul{p}) z^+} \right]
\psi^*(\ul{k}' - x \ul{p})
f_{I / I}^{(N-1)} (\ul{p} - \ul{q} \, ; \, z^+ , z^+)
\notag \\ +&
\psi^*(\ul{k}' - x \ul{p})
\left[ e^{i \Delta E^- (\ul{k}' - x \ul{p}) y^+} - e^{i \Delta E^- (\ul{k}' - x \ul{p}) z^+} \right]
\left[ - \half \right]
f_{I / I}^{(N-1)} (\ul{p} \, ; \, x^+ , z^+)
\Bigg\} \, ,
\end{align}
and the Final/Initial sector is essentially the complex conjugate:
\begin{align} \label{e:react3}
f_{F / I}^{(N)} (\ul{k} , \ul{p} \, ; \, x^+ , y^+) &= 
\int\limits_{x_0^+}^{\min[ x^+ , y^+ , R^+]} \hspace{-0.6cm} \frac{dz^+}{\lambda^+} \:
\int \frac{d^2q}{\sigma_{el}} \frac{d\sigma^{el}}{d^2 q}
\: \times \Bigg\{
\notag \\ +&
\left[ \frac{-1}{2 N_c C_F} e^{i [ \Delta E^- (\ul{k} - x \ul{p} + x \ul{q}) - \Delta E^- (\ul{k} - x \ul{p}) ] z^+} \right]
f_{F / I}^{(N-1)} (\ul{k}  , \ul{p} - \ul{q} \, ; \, z^+ , z^+)
\notag \\ +&
\left[ \frac{N_c}{2 C_F} e^{i [ \Delta E^- (\ul{k} - x \ul{p} - (1-x) \ul{q}) - \Delta E^- (\ul{k} - x \ul{p}) ] z^+} \right]
f_{F / I}^{(N-1)} (\ul{k} - \ul{q} , \ul{p} - \ul{q} \, ; \, z^+ , z^+)
\notag \\ +&
\left[ - \half \right]
f_{F / I}^{(N-1)} (\ul{k} , \ul{p} \, ; \, x^+ , z^+)
\notag \\ +& \left[ - \half - \frac{N_c}{2 C_F} \right] 
f_{F / I}^{(N-1)} (\ul{k} , \ul{p} \, ; \, z^+ , y^+)
\notag \\ +&
\left[ \frac{N_c}{2 C_F} 
e^{i [ \Delta E^- (\ul{k} - x \ul{p} - \ul{q}) - \Delta E^- (\ul{k} - x \ul{p}) ] z^+}  \right]
f_{F / I}^{(N-1)} (\ul{k} - \ul{q} , \ul{p} \, ; \, z^+ , y^+)
\notag \\ \notag \\ +&
\psi(\ul{k} - x \ul{p})
\left[ e^{-i \Delta E^- (\ul{k} - x \ul{p}) x^+} - e^{-i \Delta E^- (\ul{k} - x \ul{p}) z^+} \right]
f_{I / I}^{(N-1)} (\ul{p} - \ul{q} \, ; \, z^+ , z^+) 
\notag \\ +&
\psi(\ul{k} - x \ul{p}) 
\left[ e^{-i \Delta E^- (\ul{k} - x \ul{p}) x^+} - e^{-i \Delta E^- (\ul{k} - x \ul{p}) z^+} \right]
\left[ - \half \right]
f_{I / I}^{(N-1)} (\ul{p} \, ; \, z^+ , y^+)
\Bigg\} \, .
\end{align}
The Initial/Initial sector has already resolved both gluon splittings, and thus depends on $x^+ , y^+$ only through the minimum which sets the upper limit of the $z^+$ integral:
\begin{align} \label{e:react4}
f_{I / I}^{(N)} (\ul{p} \, ; \, x^+ , y^+) &= 
f_{I / I}^{(N)} (\ul{p} \, ; \, \min[x^+ , y^+]) = \hspace{-0.6cm}
\int\limits_{x_0^+}^{\min[ x^+ , y^+ , R^+]} \hspace{-0.6cm} \frac{dz^+}{\lambda^+} \:
\int \frac{d^2q}{\sigma_{el}} \frac{d\sigma^{el}}{d^2 q}
\: \Bigg[ f_{I / I}^{(N-1)} (\ul{p} - \ul{q} \, ; \, z^+) - 
f_{I / I}^{(N-1)} (\ul{p} \, ; \, z^+) \Bigg]\, ,
\end{align}
which is just the recursion relation (``reaction operator'') for jet broadening, see e.g. Eq.~(8.8) of Ref.~\cite{Ovanesyan:2011kn}.

The last ingredient to the recursion relations is the initial conditions, which simply correspond to the vacuum splitting (where applicable), along with the initial distribution of quark jets:
\begin{subequations} \label{e:initconds}
\begin{align}
f_{F / F}^{(0)} (\ul{k} , \ul{k}' , \ul{p} \, ; \, x^+ , y^+) &=
\psi(\ul{k} - x \ul{p})
\left[ e^{-i \Delta E^- (\ul{k} - x \ul{p}) x^+} - e^{-i \Delta E^- (\ul{k} - x \ul{p}) x_0^+} \right]
\notag \\ & \hspace{1cm} \times
\left[ e^{i \Delta E^- (\ul{k}' - x \ul{p}) y^+} - e^{i \Delta E^- (\ul{k}' - x \ul{p}) x_0^+} \right]
\psi^*(\ul{k}' - x \ul{p})
\left( p^+ \frac{dN_0}{d^2 p \, dp^+} \right) \, , 
\\ \notag \\
f_{I / F}^{(0)} (\ul{k}' , \ul{p} \, ; \, x^+ , y^+) &=
\psi^*(\ul{k}' - x \ul{p})
\left[ e^{i \Delta E^- (\ul{k}' - x \ul{p}) y^+} - e^{i \Delta E^- (\ul{k}' - x \ul{p}) x_0^+} \right]
\left( p^+ \frac{dN_0}{d^2 p \, dp^+} \right)\, , 
\\ \notag \\
f_{F / I}^{(0)} (\ul{k} , \ul{p} \, ; \, x^+ , y^+) &= 
\psi(\ul{k} - x \ul{p})
\left[ e^{-i \Delta E^- (\ul{k} - x \ul{p}) x^+} - e^{-i  \Delta E^- (\ul{k} - x \ul{p}) x_0^+} \right]
\left( p^+ \frac{dN_0}{d^2 p \, dp^+} \right) \, , 
\\ \notag \\
f_{I / I}^{(0)} (\ul{p} \, ; \, x^+ , y^+) &= \left( p^+ \frac{dN_0}{d^2 p \, dp^+} \right) \, .
\end{align}
\end{subequations}
It is interesting to note that the recursion relations between these four functions can be cast in the form of a matrix equation, with a particularly simple triangular form due to their causal structure:
\begin{align} \label{e:reactmtx}
\begin{bmatrix*}[l]
f_{F / F}^{(N)} (\ul{k} , \ul{k}' , \ul{p} \, ; \, x^+ , y^+) \\
f_{I / F}^{(N)} (\ul{k}' , \ul{p} \, ; \, x^+ , y^+) \\
f_{F / I}^{(N)} (\ul{k} , \ul{p} \, ; \, x^+ , y^+) \\
f_{I / I}^{(N)} (\ul{p} \, ; \, x^+ , y^+)
\end{bmatrix*}
= \int\limits_{x_0^+}^{\min[ x^+ , y^+ , R^+ ]} \frac{dz^+}{\lambda^+} 
\int\frac{d^2 q}{\sigma_{el}} \frac{d\sigma^{el}}{d^2 q}
\begingroup
\renewcommand*{\arraystretch}{1.4}
\begin{bmatrix}
\mathcal{K}_1^{} & \mathcal{K}_2 & \mathcal{K}_3 & \mathcal{K}_4 \\
0 & \mathcal{K}_5 & 0& \mathcal{K}_6 \\
0 & 0 & \mathcal{K}_7 & \mathcal{K}_8 \\
0 & 0 & 0 & \mathcal{K}_9
\end{bmatrix}
\endgroup
\begin{bmatrix*}[l]
f_{F / F}^{(N-1)} (\ul{k} , \ul{k}' , \ul{p} \, ; \, x^+ , y^+) \\
f_{I / F}^{(N-1)} (\ul{k}' , \ul{p} \, ; \, x^+ , y^+) \\
f_{F / I}^{(N-1)} (\ul{k} , \ul{p} \, ; \, x^+ , y^+) \\
f_{I / I}^{(N-1)} (\ul{p} \, ; \, x^+ , y^+)
\end{bmatrix*} ,
\end{align}
where the matrix of integral kernels $\mathcal{K}_{1-9}$ is an explicit representation of the reaction operator.  Note that, being complex conjugates, the Initial/Final and Final/Initial sectors are mutually exclusive and do not mix, leading to the vanishing of the $\{2,3\}$ and $\{3, 2\}$ elements of the kernel in \eq{e:reactmtx}.  We can write these integral kernels themselves explicitly using shift operators as
\begin{subequations} \label{e:reactker}
\begin{align}
\mathcal{K}_1 &= 
\left[ e^{i [ \Delta E^- (\ul{k} - x \ul{p} + x \ul{q}) - \Delta E^- (\ul{k} - x \ul{p}) ] z^+} \:
e^{i [ \Delta E^- (\ul{k}' - x \ul{p}) - \Delta E^- (\ul{k}' - x \ul{p} + x \ul{q}) ] z^+} \right]
\left[ 
e^{- \ul{q} \cdot \ul{\nabla}_{p}} \:
e^{+(z^+ - x^+) \partial_{x^+}} \:
e^{+(z^+ - y^+) \partial_{y^+}} \right]
\notag \\ &+
\left[
\frac{N_c}{C_F} 
e^{i [ \Delta E^- (\ul{k} - x \ul{p} - (1-x) \ul{q}) - \Delta E^- (\ul{k} - x \ul{p}) ] z^+} \:
e^{i [ \Delta E^- (\ul{k}' - x \ul{p}) - \Delta E^- (\ul{k}' - x \ul{p} - (1-x) \ul{q}) ] z^+}
\right]
\notag \\ & \hspace{1cm} \times
\left[
e^{- \ul{q} \cdot \ul{\nabla}_{k}} \:
e^{- \ul{q} \cdot \ul{\nabla}_{k'}} \:
e^{- \ul{q} \cdot \ul{\nabla}_{p}} \:
e^{+(z^+ - x^+) \partial_{x^+}} \:
e^{+(z^+ - y^+) \partial_{y^+}}
\right]
\notag \\ &-
\left[ 
\frac{N_c}{2 C_F} 
e^{i [ \Delta E^- (\ul{k} - x \ul{p} + x \ul{q}) - \Delta E^- (\ul{k} - x \ul{p}) ] z^+} \:
e^{i [ \Delta E^- (\ul{k}' - x \ul{p}) - \Delta E^- (\ul{k}' - x \ul{p} - (1-x) \ul{q}) ] z^+}
\right]
\notag \\ & \hspace{1cm} \times
\left[
e^{- \ul{q} \cdot \ul{\nabla}_{k'}} \:
e^{- \ul{q} \cdot \ul{\nabla}_{p}} \:
e^{+(z^+ - x^+) \partial_{x^+}} \:
e^{+(z^+ - y^+) \partial_{y^+}}
\right]
\notag \\ &-
\left[
\frac{N_c}{2 C_F}
e^{i [ \Delta E^- (\ul{k} - x \ul{p} - (1-x) \ul{q}) - \Delta E^- (\ul{k} - x \ul{p}) ] z^+} \:
e^{i [ \Delta E^- (\ul{k}' - x \ul{p}) - \Delta E^- (\ul{k}' - x \ul{p} + x \ul{q}) ] z^+}
\right]
\notag \\ & \hspace{1cm} \times
\left[
e^{- \ul{q} \cdot \ul{\nabla}_{k}} \:
e^{- \ul{q} \cdot \ul{\nabla}_{p}} \:
e^{+(z^+ - x^+) \partial_{x^+}} \:
e^{+(z^+ - y^+) \partial_{y^+}}
\right]
\notag \\ &+ 
\left[ - \half - \frac{N_c}{2 C_F} \right] 
\left[
e^{+(z^+ - x^+) \partial_{x^+}} + e^{+(z^+ - y^+) \partial_{y^+}}
\right]
\notag \\ &+
\left[
\frac{N_c}{2 C_F}
e^{i [ \Delta E^- (\ul{k} - x \ul{p} - \ul{q}) - \Delta E^- (\ul{k} - x \ul{p}) ] z^+} 
\right]
\left[
e^{- \ul{q} \cdot \ul{\nabla}_{k}} \:
e^{+(z^+ - x^+) \partial_{x^+}}
\right]
\notag \\ &+
\left[
\frac{N_c}{2 C_F}
e^{i [ \Delta E^- (\ul{k}' - x \ul{p}) - \Delta E^- (\ul{k}' - x \ul{p} - \ul{q}) ] z^+} 
\right]
\left[
e^{- \ul{q} \cdot \ul{\nabla}_{k'}} \:
e^{+(z^+ - y^+) \partial_{y^+}}
\right]  \, , 
\end{align}
\begin{align}
\mathcal{K}_2 &=
\psi(\ul{k} - x \ul{p}) 
\left[ e^{-i \Delta E^- (\ul{k} - x \ul{p}) x^+} - e^{-i \Delta E^- (\ul{k} - x \ul{p}) z^+} \right]
\notag \\ & \hspace{1cm}
\times \Bigg\{
\left[ 
\frac{-1}{2 N_c C_F} 
e^{i [ \Delta E^- (\ul{k}' - x \ul{p}) - \Delta E^- (\ul{k}' - x \ul{p} + x \ul{q}) ] z^+} 
\right]
\left[
e^{- \ul{q} \cdot \ul{\nabla}_p} \:
e^{+ (z^+ - x^+) \partial_{x^+}} \:
e^{+ (z^+ - y^+) \partial_{y^+}} 
\right]
\notag \\ & \hspace{1cm} +
\left[ 
\frac{N_c}{2 C_F} 
e^{i [ \Delta E^- (\ul{k}' - x \ul{p}) - \Delta E^- (\ul{k}' - x \ul{p} - (1-x) \ul{q}) ] z^+} 
\right]
\left[
e^{- \ul{q} \cdot \ul{\nabla}_{k'}} \:
e^{- \ul{q} \cdot \ul{\nabla}_p} \:
e^{+ (z^+ - x^+) \partial_{x^+}} \:
e^{+ (z^+ - y^+) \partial_{y^+}} 
\right]
\notag \\ & \hspace{1cm} +
\left[ - \half \right]
\left[
e^{+ (z^+ - x^+) \partial_{x^+}} 
\right] \Bigg\} \, , 
\\ \notag \\
\mathcal{K}_3 &=
\psi^*(\ul{k}' - x \ul{p}) 
\left[ e^{i \Delta E^- (\ul{k}' - x \ul{p}) y^+} - e^{i \Delta E^- (\ul{k}' - x \ul{p}) z^+} \right]
\notag \\ & \hspace{1cm}
\times \Bigg\{
\left[ 
\frac{-1}{2 N_c C_F} 
e^{i [ \Delta E^- (\ul{k} - x \ul{p} + x \ul{q}) - \Delta E^- (\ul{k} - x \ul{p}) ] z^+} 
\right]
\left[
e^{- \ul{q} \cdot \ul{\nabla}_p} \:
e^{+ (z^+ - x^+) \partial_{x^+}} \:
e^{+ (z^+ - y^+) \partial_{y^+}} 
\right]
\notag \\ & \hspace{1cm} +
\left[ 
\frac{N_c}{2 C_F} 
e^{i [ \Delta E^- (\ul{k} - x \ul{p} - (1-x) \ul{q}) - \Delta E^- (\ul{k} - x \ul{p}) ] z^+} 
\right]
\left[
e^{- \ul{q} \cdot \ul{\nabla}_k} \:
e^{- \ul{q} \cdot \ul{\nabla}_p} \:
e^{+ (z^+ - x^+) \partial_{x^+}} \:
e^{+ (z^+ - y^+) \partial_{y^+}}
\right]
\notag \\ & \hspace{1cm} +
\left[ - \half \right]
\left[
e^{+ (z^+ - y^+) \partial_{y^+}}
\right] \Bigg\}\, , 
\\ \notag \\
\mathcal{K}_4 &=
\psi(\ul{k} - x \ul{p})
\left[ e^{-i \Delta E^- (\ul{k} - x \ul{p}) x^+} - e^{-i \Delta E^- (\ul{k} - x \ul{p}) z^+} \right]
\left[ e^{i \Delta E^- (\ul{k}' - x \ul{p}) y^+} - e^{i \Delta E^- (\ul{k}' - x \ul{p}) z^+} \right]
\psi^*(\ul{k}' - x \ul{p})
\notag \\ & \hspace{1cm} \times
\left[
e^{- \ul{q} \cdot \ul{\nabla}_p} \:
e^{+ (z^+ - x^+) \partial_{x^+}} \:
e^{+ (z^+ - y^+) \partial_{y^+}}\, , 
\right]
\\ \notag \\
\mathcal{K}_5 &=
\left[ 
\frac{-1}{2 N_c C_F} 
e^{i [ \Delta E^- (\ul{k}' - x \ul{p}) - \Delta E^- (\ul{k}' - x \ul{p} + x \ul{q}) ] z^+} 
\right]
\left[
e^{- \ul{q} \cdot \ul{\nabla}_p} \:
e^{+ (z^+ - x^+) \partial_{x^+}} \:
e^{+ (z^+ - y^+) \partial_{y^+}}
\right]
\notag \\ &+
\left[ 
\frac{N_c}{2 C_F} 
e^{i [ \Delta E^- (\ul{k}' - x \ul{p}) - \Delta E^- (\ul{k}' - x \ul{p} - (1-x) \ul{q}) ] z^+} 
\right]
\left[
e^{- \ul{q} \cdot \ul{\nabla}_{k'}} \:
e^{- \ul{q} \cdot \ul{\nabla}_p} \:
e^{+ (z^+ - x^+) \partial_{x^+}} \:
e^{+ (z^+ - y^+) \partial_{y^+}}
\right]
\notag \\ &+
\left[ - \half \right] \left[ e^{+ (z^+ - x^+) \partial_{x^+}} \right]
+ \left[ - \half - \frac{N_c}{2 C_F} \right] \left[ e^{+ (z^+ - y^+) \partial_{y^+}} \right]
\notag \\ &+
\left[
\frac{N_c}{2 C_F}
e^{i [ \Delta E^- (\ul{k}' - x \ul{p}) - \Delta E^- (\ul{k}' - x \ul{p} - \ul{q}) ] z^+} 
\right]
\left[
e^{- \ul{q} \cdot \ul{\nabla}_{k'}} \:
e^{+ (z^+ - y^+) \partial_{y^+}}
\right]\, , 
\\ \notag \\
\mathcal{K}_6 &=
\psi^*(\ul{k}' - x \ul{p})
\left[ e^{i \Delta E^- (\ul{k}' - x \ul{p}) y^+} - e^{i \Delta E^- (\ul{k}' - x \ul{p}) z^+} \right]
\notag \\ & \hspace{1cm} \times \Bigg\{
\left[
e^{- \ul{q} \cdot \ul{\nabla}_p} \:
e^{+ (z^+ - x^+) \partial_{x^+}} \:
e^{+ (z^+ - y^+) \partial_{y^+}}
\right]
+ \left[ - \half \right] \left[ e^{+ (z^+ - y^+) \partial_{y^+}} \right] \Bigg\}\, , 
\\ \notag \\
\mathcal{K}_7 &=
\left[ 
\frac{-1}{2 N_c C_F} 
e^{i [ \Delta E^- (\ul{k} - x \ul{p} + x \ul{q}) - \Delta E^- (\ul{k} - x \ul{p}) ] z^+} 
\right]
\left[
e^{- \ul{q} \cdot \ul{\nabla}_p} \:
e^{+ (z^+ - x^+) \partial_{x^+}} \:
e^{+ (z^+ - y^+) \partial_{y^+}}
\right]
\notag \\ &+
\left[ 
\frac{N_c}{2 C_F} 
e^{i [ \Delta E^- (\ul{k} - x \ul{p} - (1-x) \ul{q}) - \Delta E^- (\ul{k} - x \ul{p}) ] z^+} 
\right]
\left[
e^{- \ul{q} \cdot \ul{\nabla}_k} \:
e^{- \ul{q} \cdot \ul{\nabla}_p} \:
e^{+ (z^+ - x^+) \partial_{x^+}} \:
e^{+ (z^+ - y^+) \partial_{y^+}}
\right]
\notag \\ &+
\left[ - \half \right] \left[ e^{+ (z^+ - y^+) \partial_{y^+}} \right]
+ \left[ - \half - \frac{N_c}{2 C_F} \right] \left[ e^{+ (z^+ - x^+) \partial_{x^+}} \right]
\notag \\ &+
\left[ 
\frac{N_c}{2 C_F} 
e^{i [ \Delta E^- (\ul{k} - x \ul{p} - \ul{q}) - \Delta E^- (\ul{k} - x \ul{p}) ] z^+}  
\right]
\left[
e^{- \ul{q} \cdot \ul{\nabla}_k} \:
e^{+ (z^+ - x^+) \partial_{x^+}} \:
\right]\, , 
\\ \notag \\
\mathcal{K}_8 &=
\psi(\ul{k} - x \ul{p})
\left[ e^{-i \Delta E^- (\ul{k} - x \ul{p}) x^+} - e^{-i \Delta E^- (\ul{k} - x \ul{p}) z^+} \right]
\notag \\ & \hspace{1cm} \times \Bigg\{
\left[
e^{- \ul{q} \cdot \ul{\nabla}_p} \:
e^{+ (z^+ - x^+) \partial_{x^+}} \:
e^{+ (z^+ - y^+) \partial_{y^+}}
\right]
+ \left[ - \half \right] \left[ e^{+ (z^+ - x^+) \partial_{x^+}} \right] \Bigg\}\, , 
\end{align}
\begin{align}
\mathcal{K}_9 &=
\left[
e^{- \ul{q} \cdot \ul{\nabla}_p} \:
e^{+ (z^+ - x^+) \partial_{x^+}} \:
e^{+ (z^+ - y^+) \partial_{y^+}}
\right]
+ \left[ - \half \right] \left[ e^{+ (z^+ - x^+) \partial_{x^+}} + e^{+ (z^+ - y^+) \partial_{y^+}} \right]\, .
\end{align}
\end{subequations}

The recursion relations Eqs.~\eqref{e:react1}, \eqref{e:react2}, \eqref{e:react3}, \eqref{e:react4} (or equivalently, the matrix equation Eq.~\eqref{e:reactmtx} with the reaction operator Eq.~\eqref{e:reactker}), together with the initial conditions Eqs.~\eqref{e:initconds} and the final observable Eq.~\eqref{e:reactobs}, are one of the main results of this work.  We emphasize that these results have expressed the jet substructure observable in terms of universal ingredients, the light-front wave function and the light-front energy denominator, which apply for any partonic splitting and employ exact kinematics.  The only kinematic approximations necessary to obtain this result are the eikonal approximation Eq.~\eqref{e:Aext2} for scattering in the external potential and the Gaussian averaging Eq.~\eqref{e:Gauss3}.  With this, the  calculation of any desired order in opacity is cumbersome but straightforward.  Although the output rapidly becomes so large that it is impractical to manage by hand, the algebraic computation of Eq.~\eqref{e:reactmtx} is quick and straightforward on a computer.  

We also note that the upper triangular structure of the reaction operator in  Eq.~\eqref{e:reactmtx} itself is particularly amenable to an analytic solution.  Because the reaction operator in this form has essentially already been reduced under Gauss-Jordan elimination, the coupled set of four difference (differential) equations can be reduced down to four independent ordinary difference (differential) equations.  For example, the recursion relation for the Initial/Initial sector Eq.~\eqref{e:react4} depends only on the unknown function $f_{I/I}^{(N)}$, whose solution is already well-known in the context of jet broadening (see, e.g., Eqs.~(8.10) and (8.11) of \cite{Ovanesyan:2011kn}):
\begin{align} \label{e:broadsoln}
f_{I / I}^{(N)} (\ul{p} \, ; \, x^+ , y^+) &= 
\hspace{-0.6cm} \int\limits_{x_0^+}^{\min[x^+ , y^+ , R^+]} \hspace{-0.6cm}
\frac{dz_N^+}{\lambda^+} \int\limits_{x_0^+}^{z_N^+} \frac{dz_{N-1}^+}{\lambda^+} \cdots
\int\limits_{x_0^+}^{z_2^+} \frac{dz_1^+}{\lambda^+} \,
\int d^2 r \, e^{- i \ul{p} \cdot \ul{r}} \left[ \frac{(2\pi)^2}{\sigma_{el}} \sigma(\ul{r}) - 1 \right]^N
\, \left( p^+ \frac{dN_0}{d^2 r \, dp^+} \right) ,
\end{align}
with $\sigma(\ul{r}) \equiv \int \frac{d^2 q}{(2\pi)^2} \: e^{i \ul{q} \cdot \ul{r}} \: \frac{d\sigma^{el}}{d^2 q}$ and $\left( p^+ \frac{dN_0}{d^2 r \, dp^+} \right) \equiv \int\frac{d^2 p}{(2\pi)^2} \: e^{i \ul{q} \cdot \ul{r}} \: \left( p^+ \frac{dN_0}{d^2 p \, dp^+} \right)$ the Fourier transforms of the elastic scattering cross-section and initial source distributions, respectively.  With the use of the explicit solution Eq.~\eqref{e:broadsoln}, the recursion relations for the Initial/Final and Final/Initial sectors Eqs.~\eqref{e:react2} and \eqref{e:react3} both become decoupled equations for their respective functions and can be solved similarly.  We leave the implementation of this strategy for future work.

%
\section{Results}
\label{sec:Results}
%

Having fully constructed the recursion relations Eq.~\eqref{e:reactmtx}, one immediate application we can pursue is the construction of the second order in opacity gluon-in-quark-jet distribution.  A straightforward but tedious application of Eq.~\eqref{e:reactmtx} yields the general form
\begin{align} \label{e:NLO1}
\left. x p^+ \, \frac{dN}{d^2 k \, dx \, d^2 p \, dp^+} \right|_{\ord{\chi^2}} &=
\frac{C_F}{2 (2\pi)^3 (1-x)} \int\limits_{x_0^+}^{R^+} \frac{dz_2^+}{\lambda^+} \int\limits_{x_0^+}^{z_2^+} \frac{dz_1^+}{\lambda^+} \:
\int \frac{d^2 q_1}{\sigma_{el}} \frac{d^2 q_2}{\sigma_{el}} \:
\frac{d\sigma^{el}}{d^2 q_1} \frac{d\sigma^{el}}{d^2 q_2}
\times \Bigg\{
\left( p^+ \, \frac{dN_0}{d^2 p \, dp^+} \right) \: \mathcal{N}_1 
\notag \\ & \hspace{-1cm} +
\left( p^+ \, \frac{dN_0}{d^2 (p - q_1) \, dp^+} \right) \: \mathcal{N}_2 +
\left( p^+ \, \frac{dN_0}{d^2 (p - q_2) \, dp^+} \right) \: \mathcal{N}_3 +
\left( p^+ \, \frac{dN_0}{d^2 (p - q_1 - q_2) \, dp^+} \right) \: \mathcal{N}_4 \Bigg\} ,
\end{align}
with the functions $\mathcal{N}_{1-4}$ containing the various color factors, wave functions, and cosines associated with the diagrams.  We note that the expressions to follow represent the sum of the $221$ Feynman diagrams which contribute at second order in opacity.  In the resulting expressions, we denote $\delta z_1 \equiv z_1^+ - x_0^+$ and $\delta z_2 \equiv z_2^+ - z_1^+$ for brevity, leading to:

\begin{subequations} \label{e:NLOexact}
\begin{align}
\mathcal{N}_1 &=
\notag \\ \notag \\ & \left| \psi(\ul{k} - x \ul{p}) \right|^2 \Bigg[
\frac{(C_F+N_c)^2}{C_F^2}
-\frac{N_c (C_F+N_c) }{C_F^2}\cos (\delta  z_1 \Delta  E^-(\ul{k} - x \ul{p}))
+\frac{N_c^2 }{2 C_F^2}\cos (\delta  z_2 \Delta  E^-(\ul{k} - x \ul{p}))
\notag \\ &\hspace{1cm}
-\frac{N_c (2 C_F+N_c) }{2 C_F^2}\cos ((\delta  z_1+\delta  z_2) \Delta  E^-(\ul{k} - x \ul{p}))
\Bigg]
\notag \\ &+ \psi(\ul{k} - x \ul{p}) \, \psi^*(\ul{k} - x \ul{p} - \ul{q}_1) \Bigg[
\frac{N_c (C_F+N_c) }{C_F^2}\cos (\delta  z_1 \Delta  E^-(\ul{k} - x \ul{p}))
-\frac{N_c^2 }{2 C_F^2}\cos (\delta  z_2 \Delta  E^-(\ul{k} - x \ul{p}))
\notag \\ &\hspace{1cm}
-\frac{N_c (C_F+N_c) }{C_F^2}\cos (\delta  z_1 \Delta  E^-(\ul{k} - x \ul{p})-\delta  z_1 \Delta  E^-(\ul{k} - x \ul{p} - \ul{q}_1))
\notag \\ &\hspace{1cm}
+\frac{N_c^2 }{2 C_F^2}\cos (\delta  z_1 \Delta  E^-(\ul{k} - x \ul{p} - \ul{q}_1)+\delta  z_2 \Delta  E^-(\ul{k} - x \ul{p})) 
\Bigg]
\notag \\ &+ \psi(\ul{k} - x \ul{p}) \, \psi^*(\ul{k} - x \ul{p} - \ul{q}_2) \Bigg[
-\frac{N_c^2 }{2 C_F^2}\cos (\delta  z_2 \Delta  E^-(\ul{k} - x \ul{p}))
+\frac{N_c (2 C_F+N_c) }{2 C_F^2}\cos (\delta  z_1 \Delta  E^-(\ul{k} - x \ul{p})+\delta  z_2 \Delta  E^-(\ul{k} - x \ul{p}))
\notag \\ &\hspace{1cm}
-\frac{N_c (C_F+N_c) }{2 C_F^2}\cos ((\delta  z_1+\delta  z_2) (\Delta  E^-(\ul{k} - x \ul{p})-\Delta  E^-(\ul{k} - x \ul{p} - \ul{q}_2)))
\notag \\ &\hspace{1cm}
-\frac{N_c (C_F+N_c) }{2 C_F^2}\cos ((\delta  z_1+\delta  z_2) (\Delta  E^-(\ul{k} - x \ul{p} - \ul{q}_2)-\Delta  E^-(\ul{k} - x \ul{p})))
\notag \\ &\hspace{1cm}
+\frac{N_c^2 }{2 C_F^2}\cos (-\delta  z_2 \Delta  E^-(\ul{k} - x \ul{p} - \ul{q}_2)+\delta  z_1 \Delta  E^-(\ul{k} - x \ul{p})+\delta  z_2 \Delta  E^-(\ul{k} - x \ul{p}))
\notag \\ &\hspace{1cm}
+\frac{N_c^2 }{2 C_F^2}\cos (-\delta  z_1 \Delta  E^-(\ul{k} - x \ul{p} - \ul{q}_2)-\delta  z_2 \Delta  E^-(\ul{k} - x \ul{p} - \ul{q}_2)+\delta  z_2 \Delta  E^-(\ul{k} - x \ul{p})) \Bigg]
\notag \\ &+ \psi(\ul{k} - x \ul{p} - \ul{q}_1) \, \psi^*(\ul{k} - x \ul{p} - \ul{q}_2) \Bigg[
\frac{N_c^2 }{2 C_F^2}\cos (\delta  z_2 \Delta  E^-(\ul{k} - x \ul{p}))
-\frac{N_c^2 }{2 C_F^2}\cos (\delta  z_1 \Delta  E^-(\ul{k} - x \ul{p} - \ul{q}_1)+\delta  z_2 \Delta  E^-(\ul{k} - x \ul{p}))
\notag \\ &\hspace{1cm}
-\frac{N_c^2 }{2 C_F^2}\cos (-\delta  z_1 \Delta  E^-(\ul{k} - x \ul{p} - \ul{q}_2)-\delta  z_2 \Delta  E^-(\ul{k} - x \ul{p} - \ul{q}_2)+\delta  z_2 \Delta  E^-(\ul{k} - x \ul{p}))
\notag \\ &\hspace{1cm}
+\frac{N_c^2 }{2 C_F^2}\cos (\delta  z_1 \Delta  E^-(\ul{k} - x \ul{p} - \ul{q}_1)-\delta  z_1 \Delta  E^-(\ul{k} - x \ul{p} - \ul{q}_2)-\delta  z_2 \Delta  E^-(\ul{k} - x \ul{p} - \ul{q}_2)+\delta  z_2 \Delta  E^-(\ul{k} - x \ul{p})) 
\Bigg]
\notag \\ &+ \psi(\ul{k} - x \ul{p}) \, \psi^*(\ul{k} - x \ul{p}  - \ul{q}_1 - \ul{q}_2) \Bigg[
-\frac{N_c^2 }{2 C_F^2}\cos (-\delta  z_2 \Delta  E^-(\ul{k} - x \ul{p} - \ul{q}_2)+\delta  z_1 \Delta  E^-(\ul{k} - x \ul{p})+\delta  z_2 \Delta  E^-(\ul{k} - x \ul{p}))
\notag \\ &\hspace{1cm}
+\frac{N_c^2 }{2 C_F^2}\cos (-\delta  z_1 \Delta  E^-(\ul{k} - x \ul{p} - \ul{q}_1 - \ul{q}_2)-\delta  z_2 \Delta  E^-(\ul{k} - x \ul{p} - \ul{q}_2)+\delta  z_1 \Delta  E^-(\ul{k} - x \ul{p})+\delta  z_2 \Delta  E^-(\ul{k} - x \ul{p}))
\Bigg] \, , 
\end{align}
\newpage
\begin{align}
\mathcal{N}_2 &=
\notag \\ \notag \\ & \left| \psi(\ul{k} - x \ul{p}) \right|^2 \Bigg[
-\frac{C_F+N_c}{C_F}
+\frac{N_c }{C_F}\cos (\delta  z_2 \Delta  E^-(\ul{k} - x \ul{p}))
\Bigg]
\notag \\ &+ \psi(\ul{k} - x \ul{p})  \, \psi^*(\ul{k} - x \ul{p} - \ul{q}_2) \Bigg[
-\frac{N_c }{C_F}\cos (\delta  z_2 \Delta  E^-(\ul{k} - x \ul{p}))
+\frac{N_c }{C_F}\cos (\delta  z_2 \Delta  E^-(\ul{k} - x \ul{p})-\delta  z_2 \Delta  E^-(\ul{k} - x \ul{p} - \ul{q}_2))
\Bigg]
\notag \\ &+ \psi(\ul{k} - x \ul{p})  \, \psi^*(\ul{k} - x \ul{p} + x \ul{q}_1) \Bigg[
\frac{(C_F+N_c) }{C_F^2 N_c}\cos (\delta  z_1 \Delta  E^-(\ul{k} - x \ul{p} + x \ul{q}_1))
-\frac{(C_F+N_c) }{C_F^2 N_c} 
\notag \\ &\hspace{1cm}
-\frac{1}{2 C_F^2} \cos (\delta  z_1 \Delta  E^-(\ul{k} - x \ul{p} + x \ul{q}_1)+\delta  z_2 \Delta  E^-(\ul{k} - x \ul{p}))
+\frac{1}{2 C_F^2} \cos (\delta  z_2 \Delta  E^-(\ul{k} - x \ul{p}))
\Bigg]
\notag \\ &+ \psi(\ul{k} - x \ul{p} - \ul{q}_2)  \, \psi^*(\ul{k} - x \ul{p} + x \ul{q}_1) \Bigg[
\frac{1}{2 C_F^2} \cos (\delta  z_1 \Delta  E^-(\ul{k} - x \ul{p} + x \ul{q}_1)+\delta  z_2 \Delta  E^-(\ul{k} - x \ul{p}))
\notag \\ &\hspace{1cm}
-\frac{1}{2 C_F^2} \cos (\delta  z_1 \Delta  E^-(\ul{k} - x \ul{p} + x \ul{q}_1)-\delta  z_2 \Delta  E^-(\ul{k} - x \ul{p} - \ul{q}_2)+\delta  z_2 \Delta  E^-(\ul{k} - x \ul{p}))
\notag \\ &\hspace{1cm}
-\frac{1}{2 C_F^2} \cos (\delta  z_2 \Delta  E^-(\ul{k} - x \ul{p}))
+\frac{1}{2 C_F^2} \cos (-\delta  z_2 \Delta  E^-(\ul{k} - x \ul{p} - \ul{q}_2)+\delta  z_2 \Delta  E^-(\ul{k} - x \ul{p}))
\Bigg]
\notag \\ &+ \left| \psi(\ul{k} - x \ul{p} + x \ul{q}_1) \right|^2 \Bigg[
-\frac{2 (C_F+N_c)}{C_F}
+\frac{2 (C_F+N_c) }{C_F}\cos (\delta  z_1 \Delta  E^-(\ul{k} - x \ul{p} + x \ul{q}_1))
\Bigg]
\notag \\ &+ \psi(\ul{k} - x \ul{p}) \, \psi^*(\ul{k} - x \ul{p} - (1-x) \ul{q}_1) \Bigg[
\frac{N_c (C_F+N_c) }{C_F^2} 
-\frac{N_c (C_F+N_c) }{C_F^2}\cos (\delta  z_1 \Delta  E^-(\ul{k} - x \ul{p} - \ul{q}_1 (1-x)))
\notag \\ &\hspace{1cm}
+\frac{N_c^2 }{2 C_F^2}\cos (\delta  z_1 \Delta  E^-(\ul{k} - x \ul{p} - \ul{q}_1 (1-x))+\delta  z_2 \Delta  E^-(\ul{k} - x \ul{p}))
-\frac{N_c^2 }{2 C_F^2}\cos (\delta  z_2 \Delta  E^-(\ul{k} - x \ul{p}))
\Bigg]
\notag \\ &+ \psi(\ul{k} - x \ul{p} - \ul{q}_2)  \, \psi^*(\ul{k} - x \ul{p} - (1-x) \ul{q}_1) \Bigg[
-\frac{N_c^2 }{2 C_F^2}\cos (\delta  z_1 \Delta  E^-(\ul{k} - x \ul{p} - \ul{q}_1 (1-x))+\delta  z_2 \Delta  E^-(\ul{k} - x \ul{p}))
\notag \\ &\hspace{1cm}
+\frac{N_c^2 }{2 C_F^2}\cos (\delta  z_1 \Delta  E^-(\ul{k} - x \ul{p} - \ul{q}_1 (1-x))-\delta  z_2 \Delta  E^-(\ul{k} - x \ul{p} - \ul{q}_2)+\delta  z_2 \Delta  E^-(\ul{k} - x \ul{p}))
\notag \\ &\hspace{1cm}
+\frac{N_c^2 }{2 C_F^2}\cos (\delta  z_2 \Delta  E^-(\ul{k} - x \ul{p}))
-\frac{N_c^2 }{2 C_F^2}\cos (-\delta  z_2 \Delta  E^-(\ul{k} - x \ul{p} - \ul{q}_2)+\delta  z_2 \Delta  E^-(\ul{k} - x \ul{p})) 
\Bigg] 
\notag \\ &+ \psi(\ul{k} - x \ul{p} + x \ul{q}_1)  \, \psi^*(\ul{k} - x \ul{p} - (1-x) \ul{q}_1) \Bigg[
\frac{N_c (C_F+N_c) }{C_F^2}\cos (\delta  z_1 \Delta  E^-(\ul{k} - x \ul{p} - \ul{q}_1 (1-x))-\delta  z_1 \Delta  E^-(\ul{k} - x \ul{p} + x \ul{q}_1))
\notag \\ & \hspace{1cm}
-\frac{N_c (C_F+N_c) }{C_F^2}\cos (\delta  z_1 \Delta  E^-(\ul{k} - x \ul{p} - \ul{q}_1 (1-x)))
-\frac{N_c (C_F+N_c) }{C_F^2}\cos (-\delta  z_1 \Delta  E^-(\ul{k} - x \ul{p} + x \ul{q}_1))
\notag \\ & \hspace{1cm}
+\frac{N_c (C_F+N_c) }{C_F^2} 
\Bigg]
\notag \\ &+ \left| \psi(\ul{k} - x \ul{p} - (1-x) \ul{q}_1) \right|^2 \Bigg[
-\frac{2 N_c (C_F+N_c)}{C_F^2}
+\frac{2 N_c (C_F+N_c) }{C_F^2}\cos (\delta  z_1 \Delta  E^-(\ul{k} - x \ul{p} + x \ul{q}_1-q_1))
\Bigg]
\notag \\ &+ \psi(\ul{k} - x \ul{p})  \, \psi^*(\ul{k} - x \ul{p} + x \ul{q}_1 - \ul{q}_2) \Bigg[
-\frac{1}{2 C_F^2} \cos (-\delta  z_1 \Delta  E^-(\ul{k} - x \ul{p} + x \ul{q}_1-q_2)-\delta  z_2 \Delta  E^-(\ul{k} - x \ul{p} - \ul{q}_2)+\delta  z_2 \Delta  E^-(\ul{k} - x \ul{p}))
\notag \\ & \hspace{1cm}
+\frac{1}{2 C_F^2} \cos (-\delta  z_2 \Delta  E^-(\ul{k} - x \ul{p} - \ul{q}_2)+\delta  z_2 \Delta  E^-(\ul{k} - x \ul{p}))
\Bigg] \notag
\end{align}
\newpage
\begin{align}
&+ \psi(\ul{k} - x \ul{p} + x \ul{q}_1)  \, \psi^*(\ul{k} - x \ul{p} + x \ul{q}_1 - \ul{q}_2) \Bigg[ 
\frac{N_c }{C_F}\cos (-\delta  z_2 \Delta  E^-(\ul{k} - x \ul{p} - \ul{q}_2)+\delta  z_2 \Delta  E^-(\ul{k} - x \ul{p}))
\notag \\ & \hspace{1cm}
+\frac{N_c }{C_F}\cos (-\delta  z_1 \Delta  E^-(\ul{k} - x \ul{p} + x \ul{q}_1-q_2)+\delta  z_1 \Delta  E^-(\ul{k} - x \ul{p} + x \ul{q}_1)-\delta  z_2 \Delta  E^-(\ul{k} - x \ul{p} - \ul{q}_2)+\delta  z_2 \Delta  E^-(\ul{k} - x \ul{p}))
\notag \\ & \hspace{1cm}
-\frac{N_c }{C_F}\cos (-\delta  z_1 \Delta  E^-(\ul{k} - x \ul{p} + x \ul{q}_1-q_2)-\delta  z_2 \Delta  E^-(\ul{k} - x \ul{p} - \ul{q}_2)+\delta  z_2 \Delta  E^-(\ul{k} - x \ul{p}))
\notag \\ & \hspace{1cm}
-\frac{N_c }{C_F}\cos (\delta  z_1 \Delta  E^-(\ul{k} - x \ul{p} + x \ul{q}_1)-\delta  z_2 \Delta  E^-(\ul{k} - x \ul{p} - \ul{q}_2)+\delta  z_2 \Delta  E^-(\ul{k} - x \ul{p}))
\Bigg]
\notag \\ &+ \psi(\ul{k} - x \ul{p} - (1-x) \ul{q}_1)  \, \psi^*(\ul{k} - x \ul{p} + x \ul{q}_1 - \ul{q}_2) \Bigg[ 
-\frac{N_c^2 }{2 C_F^2}\cos (-\delta  z_2 \Delta  E^-(\ul{k} - x \ul{p} - \ul{q}_2)+\delta  z_2 \Delta  E^-(\ul{k} - x \ul{p}))
\notag \\ & \hspace{1cm}
-\frac{N_c^2 }{2 C_F^2}\cos (-\delta  z_1 \Delta  E^-(\ul{k} - x \ul{p} + x \ul{q}_1-\ul{q}_2)+\delta  z_1 \Delta  E^-(\ul{k} - x \ul{p} - \ul{q}_1 (1-x))-\delta  z_2 \Delta  E^-(\ul{k} - x \ul{p} - \ul{q}_2)+\delta  z_2 \Delta  E^-(\ul{k} - x \ul{p}))
\notag \\ & \hspace{1cm}
+\frac{N_c^2 }{2 C_F^2}\cos (-\delta  z_1 \Delta  E^-(\ul{k} - x \ul{p} + x \ul{q}_1-\ul{q}_2)-\delta  z_2 \Delta  E^-(\ul{k} - x \ul{p} - \ul{q}_2)+\delta  z_2 \Delta  E^-(\ul{k} - x \ul{p}))
\notag \\ & \hspace{1cm}
+\frac{N_c^2 }{2 C_F^2}\cos (\delta  z_1 \Delta  E^-(\ul{k} - x \ul{p} - \ul{q}_1 (1-x))-\delta  z_2 \Delta  E^-(\ul{k} - x \ul{p} - \ul{q}_2)+\delta  z_2 \Delta  E^-(\ul{k} - x \ul{p}))
\Bigg]
\notag \\ &+ \psi(\ul{k} - x \ul{p})  \, \psi^*(\ul{k} - x \ul{p} - (1-x) \ul{q}_1 - \ul{q}_2) \Bigg[ 
-\frac{N_c^2 }{2 C_F^2}\cos (-\delta  z_2 \Delta  E^-(\ul{k} - x \ul{p} - \ul{q}_2)+\delta  z_2 \Delta  E^-(\ul{k} - x \ul{p}))
\notag \\ & \hspace{1cm}
+\frac{N_c^2 }{2 C_F^2}\cos (-\delta  z_1 \Delta  E^-(\ul{k} - x \ul{p} - \ul{q}_1 (1-x)-\ul{q}_2)-\delta  z_2 \Delta  E^-(\ul{k} - x \ul{p} - \ul{q}_2)+\delta  z_2 \Delta  E^-(\ul{k} - x \ul{p}))
\Bigg] 
\notag \\ &+ \psi(\ul{k} - x \ul{p} + x \ul{q}_1)  \, \psi^*(\ul{k} - x \ul{p} - (1-x) \ul{q}_1 - \ul{q}_2) \Bigg[ 
-\frac{N_c^2 }{2 C_F^2}\cos (-\delta  z_2 \Delta  E^-(\ul{k} - x \ul{p} - \ul{q}_2)+\delta  z_2 \Delta  E^-(\ul{k} - x \ul{p}))
\notag \\ & \hspace{1cm}
-\frac{N_c^2 }{2 C_F^2}\cos (-\delta  z_1 \Delta  E^-(\ul{k} - x \ul{p} - \ul{q}_1 (1-x)-\ul{q}_2)+\delta  z_1 \Delta  E^-(\ul{k} - x \ul{p} + x \ul{q}_1)-\delta  z_2 \Delta  E^-(\ul{k} - x \ul{p} - \ul{q}_2)+\delta  z_2 \Delta  E^-(\ul{k} - x \ul{p}))
\notag \\ & \hspace{1cm}
+\frac{N_c^2 }{2 C_F^2}\cos (-\delta  z_1 \Delta  E^-(\ul{k} - x \ul{p} - \ul{q}_1 (1-x)-\ul{q}_2)-\delta  z_2 \Delta  E^-(\ul{k} - x \ul{p} - \ul{q}_2)+\delta  z_2 \Delta  E^-(\ul{k} - x \ul{p}))
\notag \\ & \hspace{1cm}
+\frac{N_c^2 }{2 C_F^2}\cos (\delta  z_1 \Delta  E^-(\ul{k} - x \ul{p} + x \ul{q}_1)-\delta  z_2 \Delta  E^-(\ul{k} - x \ul{p} - \ul{q}_2)+\delta  z_2 \Delta  E^-(\ul{k} - x \ul{p}))
\Bigg]
\notag \\ &+ \psi(\ul{k} - x \ul{p} - (1-x) \ul{q}_1)  \, \psi^*(\ul{k} - x \ul{p} - (1-x) \ul{q}_1 - \ul{q}_2) \Bigg[ 
\frac{N_c^2 }{C_F^2}\cos (-\delta  z_2 \Delta  E^-(\ul{k} - x \ul{p} - \ul{q}_2)+\delta  z_2 \Delta  E^-(\ul{k} - x \ul{p}))
\notag \\ & \hspace{1cm}
+\frac{N_c^2 }{C_F^2}\cos (-\delta  z_1 \Delta  E^-(\ul{k} - x \ul{p} - \ul{q}_1 (1-x)-\ul{q}_2)+\delta  z_1 \Delta  E^-(\ul{k} - x \ul{p} - \ul{q}_1 (1-x))-\delta  z_2 \Delta  E^-(\ul{k} - x \ul{p} - \ul{q}_2)+\delta  z_2 \Delta  E^-(\ul{k} - x \ul{p}))
\notag \\ & \hspace{1cm}
-\frac{N_c^2 }{C_F^2}\cos (-\delta  z_1 \Delta  E^-(\ul{k} - x \ul{p} - \ul{q}_1 (1-x)-\ul{q}_2)-\delta  z_2 \Delta  E^-(\ul{k} - x \ul{p} - \ul{q}_2)+\delta  z_2 \Delta  E^-(\ul{k} - x \ul{p}))
\notag \\ & \hspace{1cm}
-\frac{N_c^2 }{C_F^2}\cos (\delta  z_1 \Delta  E^-(\ul{k} - x \ul{p} - \ul{q}_1 (1-x))-\delta  z_2 \Delta  E^-(\ul{k} - x \ul{p} - \ul{q}_2)+\delta  z_2 \Delta  E^-(\ul{k} - x \ul{p}))
\Bigg] \, , 
\end{align}
\newpage
\begin{align}
\mathcal{N}_3 &=
\notag \\ \notag \\ & - \left| \psi(\ul{k} - x \ul{p}) \right|^2 
\notag \\ &+ \psi(\ul{k} - x \ul{p})  \, \psi^*(\ul{k} - x \ul{p} + x \ul{q}_2) \Bigg[ 
-\frac{1}{2 C_F^2} \cos (\delta  z_2 \Delta  E^-(\ul{k} - x \ul{p} + x \ul{q}_2))
-\frac{1}{C_F N_c} 
\notag \\ & \hspace{1cm}
+\frac{(2 C_F+N_c) }{2 C_F^2 N_c}\cos (\delta  z_1 \Delta  E^-(k-x p+x q_2)+\delta  z_2 \Delta  E^-(k-x p+x q_2))
\Bigg]
\notag \\ &+ \left| \psi(\ul{k} - x \ul{p} + x \ul{q}_2) \right|^2  \Bigg[ 
-2
-\frac{N_c}{C_F}
+\frac{N_c }{C_F}\cos (\delta  z_1 \Delta  E^-(k-x p + x q_2)))
-\frac{N_c }{C_F}\cos (\delta  z_2 \Delta  E^-(k-x p + x q_2)))
\notag \\ & \hspace{1cm}
+\left(\frac{N_c}{C_F}+2\right) \cos (\delta  z_1 \Delta  E^-(k-x p + x q_2))+\delta  z_2 \Delta  E^-(k-x p + x q_2)))
\Bigg]
\notag \\ &+ \psi(\ul{k} - x \ul{p})  \, \psi^*(\ul{k} - x \ul{p} - \ul{q}_1 + x \ul{q}_2)  \Bigg[ 
\frac{1}{2 C_F^2} \cos (\delta  z_2 \Delta  E^-(k - x p + x q_2))
\notag \\ & \hspace{1cm}
-\frac{1}{2 C_F^2} \cos (\delta  z_1 \Delta  E^-(k-x p -q_1 + x q_2)+\delta  z_2 \Delta  E^-(k-p x+q_2 x))
\Bigg] 
\notag \\ &+ \psi(\ul{k} - x \ul{p} + x \ul{q}_2)  \, \psi^*(\ul{k} - x \ul{p} - \ul{q}_1 + x \ul{q}_2)  \Bigg[ 
-\frac{N_c }{C_F}\cos (\delta  z_1 \Delta  E^-(k-x p + x q_2)))
+\frac{N_c }{C_F}\cos (\delta  z_2 \Delta  E^-(k-x p + x q_2)))
\notag \\ & \hspace{1cm}
+\frac{N_c }{C_F}\cos (\delta  z_1 \Delta  E^-(k-x p + x q_2))-\delta  z_1 \Delta  E^-(k-x p -q_1 + x q_2))
\notag \\ & \hspace{1cm}
-\frac{N_c }{C_F}\cos (\delta  z_1 \Delta  E^-(k -x p -q_1 + x q_2)+\delta  z_2 \Delta  E^-(k-x p + x q_2))
\Bigg]
\notag \\ &+ \psi(\ul{k} - x \ul{p})  \, \psi^*(\ul{k} - x \ul{p} - (1-x) \ul{q}_2)  \Bigg[ 
\frac{N_c }{C_F} 
+\frac{N_c^2 }{2 C_F^2}\cos (\delta  z_2 \Delta  E^-(k-x p - q_2 (1-x)))
\notag \\ & \hspace{1cm}
-\frac{N_c (2 C_F+N_c) }{2 C_F^2}\cos (\delta  z_1 \Delta  E^-(k-x p - (1-x)  q_2)+\delta  z_2 \Delta  E^-(k- x p - (1-x)  q_2))
\Bigg]
\notag \\ &+ \psi(\ul{k} - x \ul{p} + x \ul{q}_2)  \, \psi^*(\ul{k} - x \ul{p} - (1-x) \ul{q}_2)  \Bigg[ 
\frac{N_c }{C_F} 
+\frac{N_c^2 }{2 C_F^2} \cos (\delta  z_2 \Delta  E^-(k - x p + x q_2) )
\notag \\ & \hspace{1cm}
+\frac{N_c (C_F+N_c) }{C_F^2}\cos ((\delta  z_1 + \delta z_2) \Delta  E^-(k - x p - (1-x) q_2 ) - (\delta  z_1 + \delta z_2) \Delta  E^-(k-x p + x q_2))
\notag \\ & \hspace{1cm}
-\frac{N_c^2 }{2 C_F^2}\cos ((\delta  z_1 + \delta z_2) \Delta  E^-(k - x p - (1-x) q_2 ) - \delta  z_2 \Delta  E^-(k - x p + x q_2))
\notag \\ & \hspace{1cm}
-\frac{N_c (2 C_F+N_c) }{2 C_F^2}\cos ((\delta  z_1 +\delta  z_2) \Delta  E^-(k-x p - (1-x) q_2  ))
\notag \\ & \hspace{1cm}
-\frac{N_c^2 }{2 C_F^2}\cos (-(\delta  z_1 + \delta  z_2) \Delta  E^-(k - x p + x q_2) + \delta  z_2 \Delta  E^-(k - x p - (1-x) q_2))
\notag \\ & \hspace{1cm}
+\frac{N_c^2 }{2 C_F^2}\cos ( \delta  z_2 \Delta  E^-(k - x p - (1-x) q_2) )
-\frac{N_c (2 C_F+N_c) }{2 C_F^2}\cos ((\delta  z_1 + \delta  z_2) \Delta  E^-(k - x p + x q_2))
\Bigg]
\notag \\ &+ \psi(\ul{k} - x \ul{p} - \ul{q}_1 + x \ul{q}_2)  \, \psi^*(\ul{k} - x \ul{p} - (1-x) \ul{q}_2)  \Bigg[ 
\frac{N_c^2 }{2 C_F^2}\cos (
( \delta  z_1 +\delta  z_2) \Delta  E^-(k - x p - (1-x) q_2)
-\delta  z_2 \Delta  E^-(k - x p + x q_2))
\notag \\ & \hspace{1cm}
-\frac{N_c^2 }{2 C_F^2}
\cos ( (\delta  z_1 +\delta  z_2) \Delta  E^-(\ul{k} - x \ul{p}-q_2 (1-x))
-\delta  z_1 \Delta  E^-(k-x p - q_1 + x q_2)
-\delta  z_2 \Delta  E^-(\ul{k} - x \ul{p} + x q_2) )
\notag \\ & \hspace{1cm}
-\frac{N_c^2 }{2 C_F^2}\cos ( \delta  z_2 \Delta  E^-(k - x p + x q_2) )
+\frac{N_c^2 }{2 C_F^2}\cos (\delta  z_1 \Delta  E^-(k - x  p - q_1 + x q_2) +\delta  z_2 \Delta  E^-(k - x p + x q_2 ))
\Bigg] \notag
\end{align}
\newpage
\begin{align}
&+ \left| \psi(\ul{k} - x \ul{p} - (1 - x) \ul{q}_2) \right|^2  \Bigg[ 
-\frac{N_c (2 C_F+N_c)}{C_F^2}
+\frac{N_c^2 }{C_F^2}\cos (\delta  z_1 \Delta  E^-(k - x p - (1-x) q_2 ))
\notag \\ & \hspace{1cm}
-\frac{N_c^2 }{C_F^2}\cos (\delta  z_2 \Delta  E^-(k - x p - (1-x) q_2))
+\frac{N_c (2 C_F+N_c) }{C_F^2}\cos (( \delta  z_1 +\delta  z_2) \Delta  E^-(k - x p - (1-x) q_2) )
\Bigg]
\notag \\ &+ \psi(\ul{k} - x \ul{p})  \, \psi^*(\ul{k} - x \ul{p} - \ul{q}_1 - (1-x) \ul{q}_2)  
\Bigg[ 
-\frac{N_c^2 }{2 C_F^2}\cos (\delta  z_2 \Delta  E^-(k - x p - (1-x) q_2))
\notag \\ & \hspace{1cm}
+\frac{N_c^2 }{2 C_F^2}\cos (\delta  z_1 \Delta  E^-(k - x p - q_1 - (1-x) q_2)
+\delta  z_2 \Delta  E^-(k - x p - (1-x) q_2))
\Bigg]
\notag \\ &+ \psi(\ul{k} - x \ul{p} + x \ul{q}_2)  \, \psi^*(\ul{k} - x \ul{p} - \ul{q}_1 - (1-x) \ul{q}_2)  
\Bigg[ 
\frac{N_c^2 }{2 C_F^2}\cos (- ( \delta  z_1 + \delta  z_2) \Delta  E^-(k - x p + x q_2)
+\delta  z_2 \Delta  E^-(k - x p - (1-x) q_2))
\notag \\ & \hspace{1cm}
-\frac{N_c^2 }{2 C_F^2}\cos (\delta  z_2 \Delta  E^-(k - x p - (1-x) q_2))
\notag \\ & \hspace{1cm}
-\frac{N_c^2 }{2 C_F^2}\cos (\delta  z_1 \Delta  E^-(k - x  p - q_1 - (1 - x) q_2)
- ( \delta  z_1 + \delta  z_2) \Delta  E^-(k - x p + x q_2) + \delta  z_2 \Delta  E^-(k - x p - (1-x) q_2))
\notag \\ & \hspace{1cm}
+\frac{N_c^2 }{2 C_F^2}\cos (\delta  z_1 \Delta  E^-(k - x  p - q_1 - (1-x) q_2)
+\delta  z_2 \Delta  E^-(k - x p - (1-x) q_2))
\Bigg]
\notag \\ &+ \psi(\ul{k} - x \ul{p} - (1-x) \ul{q}_2)  \, \psi^*(\ul{k} - x \ul{p} - \ul{q}_1 - (1-x) \ul{q}_2)  
\Bigg[ 
-\frac{N_c^2 }{C_F^2}\cos (\delta  z_1 \Delta  E^-(k - x p - (1-x) q_2))
\notag \\ & \hspace{1cm}
+\frac{N_c^2 }{C_F^2}\cos (\delta  z_2 \Delta  E^-(k - x p - (1-x) q_2))
\notag \\ & \hspace{1cm}
+\frac{N_c^2 }{C_F^2}\cos (\delta  z_1 \Delta  E^-(k - x p - (1-x) q_2) -\delta  z_1 \Delta  E^-(k - x p - q_1 - (1-x) q_2))
\notag \\ & \hspace{1cm}
-\frac{N_c^2 }{C_F^2}\cos (\delta  z_1 \Delta  E^-(k - x p - q_1 - (1-x) q_2) +\delta  z_2 \Delta  E^-(k - x p - (1-x) q_2))
\Bigg] \Bigg\} \, , 
\end{align}
\newpage
\begin{align}
\mathcal{N}_4 &=
\notag \\ \notag \\ &\left| \psi(\ul{k} - x \ul{p}) \right|^2 
\notag \\ &+ \psi(\ul{k} - x \ul{p})  \, \psi^*(\ul{k} - x \ul{p} + x \ul{q}_2) \Bigg[ 
\frac{1 }{C_F N_c}
-\frac{1}{C_F N_c} \cos (\delta  z_2 \Delta  E^-(k - x p + x q_2))
\Bigg]
\notag \\ &+ \left| \psi(\ul{k} - x \ul{p} + x \ul{q}_2) \right|^2 \Bigg[
2 
- 2 \cos (\delta  z_2 \Delta  E^-(k - x p + x q_2))
\Bigg]
\notag \\ &+ \psi(\ul{k} - x \ul{p})  \, \psi^*(\ul{k} - x \ul{p} - (1-x) \ul{q}_2)  \Bigg[
-\frac{N_c }{C_F}
+\frac{N_c }{C_F}\cos (\delta  z_2 \Delta  E^-(k - x p - (1-x) q_2))
\Bigg]
\notag \\ &+ \psi(\ul{k} - x \ul{p} + x \ul{q}_2)  \, \psi^*(\ul{k} - x \ul{p} - (1-x) \ul{q}_2)  \Bigg[
-\frac{N_c }{C_F}
+\frac{N_c }{C_F}\cos ( \delta  z_2 \Delta  E^-(k - x p + x q_2) )
\notag \\ & \hspace{1cm}
-\frac{N_c }{C_F}\cos (\delta  z_2 \Delta  E^-(k - x p - (1-x) q_2)-\delta  z_2 \Delta  E^-(k - x p + x q_2))
+\frac{N_c }{C_F}\cos (\delta  z_2 \Delta  E^-(k - x p -  (1-x) q_2)
\Bigg]
\notag \\ &+ \left| \psi(\ul{k} - x \ul{p} - (1-x) \ul{q}_2) \right|^2 \Bigg[
\frac{2 N_c}{C_F}
-\frac{2 N_c }{C_F}\cos (\delta  z_2 \Delta  E^-(k - x p - (1-x) q_2))
\Bigg]
\notag \\ &+ \psi(\ul{k} - x \ul{p})  \, \psi^*(\ul{k} - x \ul{p} + x \ul{q}_1 + x \ul{q}_2)  \Bigg[
\frac{1 }{2 C_F^2 N_c^2} \cos (\delta  z_1 \Delta  E^-(k - x p + x q_1 + x q_2) +\delta  z_2 \Delta  E^-(k - x p + x q_2))
\notag \\ & \hspace{1cm}
-\frac{1 }{2 C_F^2 N_c^2} \cos (\delta  z_2 \Delta  E^-(k - x p + x q_2))
\Bigg]
\notag \\ &+ \psi(\ul{k} - x \ul{p} + x \ul{q}_2)  \, \psi^*(\ul{k} - x \ul{p} + x \ul{q}_1 + x \ul{q}_2)  \Bigg[
\frac{1 }{C_F N_c} 
-\frac{1 }{C_F N_c} \cos (\delta  z_1 \Delta  E^-(k - x p + x q_1 + x q_2))
\notag \\ & \hspace{1cm}
+\frac{1 }{C_F N_c} \cos (\delta  z_1 \Delta  E^-(k - x p + x q_1 + x q_2) + \delta  z_2 \Delta  E^-(k - x p + x q_2))
-\frac{1 }{C_F N_c} \cos (\delta  z_2 \Delta  E^-(k - x p + x q_2))
\Bigg]
\notag \\ &+ \psi(\ul{k} - x \ul{p} - (1-x) \ul{q}_2)  \, \psi^*(\ul{k} - x \ul{p} + x \ul{q}_1 + x \ul{q}_2)  \Bigg[
\frac{1 }{2 C_F^2} \cos ( \delta  z_2 \Delta  E^-(k - x p + x q_2) )
\notag \\ & \hspace{1cm}
+\frac{1 }{2 C_F^2} \cos (-\delta  z_1 \Delta  E^-(k -x p + x q_1 + x q_2) +\delta  z_2 \Delta  E^-(k - x p - (1-x) q_2) -\delta  z_2 \Delta  E^-(k - x p + x q_2))
\notag \\ & \hspace{1cm}
-\frac{1 }{2 C_F^2} \cos (-\delta  z_1 \Delta  E^-(k - x p + x q_1 + x q_2)
-\delta  z_2 \Delta  E^-(k - x p + x q_2))
\notag \\ & \hspace{1cm}
-\frac{1 }{2 C_F^2} \cos (\delta  z_2 \Delta  E^-(k - x p - (1-x)  q_2)
-\delta  z_2 \Delta  E^-(k - x p + x q_2))
\Bigg]
\notag \\ &+ \left| \psi(\ul{k} - x \ul{p} + x \ul{q}_1 + x \ul{q}_2) \right|^2 \Bigg[
2 - 2 \cos (\delta  z_1 \Delta  E^-(k - x p + x q_1  + x q_2))
\Bigg]
\notag \\ &+ \psi(\ul{k} - x \ul{p})  \, \psi^*(\ul{k} - x \ul{p} - (1-x) \ul{q}_1 + x \ul{q}_2)  \Bigg[
\frac{1 }{2 C_F^2} \cos (\delta  z_2 \Delta  E^-(k - x p + x q_2))
\notag \\ & \hspace{1cm}
-\frac{1 }{2 C_F^2} \cos (\delta  z_1 \Delta  E^-(k - x p - (1-x) q_1 + x q_2)
+\delta  z_2 \Delta  E^-(k - x p + x q_2))
\Bigg] \notag
\end{align}
\newpage
\begin{align}
&+ \psi(\ul{k} - x \ul{p} + x \ul{q}_2)  \, \psi^*(\ul{k} - x \ul{p} - (1-x) \ul{q}_1 + x \ul{q}_2)  \Bigg[
-\frac{N_c }{C_F} 
+\frac{N_c }{C_F}\cos (\delta  z_1 \Delta  E^-(k - x p - (1-x)  q_1 + x q_2 ))
\notag \\ & \hspace{1cm}
-\frac{N_c }{C_F}\cos (\delta  z_1 \Delta  E^-(k - x p - (1-x) q_1 + x q_2 ) +\delta  z_2 \Delta  E^-(k - x p + x q_2))
\notag \\ & \hspace{1cm}
+\frac{N_c }{C_F}\cos (\delta  z_2 \Delta  E^-(k - x p + x q_2))
\Bigg]
\notag \\ &+ \psi(\ul{k} - x \ul{p} - (1-x) \ul{q}_2)  \, \psi^*(\ul{k} - x \ul{p} - (1-x) \ul{q}_1 + x \ul{q}_2)  \Bigg[
-\frac{N_c^2 }{2 C_F^2}\cos (\delta  z_2 \Delta  E^-(k - x p + x q_2))
\notag \\ & \hspace{1cm}
-\frac{N_c^2 }{2 C_F^2}\cos (-\delta  z_1 \Delta  E^-(k - x p - (1-x) q_1 + x q_2) +\delta  z_2 \Delta  E^-(k - x p - (1-x)  q_2) -\delta  z_2 \Delta  E^-(k - x p + x q_2))
\notag \\ & \hspace{1cm}
+\frac{N_c^2 }{2 C_F^2}\cos (\delta  z_1 \Delta  E^-(k - x p - (1-x) q_1 + x q_2) +\delta  z_2 \Delta  E^-(k - x p + x q_2))
\notag \\ & \hspace{1cm}
+\frac{N_c^2 }{2 C_F^2}\cos (\delta  z_2 \Delta  E^-(k -x p - (1-x) q_2) -\delta  z_2 \Delta  E^-(k - x p +x q_2))
\Bigg]
\notag \\ &+ \psi(\ul{k} - x \ul{p} + x \ul{q}_1 + x \ul{q}_2)  \, \psi^*(\ul{k} - x \ul{p} - (1-x) \ul{q}_1 + x \ul{q}_2)  \Bigg[
-\frac{N_c }{C_F}
+\frac{N_c }{C_F}\cos (\delta  z_1 \Delta  E^-(k - x  p - (1-x) q_1 + x q_2))
\notag \\ & \hspace{1cm}
+\frac{N_c }{C_F}\cos (\delta  z_1 \Delta  E^-(k - x p + x q_1 + x q_2))
\notag \\ & \hspace{1cm}
-\frac{N_c }{C_F}\cos (\delta  z_1 \Delta  E^-(k - x  p - (1-x) q_1 + x q_2) -\delta  z_1 \Delta  E^-(k - x p + x q_1 + x q_2))
\Bigg]
\notag \\ &+ \left| \psi(\ul{k} - x \ul{p} - (1-x) \ul{q}_1 + x \ul{q}_2) \right|^2 \Bigg[
\frac{2 N_c}{C_F}
-\frac{2 N_c }{C_F}\cos (\delta  z_1 \Delta  E^-(k - x p - (1-x) q_1 + x q_2))
\Bigg] 
\notag \\ &+ \psi(\ul{k} - x \ul{p})  \, \psi^*(\ul{k} - x \ul{p} + x \ul{q}_1 - (1-x) \ul{q}_2)  \Bigg[
\frac{1 }{2 C_F^2} \cos (\delta  z_2 \Delta  E^-(k - x p - (1-x) q_2))
\notag \\ & \hspace{1cm}
-\frac{1 }{2 C_F^2} \cos (\delta  z_1 \Delta  E^-(k - x p + x q_1 - (1-x) q_2) +\delta  z_2 \Delta  E^-(k - x p - (1-x) q_2))
\Bigg]
\notag \\ &+ \psi(\ul{k} - x \ul{p} + x \ul{q}_2)  \, \psi^*(\ul{k} - x \ul{p} + x \ul{q}_1 - (1-x) \ul{q}_2)  \Bigg[ 
\frac{1}{2 C_F^2} \cos (\delta  z_2 \Delta  E^-(k - x p - (1-x) q_2))
\notag \\ & \hspace{1cm}
+\frac{1}{2 C_F^2} \cos (\delta  z_1 \Delta  E^-(k - x p + x q_1 - (1-x) q_2) +\delta  z_2 \Delta  E^-(k - x p - (1-x) q_2) -\delta  z_2 \Delta  E^-(k - x p + x q_2))
\notag \\ & \hspace{1cm}
-\frac{1}{2 C_F^2} \cos (\delta  z_1 \Delta  E^-(k - x p + x q_1 - (1-x) q_2) +\delta  z_2 \Delta  E^-(k - x p - (1-x) q_2))
\notag \\ & \hspace{1cm}
-\frac{1}{2 C_F^2} \cos (\delta  z_2 \Delta  E^-(k - x p - (1-x) q_2) -\delta  z_2 \Delta  E^-(k - x p + x q_2))
\Bigg]
\notag \\ &+ \psi(\ul{k} - x \ul{p} - (1-x) \ul{q}_2)  \, \psi^*(\ul{k} - x \ul{p} + x \ul{q}_1 - (1-x) \ul{q}_2)  \Bigg[ 
\frac{1 }{C_F^2} 
-\frac{1 }{C_F^2} \cos (\delta  z_1 \Delta  E^-(k - x p + x q_1 - (1-x) q_2))
\notag \\ & \hspace{1cm}
+\frac{1 }{C_F^2} \cos (\delta  z_1 \Delta  E^-(k - x p - (1-x) q_2+ x q_1) +\delta  z_2 \Delta  E^-(k - x p - (1-x) q_2))
\notag \\ & \hspace{1cm}
-\frac{1 }{C_F^2} \cos (\delta  z_2 \Delta  E^-(k - x p - (1-x) q_2))
\Bigg] \notag
\end{align}
\newpage
\begin{align}
&+ \psi(\ul{k} - x \ul{p} + x \ul{q}_1  + x \ul{q}_2)  \, \psi^*(\ul{k} - x \ul{p} + x \ul{q}_1 - (1-x) \ul{q}_2)  \Bigg[ \notag \\ & \hspace{1cm}
-\frac{N_c }{C_F}\cos (
-\delta  z_1 \Delta  E^-(k - x p + x q_1 + x q_2) 
+\delta  z_1 \Delta  E^-(k - x p + x q_1 - (1-x) q_2) 
+\delta  z_2 \Delta  E^-(k - x p - (1-x) q_2) 
\notag \\ & \hspace{2cm}
-\delta  z_2 \Delta  E^-(k - x p + x q_2))
\notag \\ & \hspace{1cm}
+\frac{N_c }{C_F}\cos (\delta  z_1 \Delta  E^-(k - x p + x q_1 - (1-x) q_2)) +\delta  z_2 \Delta  E^-(k - x p - (1-x) q_2) -\delta  z_2 \Delta  E^-(k - x p + x q_2))
\notag \\ & \hspace{1cm}
+\frac{N_c }{C_F}\cos (-\delta  z_1 \Delta  E^-(k - x p + x q_1 + x q_2) +\delta  z_2 \Delta  E^-(k - x p - (1-x) q_2) -\delta  z_2 \Delta  E^-(k - x p + x q_2))
\notag \\ & \hspace{1cm}
-\frac{N_c }{C_F}\cos (\delta  z_2 \Delta  E^-(k - x p - (1-x) q_2) -\delta  z_2 \Delta  E^-(k - x p + x q_2))
\Bigg]
\notag \\ &+ \psi(\ul{k} - x \ul{p} - (1-x) \ul{q}_1  + x \ul{q}_2)  \, \psi^*(\ul{k} - x \ul{p} + x \ul{q}_1 - (1-x) \ul{q}_2)  \Bigg[ \notag \\ & \hspace{1cm}
+\frac{N_c^2 }{2 C_F^2}\cos (-\delta  z_1 \Delta  E^-(k - x p - (1-x) q_1 + x q_2)
+\delta  z_1 \Delta  E^-(k -x p  + x q_1 - (1-x) q_2) +\delta  z_2 \Delta  E^-(k - x p - (1-x) q_2) 
\notag \\ & \hspace{2cm}
-\delta  z_2 \Delta  E^-(k - x p + x q_2))
\notag \\ & \hspace{1cm}
-\frac{N_c^2 }{2 C_F^2}\cos (\delta  z_1 \Delta  E^-(k - x p - (1-x) q_2 + x q_1) +\delta  z_2 \Delta  E^-(k - x p - (1-x) q_2)  - \delta  z_2 \Delta  E^-(k - x p + x q_2))
\notag \\ & \hspace{1cm}
-\frac{N_c^2 }{2 C_F^2}\cos (-\delta  z_1 \Delta  E^-(k - x p + x q_2 - (1-x) q_1) +\delta  z_2 \Delta  E^-(k - x p - (1-x) q_2) -\delta  z_2 \Delta  E^-(k- x p + x q_2))
\notag \\ & \hspace{1cm}
+\frac{N_c^2 }{2 C_F^2}\cos (\delta  z_2 \Delta  E^-(k - x p - (1-x) q_2)
-\delta  z_2 \Delta  E^-(k - x p + x q_2))
\Bigg]
\notag \\ &+ \left| \psi(\ul{k} - x \ul{p} + x \ul{q}_1 - (1-x) \ul{q}_2) \right|^2 \Bigg[
\frac{2 N_c}{C_F}
-\frac{2 N_c }{C_F}\cos (\delta  z_1 \Delta  E^-(k - x p + x q_1 - (1-x) q_2))
\Bigg]
\notag \\ &+ \psi(\ul{k} - x \ul{p})  \, \psi^*(\ul{k} - x \ul{p} - (1-x) \ul{q}_1 - (1-x) \ul{q}_2)  \Bigg[ 
-\frac{N_c^2 }{2 C_F^2}\cos (\delta  z_2 \Delta  E^-(k - x p - (1-x) q_2))
\notag \\ & \hspace{1cm}
+\frac{N_c^2 }{2 C_F^2}\cos (\delta  z_1 \Delta  E^-(k - x p  - (1-x) q_1 - (1-x) q_2) +\delta  z_2 \Delta  E^-(k - x p - (1-x) q_2))
\Bigg]
\notag \\ &+ \psi(\ul{k} - x \ul{p} + x \ul{q}_2)  \, \psi^*(\ul{k} - x \ul{p} - (1-x) \ul{q}_1 - (1-x) \ul{q}_2)  \Bigg[ 
-\frac{N_c^2 }{2 C_F^2}\cos (\delta  z_2 \Delta  E^-(k - x p -(1-x) q_2))
\notag \\ & \hspace{1cm}
-\frac{N_c^2 }{2 C_F^2}\cos (\delta  z_1 \Delta  E^-(k - x p - (1-x) q_2 - (1-x) q_1) 
+\delta  z_2 \Delta  E^-(k - x p - (1-x) q_2) -\delta  z_2 \Delta  E^-(k - x p + x q_2))
\notag \\ & \hspace{1cm}
+\frac{N_c^2 }{2 C_F^2}\cos (\delta  z_1 \Delta  E^-(k - x p - (1-x) q_1 - (1-x) q_2) +\delta  z_2 \Delta  E^-(k - x p - (1-x) q_2))
\notag \\ & \hspace{1cm}
+\frac{N_c^2 }{2 C_F^2}\cos (\delta  z_2 \Delta  E^-(k - x p - (1-x) q_2) -\delta  z_2 \Delta  E^-(k - x p + x q_2))
\Bigg] 
\notag \\ &+ \psi(\ul{k} - x \ul{p} - (1-x) \ul{q}_2)  \, \psi^*(\ul{k} - x \ul{p} - (1-x) \ul{q}_1 - (1-x) \ul{q}_2)  \Bigg[ 
-\frac{N_c^2 }{C_F^2}
+\frac{N_c^2 }{C_F^2}\cos (\delta  z_1 \Delta  E^-(k - x p  - (1-x) q_1 - (1-x) q_2))
\notag \\ & \hspace{1cm}
-\frac{N_c^2 }{C_F^2}\cos (\delta  z_1 \Delta  E^-(k - x p - (1-x) q_1 - (1-x) q_2)
+\delta  z_2 \Delta  E^-(k - x p - (1-x) q_2))
\notag \\ & \hspace{1cm}
+\frac{N_c^2 }{C_F^2}\cos (\delta  z_2 \Delta  E^-(k - x p - (1-x) q_2))
\Bigg] \notag
\end{align}
\newpage
\begin{align}
&+ \psi(\ul{k} - x \ul{p} + x \ul{q}_1 + x \ul{q}_2)  \, \psi^*(\ul{k} - x \ul{p} - (1-x) \ul{q}_1 - (1-x) \ul{q}_2)  \Bigg[ \notag \\ & \hspace{1cm}
\frac{N_c^2 }{2 C_F^2}\cos (\delta  z_1 \Delta  E^-(k - x p - (1-x) q_1 - (1-x) q_2) -\delta  z_1 \Delta  E^-(k -x p + x q_2 + x q_1) +\delta  z_2 \Delta  E^-(k - x p - (1-x) q_2) 
\notag \\ & \hspace{2cm}
-\delta  z_2 \Delta  E^-(k - x p + x q_2))
\notag \\ & \hspace{1cm}
-\frac{N_c^2 }{2 C_F^2}\cos (\delta  z_1 \Delta  E^-(k - x p  - (1-x) q_1 - (1-x) q_2) +\delta  z_2 \Delta  E^-(k - x p - (1-x) q_2) -\delta  z_2 \Delta  E^-(k - x p + x q_2))
\notag \\ & \hspace{1cm}
-\frac{N_c^2 }{2 C_F^2}\cos (-\delta  z_1 \Delta  E^-(k - x p  + x q_1 + x q_2) +\delta  z_2 \Delta  E^-(k - x p - (1-x) q_2) -\delta  z_2 \Delta  E^-(k - x p + x q_2))
\notag \\ & \hspace{1cm}
+\frac{N_c^2 }{2 C_F^2}\cos (\delta  z_2 \Delta  E^-(k - x p - (1-x) q_2) -\delta  z_2 \Delta  E^-(k - x p + x q_2))
\Bigg]
\notag \\ &+ \psi(\ul{k} - x \ul{p} - (1-x) \ul{q}_1 + x \ul{q}_2)  \, \psi^*(\ul{k} - x \ul{p} - (1-x) \ul{q}_1 - (1-x) \ul{q}_2)  \Bigg[ \notag \\ & \hspace{1cm}
-\frac{N_c^2 }{C_F^2}\cos (-\delta  z_1 \Delta  E^-(k - x p - (1-x) q_1 + x q_2)
+\delta  z_1 \Delta  E^-(k - x p - (1-x) q_1 - (1-x) q_2) 
\notag \\ & \hspace{2cm}
+\delta  z_2 \Delta  E^-(k - x p - (1-x) q_2) 
-\delta  z_2 \Delta  E^-(k - x p + x q_2))
\notag \\ & \hspace{1cm}
+\frac{N_c^2 }{C_F^2}\cos (\delta  z_1 \Delta  E^-(k - x p  - (1-x) q_1 - (1-x) q_2) +\delta  z_2 \Delta  E^-(k - x p - (1-x) q_2) -\delta  z_2 \Delta  E^-(k - x p + x q_2))
\notag \\ & \hspace{1cm}
+\frac{N_c^2 }{C_F^2}\cos (-\delta  z_1 \Delta  E^-(k - x p  - (1-x) q_1 + x q_2) +\delta  z_2 \Delta  E^-(k - x p - (1-x) q_2) -\delta  z_2 \Delta  E^-(k - x p + x q_2))
\notag \\ & \hspace{1cm}
-\frac{N_c^2 }{C_F^2}\cos (\delta  z_2 \Delta  E^-(k - x p - (1-x) q_2) -\delta  z_2 \Delta  E^-(k - x p + x q_2))
\Bigg]
\notag \\ &+ \psi(\ul{k} - x \ul{p} + x \ul{q}_1 - (1-x) \ul{q}_2)  \, \psi^*(\ul{k} - x \ul{p} - (1-x) \ul{q}_1 - (1-x) \ul{q}_2)  \Bigg[ 
-\frac{N_c^2 }{C_F^2}
+\frac{N_c^2 }{C_F^2}\cos (\delta  z_1 \Delta  E^-(k - x p + x q_1 - (1-x) q_2))
\notag \\ & \hspace{1cm}
-\frac{N_c^2 }{C_F^2}\cos (\delta  z_1 \Delta  E^-(k - x p  - (1-x) q_1 - (1-x) q_2) -\delta  z_1 \Delta  E^-(k - x p - (1-x) q_2 + x q_1))
\notag \\ & \hspace{1cm}
+\frac{N_c^2 }{C_F^2}\cos (\delta  z_1 \Delta  E^-(k - x p - (1-x) q_2 - (1-x) q_1))
\Bigg]
\notag \\ &+ \left| \psi(\ul{k} - x \ul{p} - (1-x) \ul{q}_1 - (1-x) \ul{q}_2) \right|^2 \Bigg[
\frac{2 N_c^2}{C_F^2}
-\frac{2 N_c^2 }{C_F^2}\cos (\delta  z_1 \Delta  E^-(k - x p - (1-x) q_1 - (1-x) q_2))
\Bigg] \, .
\end{align}
\end{subequations}

The exact results Eqs.~\eqref{e:NLO1} and \eqref{e:NLOexact} at second order in the opacity expansion are the second major result of this work, presented here for the first time.  Although there is no existing calculation in the literature against which this exact result can be compared, we can compare with Ref.~\cite{Gyulassy:2000er} under the broad source and small-x approximations, obtaining
\begin{align}
x p^+ & \left. \frac{dN}{d^2 k \, dx \, d^2 p \, dp^+} \right|_{\ord{\chi^2}} 
\overset{(x \ll 1)}{\approx} \frac{C_F}{(2\pi)^3 (1-x)} 
\int\limits_{x_0^+}^{R^+} \frac{dz_2^+}{\lambda_g^+} 
\int\limits_{x_0^+}^{z_2^+} \frac{dz_1^+}{\lambda_g^+} 
\int\frac{d^2 q_1}{\sigma_{el}} \frac{d^2 q_2}{\sigma_{el}} \:
\frac{d\sigma^{el}}{d^2 q_1} \frac{d\sigma^{el}}{d^2 q_2} 
\times \frac{1}{\mu^2} \, \left( p^+ \frac{dN_0}{dp^+} \right)
\notag \\ &\times \Bigg\{
\left[ 
\psi(\ul{k}) \, \psi^* (\ul{k} - \ul{q}_1) 
- \left| \psi(\ul{k} - \ul{q}_1) \right|^2 
\right]
\left[ 
1 
- \cos \left( \Delta E^- (\ul{k} - \ul{q}_1) \delta z_1 \right) 
\right]
\notag \\ &+
\left[ 
\psi(\ul{k}) \, \psi^* (\ul{k} - \ul{q}_2) 
- \left| \psi(\ul{k} - \ul{q}_2) \right|^2 
\right]
\left[ 
\cos \left( \Delta E^- (\ul{k} - \ul{q}_2) \delta z_2 \right) 
- \cos \left( \Delta E^- (\ul{k} - \ul{q}_2) (\delta z_1 + \delta z_2) \right) 
\right]
\notag \\ &-
\left[ 
\psi(\ul{k} - \ul{q}_2) \, \psi^* (\ul{k} - \ul{q}_1 - \ul{q}_2) 
- \left| \psi(\ul{k} - \ul{q}_1 - \ul{q}_2) \right|^2 
\right]
\left[
1
- \cos\left( \Delta E^-(\ul{k} - \ul{q}_1 - \ul{q}_2) \delta z_1 \right)
\right]
\notag \\ &-
\left[ 
\psi(\ul{k}) \, \psi^* (\ul{k} - \ul{q}_1 - \ul{q}_2) 
- \psi(\ul{k} - \ul{q}_2) \, \psi^* (\ul{k} - \ul{q}_1 - \ul{q}_2) 
\right]
\left[
\cos\left( \Delta E^-(\ul{k} - \ul{q}_2) \delta z_2 \right) \right.
\notag \\ & \hspace{1cm}
- \left. \cos\left( \Delta E^-(\ul{k} - \ul{q}_1 - \ul{q}_2) \delta z_1
+ \Delta E^- (\ul{k} - \ul{q}_2) \delta z_2 \right)
\right] \Bigg\} \, ,
\end{align}
where we have again introduced the gluon mean free path $\lambda_g^+ = \frac{C_F}{N_c} \lambda^+$.  Introducing the notation
\begin{subequations}
\begin{align}
\ul{C}_1 &\equiv \frac{\ul{k} - \ul{q}_1}{(k - q_1)_T^2}  \, ,  \\
\ul{C}_2 &\equiv \frac{\ul{k} - \ul{q}_2}{(k - q_2)_T^2}  \, , \\
\ul{C}_{(12)} &\equiv \frac{\ul{k} - \ul{q}_1 - \ul{q}_2}{(k - q_1 - q_2)_T^2}  \, , \\
\ul{B}_1 &\equiv \frac{\ul{k}}{k_T^2} - \frac{\ul{k} - \ul{q}_1}{(k - q_1)_T^2}  \, , \\
\ul{B}_2 &\equiv \frac{\ul{k}}{k_T^2} - \frac{\ul{k} - \ul{q}_2}{(k - q_2)_T^2}  \, , \\
\ul{B}_{2(12)} &\equiv \frac{\ul{k} - \ul{q}_2}{(k - q_2)_T^2} - \frac{\ul{k} - \ul{q}_1 - \ul{q}_2}{(k - q_1 - q_2)_T^2}  \, , \\
\omega_1 &\equiv \frac{(k - q_1)_T^2}{2 x p^+} = - \Delta E^- (\ul{k} - \ul{q}_1) \, ,  \\
\omega_2 &\equiv \frac{(k - q_2)_T^2}{2 x p^+} = - \Delta E^- (\ul{k} - \ul{q}_2)  \, , \\
\omega_{(12)} &\equiv \frac{(k - q_1 - q_2)_T^2}{2 x p^+} = - \Delta E^- (\ul{k} - \ul{q}_1 - \ul{q}_2) \, , 
\end{align}
\end{subequations}
and the small-$x$ limit of the wave functions Eq.~\eqref{e:LFWFsq} gives
\begin{align} \label{e:NLOsmallx}
x p^+ & \left. \frac{dN}{d^2 k \, dx \, d^2 p \, dp^+} \right|_{\ord{\chi^2}} 
\overset{(x \ll 1)}{\approx} \frac{\alpha_s C_F}{\pi^2} 
\int\limits_{x_0^+}^{R^+} \frac{dz_2^+}{\lambda_g^+} 
\int\limits_{x_0^+}^{z_2^+} \frac{dz_1^+}{\lambda_g^+} 
\int\frac{d^2 q_1}{\sigma_{el}} \frac{d^2 q_2}{\sigma_{el}} \:
\frac{d\sigma^{el}}{d^2 q_1} \frac{d\sigma^{el}}{d^2 q_2} 
\times \frac{1}{\mu^2} \, \left( p^+ \frac{dN_0}{dp^+} \right)
\notag \\ &\times \Bigg\{
2(\ul{B}_1 \cdot \ul{C}_1)
\left[ 1 - \cos \left( \omega_1\delta z_1 \right) \right]
+ 2(\ul{B}_2 \cdot \ul{C}_2)
\left[ \cos \left( \omega_2 \delta z_2 \right) - 
\cos \left( \omega_2 (\delta z_1 + \delta z_2) \right) \right]
\notag \\ &-
2(\ul{B}_{2(12)} \cdot \ul{C}_{(12)})
\left[ 1 - \cos\left( \omega_{(12)} \delta z_1 \right) \right]
- 2(\ul{B}_{2} \cdot \ul{C}_{(12)})
\left[
\cos\left( \omega_2 \delta z_2 \right) \right.
- \left. \cos\left( \omega_{(12)} \delta z_1 + \omega_2 \delta z_2 \right)
\right] \Bigg\} \, ,
\end{align}
in perfect agreement with Eq.~(F8) of Ref.~\cite{Gyulassy:2000er}.  The agreement of  Eq.~\eqref{e:NLOsmallx} with the literature is an important cross-check, because the second-order result is a test of all the elements of the reaction operator Eq.~\eqref{e:reactker}.  Although this validation of the second-order result only holds under the small-$x$ and broad source approximations, together with the validation Eq.~\eqref{e:LOexact} of the first-order result under exact kinematics it represents a fairly robust test of the full recursion relations Eqs.~\eqref{e:reactmtx}.

%
\section{Outlook and Conclusions}
\label{sec:Concl}
%

In the past several years, the QCD community has taken important steps toward the realization of  an electron-ion collider in the near future. For this endeavor to be successful, the scientific potential of the new facility must be fully explored. One area that can benefit from new developments is the investigation of cold nuclear effects on hadron and jet production in semi-inclusive deep inelastic scattering.  A dedicated program at an EIC can shed new light on the transport properties of large nuclei across  different energy regimes and complement the extensive studies of jet quenching and hadronization in relativistic heavy ion collisions.   It can also unify communities and directions of research in high energy nuclear physics that have so far followed separate paths.  

To this end, we set out to calculate the quark branching (gluon emission off of an energetic quark) beyond the soft gluon approximation and to an arbitrary order in the correlation between the multiple interactions of the parton system in the medium (opacity).  To complete this task required the development of a new theoretical  framework which is easily generalizable to all in-medium splitting  processes. In this work, we established the new formalism by  constructing a new recursion relation Eq.~\eqref{e:reactmtx} and Eq.~\eqref{e:reactker} in the opacity expansion approach to jet-medium interactions.  This equation, which has been derived using only the eikonal approximation to the medium scattering, can be used to calculate the medium modification of jet substructure to any finite order in opacity.  The fundamental ingredients, the light-front wave functions Eq.~\eqref{e:LFWF} and Eq.~\eqref{e:LFWFsq} and energy denominators Eq.~\eqref{e:Edenom1}, are universal and easily generalized to other partonic splitting processes or to include mass effects.   With this recursion relation at hand,  it is straightforward (although cumbersome) to calculate the medium-induced parton splitting kernel input to  jet substructure observables to any finite order in opacity with exact kinematics.

We have also applied this recursion relation to calculate the exact gluon-in-quark-jet distribution at second order in opacity Eq.~\eqref{e:NLO1} and  Eq.~\eqref{e:NLOexact}, presented here for the first time.  The validation of this result against the known soft-gluon limit, together with multiple validations of the exact first order in opacity result~Eq.~\eqref{e:LOexact}, comprises a fairly robust test of the kernel Eq.~\eqref{e:reactker}.  These new results are fully ready to be incorporated into existing phenomenology \cite{Kang:2014xsa,Chien:2015vja,Chien:2015hda,Chien:2016led,Kang:2016ofv,Kang:2017frl,Li:2017wwc} for calculations ranging from inclusive light and  heavy hadron production to jet substructure. These second-order-in-opacity results beyond the soft-gluon approximation may improve these calculations in significant ways, especially for  more differential observables such as the production of hadrons inside jets  \cite{Chien:2015ctp,Kang:2016ehg},  by potentially leading to a smoother behavior of the splitting kernels as a function of the kinematic variables and reducing the regions of phase space where the distributions can be negative at any order in opacity~\cite{Gyulassy:2000er}.  While undeniably Eqs.~\eqref{e:NLO1} and~\eqref{e:NLOexact} are quite lengthy, nevertheless their incorporation in numerical code is straightforwad. We defer this study to future work but note that our preliminary results are very encouraging.  While these methods rely on the ability to truncate the opacity expansion at finite order,  the particularly simple triangular structure of the recursion kernel in Eq.~\eqref{e:reactmtx} makes it possible to decouple the system of equations into 4 independent difference (differential) equations.  In future work, we will attempt to solve this system sequentially to obtain a fully resummed expression for the jet substructure distributions.  If successful, this approach will be valid for any value of the opacity. 

It will also be interesting to study numerically the approach of these distributions to the known analytic limits.  For instance, the soft-gluon limit Eq.~\eqref{e:NLOsmallx} of the second-order in opacity distribution is tremendously more compact than the full result Eq.~\eqref{e:NLOexact}; a systematic study of the error involved in making such an approximation could be used to increase calculational efficiency.  Similarly, by systematically extending the opacity expansion to higher orders, it will be possible to explore the matching onto the continuous path-integral descriptions of the medium interactions.  Such a study could yield quantifiable, reliable comparisons of the different descriptions of the medium, along with a better understanding of the boundary between the region of applicability of finite-opacity versus resummed methods.

Looking into the future, one might consider the possibility of evaluating $1 \rightarrow 3$ splitting processes, such as two gluon emission, to higher orders in opacity beyond what was achieved in Ref.~\cite{Fickinger:2013xwa}.  While the calculation will necessarily be technically involved, we do not foresee conceptual difficulties. However,  at present, there are no known exact applications of   $1 \rightarrow 3 $ branchings in heavy ion collisions and cascades of the lowest order binary splitting processes have served as a decent approximation. In light of this, it will be prudent to first focus on the development of $\ord{\alpha_s^2}$ parton splitting phenomenology, prior to attacking the higher orders in opacity derivation.
 
Finally, coming back to our original motivation, it will be important to incorporate these results into the detailed phenomenology of cold nuclear matter effects expected at a EIC.  The use of jet tomography as a probe of QCD matter has been rightly studied in detail for use in heavy-ion collisions, but the jet component of the science program for a future EIC has not received nearly as much attention.  In spite of this, the formalism for the how a jet couples to a QCD medium is universal, and the architecture developed for the heavy-ion program at RHIC and the LHC has the potential to add considerable strength to an EIC science program.  Because of this, we believe the new results presented here constitute an important advance in the study of jets in both hot and cold nuclear environments.


%
\section*{Acknowledgments}
M.S. wishes to thank Mauricio Martinez Guerrero for useful discussions.  This material is based upon work supported by
the U.S. Department of Energy, Office of Science, Office of Nuclear
Physics under DOE Contract No. DE-SC0012704 and the DOE Early Career Program. \\
%


\providecommand{\href}[2]{#2}\begingroup\raggedright\endgroup

\end{document}